\documentclass{article}

\RequirePackage{amsthm,amsmath}
\usepackage{amsfonts}

\usepackage{graphicx}
\usepackage{color}
\usepackage{amsfonts}

\begin{document}

 \large

\newcommand{\al}{\mbox{$\alpha$}}
\newcommand{\be}{\mbox{$\beta$}}
\newcommand{\ep}{\mbox{$\epsilon$}}
\newcommand{\gam}{\mbox{$\gamma$}}
\newcommand{\sig}{\mbox{$\sigma$}}

\DeclareRobustCommand{\FIN}{%
  \ifmmode 
  \else \leavevmode\unskip\penalty9999 \hbox{}\nobreak\hfill
  \fi
  $\bullet$ \vspace{5mm}}

\newcommand{\calA}{\mbox{${\cal A}$}}
\newcommand{\calB}{\mbox{${\cal B}$}}
\newcommand{\calC}{\mbox{${\cal C}$}}

\newcommand{\muas}{\mbox{$\mu$-a.s.}}
\newcommand{\Nat}{\mbox{$\mathbb{N}$}}
\newcommand{\Rea}{\mbox{$\mathbb{R}$}}
\newcommand{\Prob}{\mbox{$\mathbf{P}$}}

\newcommand{\nin}{\mbox{$n \in \mathbb{N}$}}
\newcommand{\suc}{\mbox{$\{X_{n}\}$}}
\newcommand{\sucP}{\mbox{$\mathbf{P}_{n}\}$}}

\newcommand{\Pd}{\mbox{$\mathcal{P}_2(\mathbb{R}^d)$} }
\newcommand{\subPd}{\mbox{$_{\mathcal{P}_2(\mathbb{R}^d)}$} }
\newcommand{\Wd}{\mbox{$\mathcal{W}_2$}}
\newcommand{\Wdd}{\mbox{$\mathcal{W}_2^2$}}
\newcommand{\Read}{\mbox{$ \mathbb{R}^d$} }
\newcommand{\Exp}{\mbox{$ \mathbf{E}$}}
\newcommand{\muo}{\mbox{$ \mu_\omega $}}

\newcommand{\Aal}{\mbox{$ {\cal T}^\alpha_\mathbf{P} $}}
\newcommand{\calX}{\mbox{$ {\cal X}$}}
\newcommand{\calQ}{\mbox{$ {\cal Q}$}}
\newcommand{\calQX}{\mbox{$ {\cal Q}^{\cal X}$}}
\newcommand{\cBall}{\mbox{$\overline{B}_\mathcal{W}($}}
\newcommand{\nuP}{\mbox{$\nu_{\mathbf{P}}$}}
\newcommand{\nuPn}{\mbox{$\nu_n^\omega$}}
\newcommand{\Hvec}{\mbox{\bf H}}

\newcommand{\conv}{\rightarrow}
\newcommand{\convn}{\rightarrow_{n\rightarrow \infty}}
\newcommand{\convp}{\rightarrow_{\mbox{c.p.}}}
\newcommand{\convs}{\rightarrow_{\mbox{a.s.}}}
\newcommand{\convw}{\rightarrow_w}
\newcommand{\convd}{\stackrel{\cal D}{\rightarrow}}

\newtheorem {Prop}{Proposition} [section]
 \newtheorem {Lemm}[Prop] {Lemma}
 \newtheorem {Theo}[Prop]{Theorem}
 \newtheorem {Coro}[Prop] {Corollary}
 \newtheorem {Nota}[Prop]{Remark}
 \newtheorem {Ejem}[Prop] {Example}
 \newtheorem {Defi}[Prop]{Definition}
 \newtheorem {Figu}[Prop]{Figure}
 \newtheorem {Tabla}[Prop]{Table}

\title{\sc  Robust clustering tools based on optimal transportation.\footnote{Research partially supported by the
Spanish Ministerio de Econom\'{\i}a y Competitividad and FEDER, grants  
MTM2014-56235-C2-1-P, MTM2014-56235-C2-2, and  by Consejer\'{\i}a de Educaci\'on de la Junta de Castilla y Le\'on, grant VA212U13.}}

\author{E. del Barrio$^{1}$, J.A. Cuesta-Albertos$^{2}$, C. Matr\'an$^{1}$ \\ and A. Mayo-\'Iscar$^{1}$ \\
$^{1}$\textit{Departamento de Estad\'{\i}stica e Investigaci\'on Operativa and IMUVA,}\\
\textit{Universidad de Valladolid} \\ $^{2}$ \textit{Departamento de
Matem\'{a}ticas, Estad\'{\i}stica y Computaci\'{o}n,}\\
\textit{Universidad de Cantabria}}
\maketitle

\begin{abstract}
A robust clustering method for probabilities in Wasserstein 
space is introduced. This new `trimmed $k$-barycenters' approach relies on recent 
results on barycenters in Wasserstein space that allow intensive computation, 
as required by clustering algorithms. The possibility of trimming the most discrepant 
distributions results in a gain in stability and robustness, highly convenient in this 
setting. As a remarkable application we consider a parallelized estimation setup in which each of $m$ units processes a portion of the data, producing an estimate of $k$-features, encoded as $k$ probabilities. We prove that the trimmed $k$-barycenter of the $m\times k$ estimates produces a consistent aggregation. We illustrate
the methodology with simulated and real data examples. These include clustering populations by age distributions
and  analysis of cytometric data.
\end{abstract}

\noindent {\small \textsc{AMS Subject Classification:} Primary:  62H30, 62G35. Secondary: 62G20, 62P99. }

\noindent {\small \textsc{Keywords:} Cluster prototypes, $k$-barycenter, trimmed barycenter,  
robust aggregation, Wasserstein distance, Monge-Kantorovich problem, transport maps, 
trimmed distributions, parallelized inference, bragging, subragging, trimmed $k$-means algorithm.}

\section{Introduction.}

Cluster Analysis belongs to the class of statistical procedures which are  most required by practitioners. Even being simple to describe its scope, the fine details involved in the admissible shapes for clusters, the determination of the number of clusters and the habitual unexistence of exact algorithms to get the solution to relatively basic problems are    difficulties intrinsic to the theory. But today, the enormous sizes of data sets and the increasing interest in structured  complex data have also increased the interest and the inherent  difficulties of the theory. On the whole, these facts make cluster analysis a challenging theory demanding new tools for the statistical analysis (see e.g. Hennig et al. \cite{Hennig} for the current state of the art and a panoramic view of the theory). In particular, data available as 
probability distributions are the focus of several disciplines such as demography or 
weather forecasting. In such cases, the natural space to describe or analyze the data must 
preserve their intrinsic structure, leading to consider abstract spaces where the elements are probability distributions. Although this would suffice to justify developing cluster analysis on these spaces, meta-analysis or aggregation of cluster analyses as well as parallelization of some cluster procedures can be also addressed from such a perspective, giving an unexplored added value to the theory. In fact, a main goal of this paper is to provide some  tools for clustering in Euclidean spaces that arise by resorting to clustering in suitable metric spaces constituted by probability distributions. 

By the way, $k$-means is an unquestionable reference in the clustering framework  (see e.g. \cite{Hennig} to get a general perspective, in particular  chapter 3 by B. Mirkin and chapter 5 by P. Awasthi and M.F. Balcan), being  the simplest partitional  clustering procedure generalizable to metric spaces.  
$k$-means in abstract spaces have been considered in Sverdrup-Thygeson \cite{Sverdrup}, 
Cuesta-Albertos and Matr\'an \cite{Cues88}, P\"arna \cite{Parna2} and \cite{Parna}, 
Luschgy and Pag\`es \cite{Luschgy} or Lember \cite{Lember}, although 
their settings do not properly cover the current objective. For a general metric space $(E,d)$,  a $k$-mean or $k$-barycenter of the 
points $e_1,\dots,e_n$ in $E,$  would be any set $\{\bar m_1,\dots,\bar m_k\}\subset E$ verifying 
\begin{equation}\label{preintroduccion}
\frac 1 {n}\sum_{i=1}^n \min_{j \in \{1,\dots,k\}} d^2(e_i,\bar m_j) \leq \frac 1 {n}\sum_{i=1}^n \min_{j \in \{1,\dots,k\}} d^2(e_i, m_j) \mbox{ for every set } \{ m_1,\dots,m_k\}\subset E.
\end{equation}
 The 
$k$-means procedure shares many of the
merits of the mean, as well as its drawbacks,   showing a very bad behaviour in presence of outliers or even 
bridge-points between clusters (see Cuesta-Albertos 
et al. \cite{Cuesta97}). Trimmed $k$-means were introduced in  \cite{Cuesta97} in Euclidean spaces as a way of robustifying $k$-means.
Given a trimming level $\alpha \in (0,1)$ and the set of $n$ 
points $e_1,\dots,e_n$ in $E$, trimmed $k$-means (see Definition \ref{trimmedbary}) 
look for a set $\{\bar m_1,\dots,\bar m_k\}\subset E$ and a partition, $\mathcal C_0, \mathcal C_1,\dots,\mathcal C_k$, of $E$ 
that mimimize the \textit{trimmed dispersion}, namely,
\begin{equation}\label{introduccion}
\frac 1 {n-[n\alpha]}\sum_{j=1}^k \sum_{e_i \in \mathcal C_j}d^2(e_i,\bar m_j),
\end{equation}
where $\mathcal C_0$, the set of trimmed elements, has  $[n\alpha]$ elements.
A minimizing set $\{\bar m_1,\dots,\bar m_k\}$ will be called a \textit{trimmed $k$-barycenter} of $e_1,\dots,e_n$. 
We note that trimmed $k$-means in linear functional spaces have been considered in Cuesta-Albertos and Fraiman \cite{Cuesta-Fraiman} 
and in Garc\'{\i}a-Escudero and Gordaliza \cite{splines}. Our framework, in contrast, focuses on the case where each of $m$ units, possibly after some preprocessing, 
gives an estimated distribution, resulting in a meta-sample 
$\hat P_1,\dots,\hat P_m$ of distributions or sets of distributions. These objects naturally live in $\Pd$, the set of 
probabilities on $\mathbb{R}^d$ with finite second moment, which we endow with the $L_2$-{\it Wasserstein  distance},
defined by 
\begin{equation} \label{int1}
\Wd (P,Q):=\inf \left\{\left(\Exp\|U-V\|^2\right)^{1/2}: \  \mathcal{L}(U)=P, \  \mathcal{ L}(V)=Q\right\},
\end{equation}
where we use $\mathcal{L}(X)$ to denote the distribution law of a r.v. $X$ and $\Exp(X)$ for its mean.

In this work we consider $k$-barycenters and trimmed $k$-barycenters in this 
metric space, the $L_2-$Wasserstein space. This space has deserved notable interest by its connection with the  celebrated 
Monge-Kantorovich transport problem, and provides a suitable setting for statistical analyses of structured data such 
as histograms, density functions or probabilities. For $k=1$, existence, uniqueness and characterizations of (1-)barycenters (or Fr\'echet means) in this space
have been considered in Agueh and Carlier \cite{Agueh10}, while 
consistency results can be found in Le Gouic and Loubes \cite{Le Gouic} 
or Bigot and Klein \cite{Bigot} and trimmed barycenters have been introduced in \'Alvarez-Esteban et al  \cite{preprint2}.
Some recent additional references that resort to Wasserstein spaces with an statistical
motivation are \cite{Tasio}, \cite{Carlier2}, \cite{Depth}.
 Being of indubitable mathematical interest, this approach could be considered just as the introduction of an additional abstract space where clustering of probabilities could be carried with more or less success. However, as already announced, through the paper we will stress mainly on other applications of the theory: We will address the consensus or aggregation of model-based cluster analyses through a kind of cluster prototypes, where the prototypes are probability distributions. 

Focusing on the aggregation of structured data, there can hardly be any doubt 
(beyond its lack of robustness) about the good properties of the mean as a summary 
of a set of elements in a linear space. However, when we move to a shaped space 
the use of the mean can be unfeasible or produce undesirable effects (for example, the point-wise mean of several normal densities is no longer a normal density and the normal shape could be largely distorted). 
This kind of problem has 
been already pointed out in the statistical setting by several authors, notably by 
Kneip and Gasser \cite{Gasser}, and more recently, proposing solutions based on 
Wasserstein spaces by Boissard et al. \cite{Bois15}.

Consensus procedures in Cluster Analysis have a long story, but generally suffer 
from the lack of feasible ways to measure the similarity between several clustering proposals 
(see e.g. ``A survey of Consensus Clustering" by Ghosh and Acharya in \cite{Hennig}). The use of the $L_2$-Wasserstein distance allows 
to measure in a non-heuristic way how close the consensus is to each base 
solution. Also, the trimmed nature of our  aggregation process provides 
robustness to the proposal. Trimming procedures have always been 
at the core of Robust Statistics as an easily understandable way of limiting 
the effects of (in some sense) extreme observations.


A main difficulty avoiding a broader use of Wasserstein metrics in applications is of 
computational nature. Computation of Wasserstein distances and barycenters for 
multivariate probabilities can be a hard task, although the efforts made in the last years predict better future. Cuturi and Doucet \cite{Cuturi}, Benamou et al 
\cite{Benamou}, Carlier, Oberman and Oudet \cite{Carlier} or Anderes, Borgward and Miller \cite{Anderes} develop optimization 
procedures for such a hard goal that often involve intensive computation and are rather time-consuming,
a major drawback if one is planning to use these procedures for a distance based
clustering methodology. However, when the probabilities share a common shape, that is, 
when they belong to a location-scatter family, as it frequently happens in model-based 
clustering,  we can now resort to a very efficient algorithm introduced in \'Alvarez-Esteban et al. 
\cite{preprint} to compute barycenters. This will result in a feasible computation of $k$-barycenters 
as well as the robust trimmed $k$-barycenters of our proposal in general dimension. 

Here we explore the use of trimmed $k$-barycenters in the Wasserstein space, providing theoretical support 
as well as showing the feasibility and suitability of the approach to get an understandable clusterization.  
As our main goal, we will consider the adaptation of trimmed $k$-barycenters to allow 
aggregation of clusters. The idea is simple. Often clustering procedures are related to shapes 
that allow the interpretation in terms of, say, Gaussian distributions. Thus, under a well-clusterized 
problem, each unit would produce $k$ Gaussian distributions, having, as a result, a total of $m \times k$ Gaussian
distributions. Now, these can be naturally clustered in Wasserstein space into $k$ groups, and 
the $k$ probabilities obtained through the trimmed $k$-barycenter (which, remarkably, will 
also be Gaussian) would be the consensus representation of the reports of the different units.
{We give a result on the consistency of this procedure (see Theorem \ref{consist2}). We stress that 
this seems to be the first consistency result on parallelization in the $k$-means setting, 
being valid in very general metric spaces.} The procedure allows adaptations to cover other settings, 
but to simplify the exposition we will consider the aggregation just in a parallelized setup.

The remaining sections of this paper are organized as follows. Section 2 introduces $k$-barycenters and trimmed $k$-barycenters in
Wasserstein space, including relevant results about existence, consistency and error bounds for aggregation based on
these trimmed $k$-barycenters. Computational issues are discussed in Section 3. It is well known that in Euclidean spaces, the usual $k$-means type algorithms produce iterations that converge to some stationary point, which coupled with a moderate number of random starts provide probabilistic guarantee of convergence to the global minimizer. We prove that the geometry of Wasserstein spaces allows to get the same conclusion in this setup. A real data application to 
population clustering by age distributions is described in Section 4. Section 5 discusses several applications of 
trimmed $k$-barycenters to model-based clustering. It covers aggregation issues, like parallelization and resampling procedures, but it also includes improvements on initialization steps for clustering algorithms and exploratory tools, showed on a troubling data set of cytometries. Finally, we include an Appendix with a short account of some 
relevant facts related to Wasserstein spaces as well as proofs for
the main results in this paper.

\section{Trimmed $k$-barycenters in Wasserstein space}

In this section we present some relevant results about $k$-barycenters and trimmed $k$-barycen\-ters in Wasserstein space.
Recall that it is the space \Pd of probabilities on $\mathbb{R}^d$ with finite second moment equipped with the metric 
$\mathcal{W}_2$ defined in (\ref{int1}). This is a complete and separable metric space. Further details about it can 
be found in the Appendix. For convenience we consider a general setup generalizing  the one in 
 (\ref{preintroduccion}), based on a sample distribution $\mu$ (giving uniform mass to $\{e_1,\dots,e_n\}$), and consider (Borel) probabilities, $\mu$, over $\Pd$ such that
\begin{equation}\label{W2space}
\int_{\Pd}\mathcal{W}_2^2(P,Q)\mu(dP) < \infty, \ \mbox{ for some (hence, for every) } Q \in \Pd.
\end{equation}
We write $W_2(\Pd)$ for the set of such $\mu$'s. Now, 
a $k$-barycenter of $\mu\in W_2(\Pd)$ is a set $\{\bar M_1,\dots,\bar M_k\} \subset \Pd $ such that
\begin{equation}\label{kbar1}
\int \min_{i=1,\dots,k}\Wdd(P,\bar M_i)\mu(dP)\leq \int \min_{i=1,\dots,k}\Wdd(P,Q_i)\mu(dP) 
\end{equation}
for any set $\{Q_1,\dots,Q_k\} \subset \Pd$. Existence of a $k$-barycenter for any $\mu\in W_2(\Pd)$ 
is proved in Theorem \ref{existkbar} in the Appendix. In fact, (\ref{k-var}) and (\ref{k-bar}) there
show that the minimal value (the left-hand side) in (\ref{kbar1}), to be denoted by $V_k(\mu)$ in the sequel, has the 
meaning of a dispersion measure with respect to an optimal $k$-set. Simple conditions 
guaranteeing uniqueness of $k$-barycenters are not available even for distributions 
on the real line. In contrast, uniqueness of $k$-barycenters is often used as a natural 
assumption to state consistency results. Alternatively, consistency results are sometimes 
stated through limit points of convergent subsequences, as in \cite{Le Gouic}. Under this type of assumption
we prove in Theorem \ref{consistencykbar} consistency of $k$-barycenters.  

To introduce a trimmed version of the $k$-barycenter we proceed as in \cite{Cuesta97} 
(see also \cite{preprint2} for trimmed barycenters in Wasserstein space)
and consider the following abstract definition.
Given a level $0\leq\alpha\leq 1$ and a probability $P$  on a measurable space $(\Omega,\sigma)$,  
a probability $P^*$ on $\sigma$, is an $\alpha$-trimming of $P$ if there exists a measurable 
function $\tau:\Omega \to \Rea$ such that $0\leq \tau(\omega) \leq 1$ for every $\omega \in \Omega$ 
and $P^*(A) =\frac 1 {1-\alpha} \int_A \tau(\omega)P(d\omega)$ for every $A\in \sigma$. 
{Such a} function is often called an $\alpha$-trimming function.
In the sequel, the set of all $\alpha$-trimmings of $P$  will be denoted by ${\cal T}_\alpha (P)$. 
Note that a hard $0-1$ trimming of a probability $P$ (the case when $\tau\in\{0,1\}$) corresponds to 
the conditional probability $P$  given a set $A$, with $P(A)= 1-\alpha$, but we are also including 
the possibility of partial trimming of any point $\omega$ through the trimming function $\tau(\omega)\in[0,1]$. 
Also note that no-trimming  is included in our definition as a trimming function (take $\tau\equiv 1-\alpha$), 
thus $P$ is a trimmed version of itself for any $\alpha$. 
We can now define trimmed $k$-barycenters in Wasserstein space.
\begin{Defi}\label{trimmedbary}
An ($\alpha$-)trimmed $k$-barycenter of $\mu\in W_2(\Pd)$ is any set 
$ {\bf \bar M}=\{\bar M_1,\dots,\bar M_k\}\subset \Pd $ such that for some $\mu^\alpha\in {\cal T}_\alpha(\mu):$
\begin{eqnarray}\label{kbar}
V_{k,\alpha}(\mu)&:=&\int \min_{i\in\{1,\dots,k\}}\Wdd(P,\bar M_i)\mu^\alpha(dP)\\
\nonumber &=& \inf\left\{\int \min_{i\in\{1,\dots,k\}}\Wdd(P,Q_i)\mu^*(dP) : \ \ \mu^*\in {\cal T}_\alpha (\mu),\{Q_1,\dots,Q_k\} \subset \Pd \right\}.
\end{eqnarray}
\end{Defi}
The related trimmed probability, $\mu^\alpha$, will be called an \textit{optimally trimmed probability} and its corresponding 
trimming function, $\tau_{\mu}^\alpha$, an \textit{optimal trimming function} (associated to  $ {\bf \bar M}$). 
We will also refer to the minimum value in (\ref{kbar}), $V_{k,\alpha}(\mu)$, as the ($\alpha$-)trimmed $k$-variation of $\mu$. 
Note that in terms of trimming functions we have
\begin{eqnarray}\label{kbar2}
V_{k,\alpha}(\mu)=\frac 1 {1-\alpha}\int \min_{i\in\{1,\dots,k\}}\Wdd(P,\bar M_i)\tau_{\mu}^\alpha(P)\mu(dP).
\end{eqnarray}
We collect in the following proposition some main facts concerning  trimmed $k$-barycen\-ters, including 
existence and a simple characterization of optimal trimming functions.
Existence arises from easy modifications to the arguments in \cite{preprint2} to prove existence 
of trimmed barycen\-ters once existence of barycenters is known. In this case we need the support of the result concerning  existence of 
$k$-barycenters, that we include as Theorem \ref{existkbar} in the Appendix. All other
claims can be proved following available proofs for trimmed $k$-means (in \cite{Cuesta97}) 
with minor changes (as in the proofs in \cite{preprint2} for trimmed barycenters).
For a nicer statement we denote the (generalized) open ball in $\Pd$ centered at ${\bf M}=\{M_1,\ldots,M_k\}$ by 
$B({\bf M},r):=\cup_{i=1}^kB(M_i,r),$ (and write $\overline B({\bf M},r)$ for its closure) and set
\begin{eqnarray*}
r_\alpha ({\bf M})=\inf \left\{ r\geq 0 : \ \mu(B({\bf M},r))\leq 1-\alpha \leq \mu(
\overline{B}({\bf M},r))\right\}.
\end{eqnarray*}

\begin{Prop}\label{caract}
Given $\alpha\in (0,1)$ and $\mu\in W_2(\Pd)$, there exists a trimmed $k$-barycenter, 
$ {\bf \bar M}=\{\bar M_1,\dots,\bar M_k\}$ of $\mu$. Furthermore,
\begin{itemize}
\item[i)]
If $\tau_\mu^\alpha$ is an  optimal trimming function associated to $ {\bf \bar M}$, 
then $$I_{B({\bf \bar{M}},r_\alpha ({\bf \bar{M}}))}\leq
\tau_\mu^\alpha \leq I_{\overline{B}({\bf \bar{M}},r_\alpha ({\bf \bar{M}}))}\mbox{ }\mu- \mbox{a.s.}$$
\item[ii)]
If $0<\alpha<\beta<1$, then $V_{k,\alpha}(\mu)\geq V_{k,\beta}(\mu)$, with equality if and only if there 
is a common solution, $ {\bf \bar M}$, to the $\alpha$ and $\beta$ trimmed $k$-barycenter problems, 
that additionally should satisfy $r_\alpha ({\bf \bar{M}})=r_\beta ({\bf \bar{M}})$ and $\mu[\bar{B}({\bf M},r_\alpha ({\bf M}))^C]=0.$ 
\item[iii)]
$V_{k,\alpha}(\mu)\geq V_{k+1,\alpha}(\mu)$ and the inequality is strict unless  $V_{k,\alpha}(\mu)=0$.
\end{itemize}
\end{Prop}
Note that the consideration of trimmed probabilities allows to  guarantee the existence of trimmed $k$-barycenters without 
any integrability condition on $\mu$ (which would be necessary to guarantee existence of $k$-barycenters). 
We note also that Proposition \ref{caract} is the key link between quantizers (the trimmed $k$-barycenters)  
and clustering. Item i) establishes that an optimal trimming function,  say $\tau_\mu^\alpha$, associated to a 
trimmed $k$-barycenter, $\bar{\bf M}$, is essentially an indicator set of the union of $k$ balls with the same radii. 
Moreover, the $k$-set $\bar{\bf M}$ induces a partition of the set $\overline{B}(\bar{\bf M},r_\alpha (\bar{\bf M}))$ 
into $k$ clusters, $C_1,\dots,C_k$, with $C_i$ consisting of those probabilities in $\overline{B}(\bar{\bf M},r_\alpha 
(\bar{\bf M}))$ which are closer to $\bar M_i$, $i=1,\dots,k$ (probabilities equidistant to several $M_i$'s can be 
arbitrarily assigned without changing the value $V_{k,\alpha}(\mu)$). This induces a decomposition 
of the trimmed $k$-variation of $\mu$ as
\begin{equation}\label{descomp}
V_{k,\alpha}(\mu)=\frac 1 {1-\alpha}\sum_{i=1}^k \int_{C_i}\Wdd(P,\bar M_i)\tau_{\mu}^{\alpha}(P)\mu(dP).
\end{equation}
From this expression we see that each $\bar M_i$ in the optimal $k$-set must 
be the  barycenter of its cluster, namely, the barycenter of $\mu^*$, given by $d\mu^*=\frac { \tau_\mu^\alpha}{1-\alpha}d\mu$, 
conditioned to $C_i$. We note also that
items \textit{ii)} and \textit{iii)} establish that, apart from some degenerate cases, the 
$\alpha$-trimmed $k$-variation, $V_{k,\alpha}(\mu)$, decreases by 
increasing $\alpha$ or $k$.

Trimmed $k$-barycenters share the consistency properties of trimmed $k$-means. This is just a  minimal requirement for model-based clustering procedures. A proof of the 
next results can be obtained combining the arguments used in  \cite{preprint2} to prove consistency of 
trimmed barycenters with Theorem \ref{consistency} in the Appendix, about consistency of 
untrimmed $k$-barycenters. Convergence of sets of $k$-barycenters
in the following statements must be understood in the Hausdorff distance,
namely, $$d_H(A,B)=\max\Big(\sup_{P\in A} \inf_{Q\in B} \mathcal{W}_2 (P,Q), \sup_{Q\in B} \inf_{P\in A} \mathcal{W}_2 (P,Q) \Big).$$

\begin{Theo}\label{consistencytypeLoubes}
Assume that $(\mu_n)_n, \mu \in W_2(\Pd)$ are such that  $\mu_n \convw\mu$. 
For a fixed $\alpha \in(0,1)$, let ${\bf \bar M}_n$ be any $\alpha$-trimmed $k$-barycenter 
of $\mu_n$. Then  the  trimmed clusterized variations of $\mu_n$  converge, namely, 
$V_{k,\alpha}(\mu_n)\conv V_{k,\alpha}(\mu),$  the sequence  $({\bf \bar M}_n)_n$ is precompact  
and any limit is a trimmed $k$-barycenter of $\mu$. { If}  
$\mu$ has a unique trimmed $k$-barycenter, ${\bf \bar M}$, then  $d_H({\bf \bar M}_n, {\bf \bar M})\to 0$.
\end{Theo}

\begin{Theo}\label{coro} Consider $\alpha \in (0,1)$ and assume that $\mu\in W_2(\Pd)$ has a unique $\alpha$-trimmed $k$-barycenter, $ {\bf \bar M}$. 
If $\mu_n$ is the empirical measure giving mass $1/n$ to probabilities $P_1,\ldots,P_n$ obtained as independent realizations of $\mu$, 
then the trimmed $k$-barycenters and trimmed clusterized variations are strongly consistent, that is, $V_{k,\alpha}(\mu_n)\conv V_{k,\alpha}(\mu)$ 
a.s. and, if ${\bf \bar M}_n$ is any trimmed $k$-barycenter of $\mu_n$, then $d_H({\bf \bar M}_n,{\bf \bar M})\to 0$ a.s..
\end{Theo}

A main application of trimmed $k$-barycenters in Wasserstein space
concerns aggregation of clustering procedures, either in a parallelization or distributed inference  setup
or through the use of subagging or other resampling  strategies to allow or improve computation.
The parallelization setup refers to the case in which 
data come from $m$ units $U_j$, $j=1,\dots,m$. Each unit processes 
its own samples, that we assume to consist of $n_j$ i.i.d. observations from some $P$.
We write ${\hat P}_j$ for the empirical measure observed by the $j$-th unit. Through the use of
some statistical engine (think of a mixture estimation method, for instance) the $j$-th unit 
produces the $k$-features $\mathbb F(  { P}_j)=\{\hat N_1^j,\dots,\hat N_k^j\}, \ 
j=1,\dots,m$, consisting of $k$ distributions in $\Pd$. We write $\mathbb{F}(P)=\{N_1,\dots,N_k\}$ 
for the $k$-feature associated to $P$. We assume that $\mathbb{F}(P)$ is uniquely defined and
consists of $k$ different elements. We write
\begin{equation}\label{contmodulus}
r(\eta):=\sup_{Q\in B(P,\eta)} d_H(\mathbb{F}(P),\mathbb{F}(Q))
\end{equation}
for the modulus of continuity of $\mathbb{F}$ at $P.$
We assume also that  the involved random elements are defined on the same probability space $(\Omega,\mathcal{F},\mbox{Pr})$. 
Then the following result shows that trimmed $k$-barycenters in Wasserstein space can be used for consistent aggregation of
the $k$-features. To avoid technicallities we consider only the case when $\alpha k m$ is  integer.

\begin{Theo}\label{consist2}
Assume that $\mathbb{F}$, $P$ and $\hat{P}_j$  are as above and that samples from different units are independent. 
Assume further that $\alpha\in (0,\frac 1{2k})$ and $\alpha k m$ is an integer.  Set $H=2\big(1+k\sqrt{\frac{1-\alpha}{1-(k+1)\alpha}} \big)$. 
If $\eta>0$ is such that 
\begin{equation}\label{separationcondition}
r(\eta)<\frac 1 H \min_{i\ne i'} \mathcal{W}_2(N_i,N_{i'}),
\end{equation}
$\mbox{\em Pr}(\mathcal{W}_2(\hat{P}_j,P)>\eta)<\alpha/2$, $j=1,\ldots,k$, and  ${\bf M}_{n_1,\dots,n_m;m}$ is an 
$\alpha$-trimmed $k$-barycenter of the set $\cup_{j=1}^m\{\hat N_1^j,\dots,\hat N_k^j\}$ 
(with equal weights $\frac 1{mk}$), then 
\begin{equation}\label{cotasup}
\mbox{\em Pr}\Big[d_H\left({\bf M}_{n_1,\dots,n_m;m},\{N_1,\dots,N_k\}\right)>{\textstyle \frac{r(\eta)H}{2}} \Big] 
\leq k e^{-\frac {\alpha^2} 2 m},\quad m\geq 1.
\end{equation}
\end{Theo}

We see from Theorem \ref{consist2} that the aggregation procedure based on trimmed $k$-barycen\-ters 
shows a stable behavior with respect to variations in the trimming size and that, in this setup, aggregation does not 
introduce any (asymptotic) bias. We note also that 
$\mathcal{W}_{2}({\hat P}_j , P)\to 0 \mbox{ a.s.},$ as the sample sizes $n_j\to \infty$
(this follows from the strong law of large numbers and (\ref{convergences}), see the Appendix for details). In particular, for any
$\eta>0$,
$\mbox{Pr}[\mathcal{W}_{2}(\mathbb {\hat P}_j , \mathbb P)>\eta]<\frac{\alpha}{2}, \ j=1,\dots, k$,
provided the sample sizes, $n_j$, are not too small. Also, if $\mathbb{F}$ is continuous at $P$ then $r(\eta)\to 0$ as $\eta\to 0$
and we can ensure that (\ref{separationcondition}) holds by taking $\eta$ small enough. 
Condition (\ref{separationcondition}) concerns the degree of separation among the $N_i$ and shows that
consistent aggregation is simpler when the $N_i$'s are well separated. The upper bound 
(\ref{cotasup}) can be made more precise with further assumptions on $P$ and $\mathbb{F}$. For instance, if 
$d\geq 3$ and $P$ satisfies some regularity requirements (see \cite{DobricYukich95}) then $\mathcal{W}_2
(\hat{P}_j,P)=O_{\mbox{\scriptsize Pr}}(n_j^{-1/d})$. Hence, for some constant $C=C(P)$ we have
$\mbox{Pr}[\mathcal{W}_{2}(\mathbb {\hat P}_j , \mathbb P)>CN^{-1/d}]<\frac{\alpha}{2}$ with $N=\min(n_1,\ldots,n_k)$.
If, further, $r(\eta)=O(\eta^{\beta})$ for some $\beta>0$, then, for some constant $\tilde{C}=\tilde{C}(P,\mathbb{F})$ 
we conclude that, with probability
at least $1-ke^{-\frac{\alpha^2}2 m}$,
$$d_H\left({\bf M}_{n_1,\dots,n_m;m},\{N_1,\dots,N_k\}\right)\leq \tilde{C} N^{-\frac \beta d},$$
and we see the influence of the number of units, the sample sizes and the smoothness of $\mathbb{F}$ on
the quality of aggregation.

The scope of Theorem \ref{consist2} can be enlarged assuming that the parent distributions that 
produce the data processed by the units are slightly different. In this case, under continuity
of the statistical engine $\mathbb{F}$, we could assume that the $k$-features associated to the 
parent distributions are within some small, $r_0$, $d_H$-distance from $\{N_1,\ldots,N_k\}$ and 
obtain a similar upper bound with $\frac{r(\eta)H}{2}$ replaced by $\frac{r(\eta)H}{2}+r_0$.

\begin{Nota}\label{pesos}{\em
Often, the statistical engines providing the k-set of features also give associated weights that are necessary to define the clusters (and should be not confused with the weights appearing in the definition of the k-barycenter). This is also the case for the procedure  TCLUST (introduced in \cite{TCLUST}), that we use in the applications in Section \ref{loc-scatsection}. Let us to introduce our proposal for the estimates of the weights associated to the aggregated solution obtained through the trimmed $k$-barycenter.
 
Since our trimming procedure will discard the most discrepant distributions reported by the units, 
 its effect will be notably apparent just for the less clearly defined clusters. This suggests that we should not estimate the weights  merely through the average of the  weights associated to each barycenter from those corresponding to the distributions reported by the units. This assignment would produce an overload effect on the sharpest barycenters. However, we can consider the average rescaled by resorting just to the weights  which rely on untrimmed distributions. 
 More precisely, additionally to the already  introduced notation, let $\pi_i^j$ the weight that unit $U_j$ reported for $\hat N_i^j$ and let ${\bf M}_{n_1,\dots,n_m;m}=\{ \bar N_1, \dots,\bar N_k\}$ be the $\alpha$-trimmed $k$-barycenter  of $\{\hat N_i^j, i=1,\dots,k, \ k=1,\dots,m\}$. For those $\hat N_t^j$ non-trimmed in the trimming process to get $\{ \bar N_1, \dots,\bar N_k\}$,   define  $g^t_j := \arg \min_{s = 1, . . . , k}  {\cal W}_2(\bar N_s, \hat N^i_j )$ and   $\pi_i^*:= \mbox{\bf mean} \{\pi_t^j : g_t^j = i\}, \ i=1,\dots,k.$
 Our estimates for the weights associated to the $k$-barycenters are:
 \begin{equation}\label{pesos}
 \bar \pi_i=\pi_i^*\left(\sum_{r=1}^k \pi_r^*\right)^{-1}, \ i=1,\dots,k.
 \end{equation}
On the basis offered by Theorem \ref{consist2} and  an additional consistency assumption for the weights reported by the procedure, it is easy to show the consistency of these aggregated estimates to the same limit weights. }
\end{Nota}

As already noted, computation of Wasserstein distances or barycenters can be a hard task and of course this applies
also to trimmed $k$-barycenters. We close this section with a result that can simplify this problem. It concerns
location-scatter families, that is,  families of distributions on $\mathbb{R}^d$ that
can be obtained from positive definite affine transformations from a standard representative.
More precisely, given $\bf{X}_0$, a random vector with probability law  $P_0 \in \mathcal{P}_{2,ac}(\Read)$ (the subset of $\mathcal P_2(\Read)$ 
containing the absolutely continuous distributions) and ${\cal M}_{d\times d}^+$ the set of $d\times d$ symmetric positive definite matrices, 
the set $$\mathcal{F}(P_0):=\{\mathcal{L}(A{\bf X}_0+m): A\in {\cal M}_{d\times d}^+, m\in\Read\},$$ 
is a {\it location-scatter family}. It is easy to check (see \cite{preprint2}) that these families can be re-parameterized 
in terms of the vector of means and the covariance matrix and also that we can assume w.l.o.g. (as we do in the sequel) 
that $P_0$ is centered and has the identity $I_d$ as covariance matrix. These families are often involved in model-based 
clustering procedures, particularly when we are looking for elliptically shaped clusters (elliptical families belong to this 
class, but the definition includes the possibility of non-elliptical families). Location-scatter families are closed
for Wasserstein barycenters, that is, if $\mu\in W_2(\Pd)$ is supported in $\mathcal{F}(P_0)$ then
the barycenter of $\mu$ belongs to $\mathcal{F}(P_0)$ (Theorem 3.11 in \cite{preprint2}). From this fact
and the comments following equation (\ref{descomp}) we obtain the following result.

\begin{Prop}\label{kbarnor}
Let $\mu$ be a probability on $\Pd$ which is supported in the location-scatter family 
$\mathcal{F}(P_0)$, for any $P_0 \in \mathcal{P}_{2,ac}(\Read)$. Then any 
trimmed $k$-barycenter of $\mu$ is a set of $k$ probabilities  that belong to  $\mathcal{F}(P_0)$.
\end{Prop}

\section{Computation of trimmed $k$-barycenters in Wasserstein space}\label{computation}

We discuss now the problem of (approximate) computation of trimmed $k$-barycenters 
of a set of probabilities $P_1,\dots,P_r\in\Pd$ with weights $w_1,\dots,w_r$. This corresponds to the case
where $\mu\in W_2(\Pd)$ is concentrated on the finite set $\{P_1,\dots,P_r\}$ with probabilities $w_i$, $i=1,\ldots,
r$ and covers the case of empirical trimmed $k$-barycenters considered in Theorem \ref{coro}. Hence, our goal is to 
compute $k$ probabilities, $\bar P_1,\dots, \bar P_k\in \Pd$, and  
weights $\bar{w}=(\bar w_1,\dots, \bar w_r)\in \mathcal{C}_\alpha(w)$ such that
\begin{eqnarray}\label{e2}
 \sum_{i=1}^r \bar w_i \min_{j=1,\dots,k}\Wdd(P_i,\bar P_j)= \min_{Q_i\in\Pd, w^*\in \mathcal{C}_\alpha(w) } \sum_{i=1}^r w_i^* \min_{j=1,\dots,k}\Wdd(P_i,Q_j),
\end{eqnarray}
where $\mathcal{C}_\alpha(w)=\{ w^*=(w_1^*,\ldots,w_r^*): 0\leq w^*_i\leq w_i/(1-\alpha),\, \sum_{i=1}^r w_i^*=1\}$. 
Below, we present an iterative procedure for computing a solution to (\ref{e2}). 
In fact, it is a suitable  adaptation of available algorithms for obtaining  trimmed $k$-means, 
with updates of weights and distances in each concentration step. Of course its utility is conditioned by  
feasibility  of the computation of Wasserstein distances and barycenters, which we consider later.

\medskip

\noindent\textbf{Algorithm (trimmed $k$-barycenters with weights):}\medskip

\begin{enumerate}

\item
{\it Random start: }Take $n = 0$ and draw $k$ random initial centers $P_1^n,\ldots,P_k^n$,
(say from the original sample).

\item
{\it Concentration step: }

\begin{enumerate}

\item
For every $i = 1, \ldots , r$, compute the values 
\begin{equation*}
d^n_i = \min_{j=1,\ldots,k} {\cal W}_2^2(P_j^n, P_i),
\hspace{1cm}
g_i^n = \arg\min_{j=1,\ldots,k} {\cal W}_2^2(P_j^n, P_i).
\end{equation*}

\item
Consider the permutation $((1), \ldots , (r))$ such that $d^n_{(1)} \leq \ldots \leq d^n_{(r)}$.

\item
Set $h_n = \inf \{h\in \Nat:  \sum_{i\leq h} w_{(i)} \geq  1- \alpha\}$ and define
\[
\delta_{(i)}^n =\left\{
\begin{array}{ll}
w_{(i)} & \mbox{ if } i < h_n
\\
1-\alpha - \sum_{i\leq h_n} w_{(i)} & \mbox{ if } i = h_n
\\
0 & \mbox{ if } i > h_n,
\end{array}
\right.
\]
and take the weights $w_{(i)}^n=\delta_{(i)}^n/(1-\alpha)$.
\end{enumerate}

\item
Set $n=n+1$ and
update the centers taking as new center $P^{n+1}_j$ the 
barycenter of the probabilities in $\{P_i: g_i^n=j\}$ with  weights $w_{(i)}^n$.

\item
Repeat steps 2 and 3 until $g^n_{(i)}=g^{n+1}_{(i)}$ for $i \leq h_n$ and $w^{n+1} _{(i)}=w^n _{(i)}$.

\item
Repeat several times steps 1 to 4 and keep the best solution in the sense of minimizing the objective function given in (\ref{e2}).

\end{enumerate}

Even in its simpler version of $k$-means, it is well known that minimizing an objective function like (\ref{e2}) is an 
NP-hard problem and that with a greedy algorithm such as $k$-means only convergence to a local minimum is guaranteed.  However, if a large amount 
of random initializations are considered, there is enough empirical evidence of the nice behavior of this kind of algorithms 
suggesting convergence to the global optimum. 
In the Appendix, we will provide evidence that steps 2 to 4 of the algorithm will produce stationary partitions (no loops in the process may occur). This is a well known fact for the non-trimmed $k$-means algorithm on non-weighted data sets on Euclidean spaces (see e.g. page 38 in \cite{Hennig}), but some extra effort is needed in our current setting. Broadly speaking, we need to assure that the barycenter of $\{P_1,\dots,P_r\}$ with weights $\{w_1,\dots,w_r\}$ changes as soon as we trim one of the involved probabilities. This goal will be carried through Propositions \ref{Prop.cont}, \ref{Prop.discret} and \ref{convalg}, covering the case of absolutely continuous probabilities, but also that of discrete probabilities with finite supports.
By now, let us begin stressing some distinctive facts about the presented algorithm:
\begin{enumerate}
\item
In Step 2.(b) we obtain a permutation such that $d^n_{(1)}\leq \ldots \leq d^n_{(r)}$ in which the ties are not broken arbitrarily. The way in which they are broken is irrelevant, but it is important to fix one. In this case, we have chosen  the same order as in the initial sample.

The effect of this previously selected order is to determine only a point which is going to be partially trimmed. Notice that this has no effect in the value of the objective function. It is obvious that if $d_{(i)}^n < d_{(h_n)}^n$, then the probability $P_{(i)}$ should not be trimmed, and that if $d_{(i)}^n > d_{(h_n)}^n$, then the probability $P_{(i)}$ should be fully trimmed. Thus, the only concern is how to split the amount of trimming  $1 - \alpha -   \sum_{ i < h_n}w_{(i)}$ between the probabilities whose indices satisfy that $d_{(i)}^n = d_{(h_n)}^n$. However, since all the distances $d_{(i)}^n$ for the indices in this group coincide, the way in which these probabilities are trimmed does not affect the value of the objective function.

Anyway, it is important to take into account that in practice it is quite strange to have two identical values for $d_{(i)}^n$.

\item
The reason to choose the stopping criteria based on the stationarity of the partition instead of that one of the $k$-barycenters, is due to  the possibility of lack of uniqueness of the barycenters, that would lead to the possibility of getting a partition with more than one possibility for the $k$-barycenters and, consequently, to a non-ending  algorithm.

However, as already noted, if the involved probabilities are absolutely continuous, then the barycenter is unique, and, consequently, the stationarity of the  partition and that of the $k$-barycenters would be equivalent.
\end{enumerate}

The algorithm can be efficiently improved to address several kinds of problems, and this is the case for aggregation. Concerning  the first step of initializations, as we  
will discuss later, for aggregation problems  random initializations can be advantageously substituted by handling 
the solutions provided by the units as initializations. Also the use of weights can be highly recommended to 
improve the merging effect in parallelized schemes when the sizes of the subsamples are highly inhomogeneous.
  
Steps 2 and 3 of the algorithm above involve a large amount of computations of distances and barycenters in Wasserstein space. This limits
a practical use of the algorithm to cases where these computations can be efficiently done.  Of course, as we will show next, it covers the case of probability measures on the real line. Remarkably, it also covers the case of location-scatter families, which play a pivotal role in statistical applications. In the model-based cluster analysis setup, our algorithm mainly addresses the problem of clustering based on elliptical shapes. Moreover, recent progress in the computation of optimal transportation through regularization/discretization ideas is enlarging the range of problems which could be addressed with our approach (recall  \cite{Cuturi}, \cite{Benamou},  \cite{Anderes}).
 
For probabilities on $\Rea,$ Wasserstein distance is simply the $L_2$-distance between quantile functions (see (\ref{casoreal}) in the Appendix).
Furthermore, if $F_1^{-1},\ldots,F_r^{-1}$ are the quantile functions associated to the probabilities $P_1,\ldots , P_r$
then $\sum_{i=1}^r w_r F^{-1}_i $
is the quantile of the barycenter of $P_1,\ldots,P_r$ with weights $w_1,\ldots,w_r$ (see \cite{Agueh10}). 
This allows to use the algorithm to compute trimmed $k$-barycenters. We illustrate this application through the analysis of a dataset 
of population densities in Section \ref{pyramids}.
 
In higher dimension there is no simple general way to compute Wasserstein distances and barycenters. 
However, an important exception to this claim is given by the case where $P_1,\ldots,P_r$ are
probabilities in the same location-scatter family as introduced in Section 2. A relevant fact  
(see Theorem 2.3 in \cite{preprint}) is that distances between distributions $P, Q \in \mathcal{F}(P_0)$,  
depend only on their means $m_P, m_Q$ and covariance matrices $\Sigma_P, \Sigma_Q$ and can be computed through 
\begin{equation}\label{distnormales}
\Wdd(P,Q)= \|m_P-m_Q\|^2+ trace\left(\Sigma_P+\Sigma_Q-2\left(\Sigma_P^{1/2}\Sigma_Q\Sigma_P^{1/2}\right)^{1/2}\right).
\end{equation}
On the other hand, by Proposition \ref{kbarnor}, we know that in these families the barycenters 
also belong to the family. Hence, if in the random start step of the trimmed $k$-barycenter algorithm we
choose initial centers within the family $\mathcal{F}(P_0)$, then all the subsequent centers will belong to
$\mathcal{F}(P_0)$ and all distances can be computed using (\ref{distnormales}). Moreover, for the computation of updated
centers in Step 3 it suffices to use the facts that (see \cite{preprint, preprint2})
if $\mu$ is the probability  giving weights $w_1,\dots,w_r$ to the probabilities 
$\mathbb{P}_{m_1,\Sigma_1},\dots,\mathbb{P}_{m_r,\Sigma_r}\in \mathcal{F}(P_0)$ then,
\begin{itemize}

\item[i)] its barycenter  is the probability  $\mathbb{P}_{\bar m,\bar \Sigma} \in \mathcal{F}(P_0),$ where $\bar m=\sum_{i=1}^r w_im_i$ and $\bar \Sigma$ is the 
unique positive definite matrix satisfying  
\begin{equation}\label{ecuacion}
\sum_{i=1}^r w_i\left(\bar\Sigma^{1/2}\Sigma_i\bar\Sigma^{1/2}\right)^{1/2}=\bar\Sigma.
\end{equation}
\item[ii)] starting from any  positive definite matrix
$S_0$, the iterative procedure
\begin{equation}\label{recur}
S_{n+1}=S_{n}^{-1/2}\Big(\sum_{i=1}^r w_i(S_{n}^{1/2}\Sigma_i S_{n}^{1/2})^{1/2} \Big)^2 S_{n}^{-1/2},
\end{equation}
converges to the solution of (\ref{ecuacion}), $\bar \Sigma = \lim_{n \to \infty} S_n$,
\item[iii)] the (generalized) variance of $\mu$,  $\mbox{Var}(\mu):=\sum_{i=1}^r w_i\Wdd(\mathbb{P}_{m_i,\Sigma_i},\mathbb{P}_{\bar m,\bar \Sigma}),$ takes the value
\begin{equation*}
\mbox{Var}(\mu)=
\sum_{i=1}^r w_i(\|m_i\|^2+\mbox{trace}(\Sigma_i))-(\|\bar m\|^2+\mbox{trace}(\bar \Sigma)).
\end{equation*}
\end{itemize}

Section \ref{loc-scatsection} will explore some possibilities of the new clustering methodology in this setup of location-scatter families,
including the parallelization setup of Theorem \ref{consist2}. Our choice for $\mathbb{F}$ is the TCLUST algorithm for
model based clustering introduced in \cite{TCLUST}. 
Although, to simplify the exposition, Theorem \ref{consist2} assumed equal weights 
for the distributions of the different units, it is natural to be more confident with the reports based on larger samples. Thus, we propose 
to compute trimmed $k$-barycenters with weights proportional to sample sizes, that is,
$$\omega_i^j:=\frac{n_j}{\sum_{j=1}^mn_j}, i=1,\dots,k, \ j=1,\dots,m. $$ For the initialization steps we will consider every set $\{N_i^1,\dots,$ $N_i^k\}$ for $i=1,\dots,m$. 

\begin{Nota}{\em 
By clustering the $k\times m$ distributions obtained from the $m$ units, the parallelization procedure of Theorem \ref{consist2} avoids 
to address the label correspondence problem. In a well separated configuration of the true $k$-feature, it should 
be expected that any of the $k$ distributions reported by every unit has an unambiguous classification. In such cases, trimming 
has only a mild impact and the procedure can even be seen as an alignment-by-groups tool. 

When some clusters are not well separated or their relative sizes are very different, parallelization or analysis 
with distinct procedures will generally produce serious troubles for getting a consensus. We want to stress that, in 
such cases, trimming following our scheme would mainly act on the troubling clusters, eliminating the more discrepant 
distributions. In some way this is consistent with the idea of consensus by voting. Since the trimming procedure is 
not based on a labeling of the $k\times m$ distributions according to the reporting units, the real trimming size on 
the troubling clusters will be notably increased allowing  to get the consensus just among the more similar distributions 
in the meta-sample. In contrast, the proposal resorts to that labeling for the initialization steps of the algorithm. 
For moderate values of $m$, the number of units processing the data, it would be convenient to try all the 
solutions reported by the units as an initialization for the algorithm. In this way the final solution will be 
the best (wide) consensus that any initial proposal could produce through the `negotiation'.

Finally, we note that by varying the trimming level $\alpha$ we can obtain a picture of the stability of the 
solution provided by the method. In fact, a joint analysis of the influence of $\alpha$ and $k$ 
on the output of the procedure could lead to reconsider the number of clusters present in the sample.
This strategy was introduced in Garc\'{\i}a-Escudero et al \cite{tools} and can be suitably adapted to this setup.}
\end{Nota}

\vspace{-.5cm}\section{Clustering probabilities on the real line}\label{pyramids}

\begin{center}
\begin{figure}[htb]
\includegraphics[width=5.28cm]{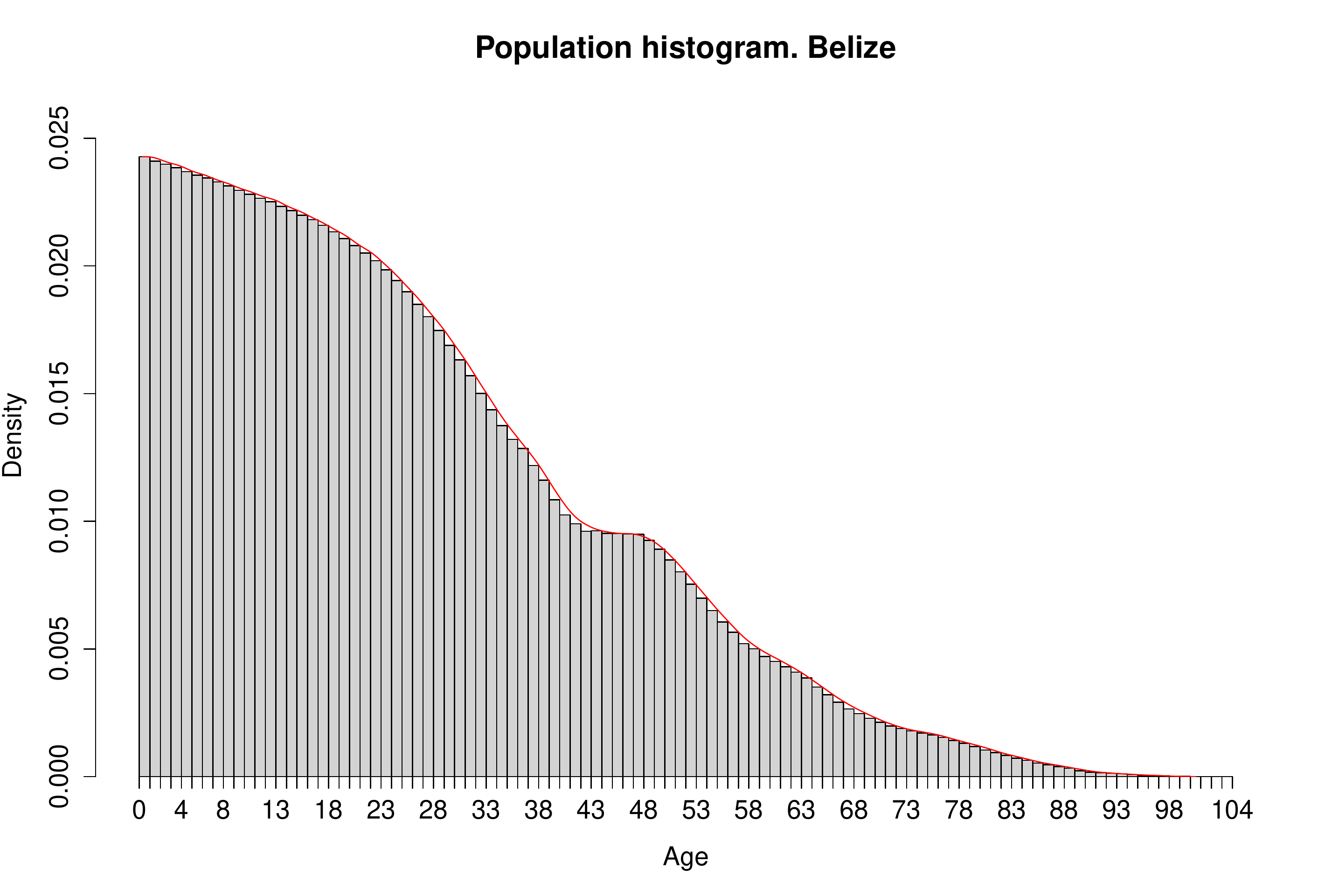}
\includegraphics[width=5.28cm]{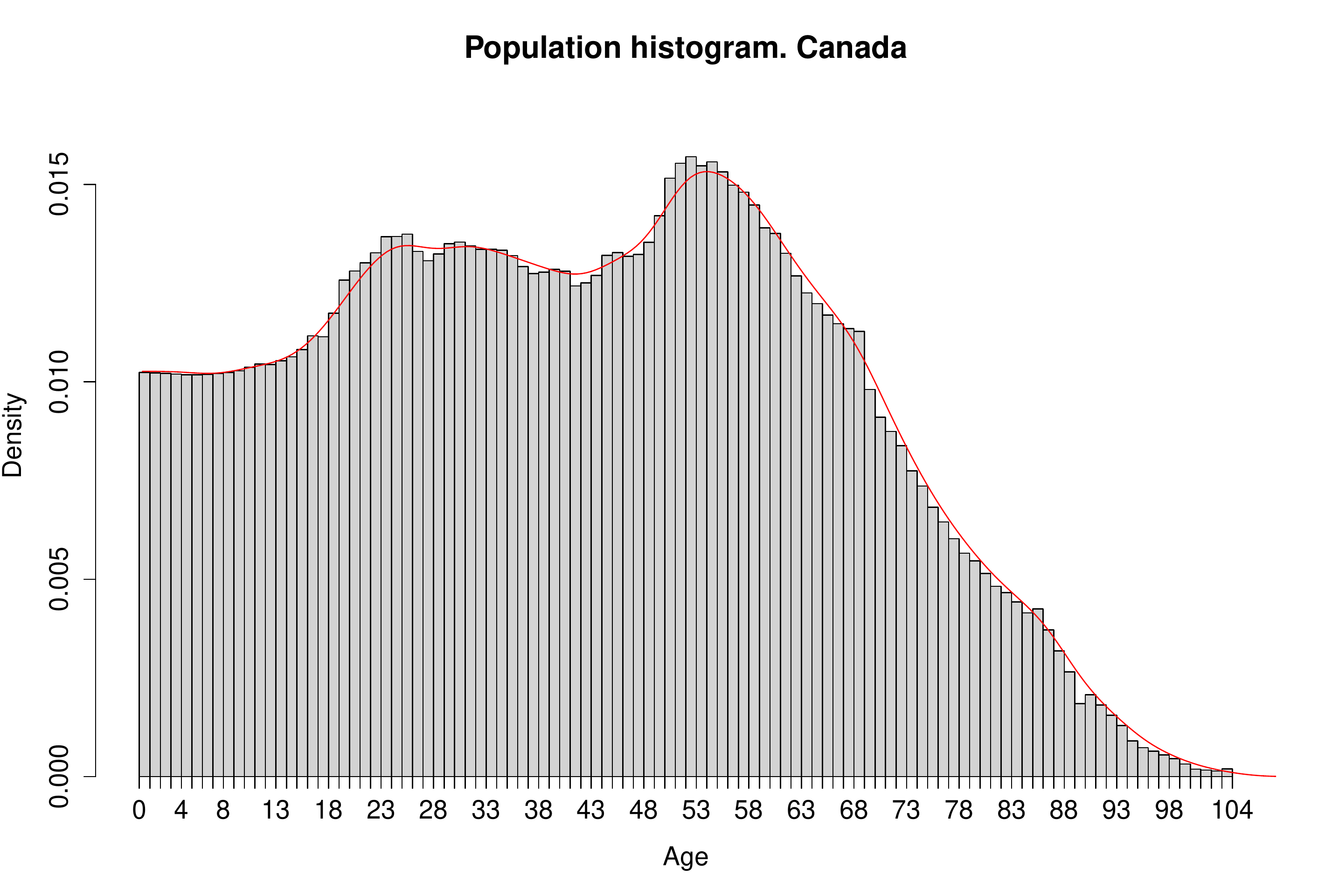}
\includegraphics[width=5.28cm]{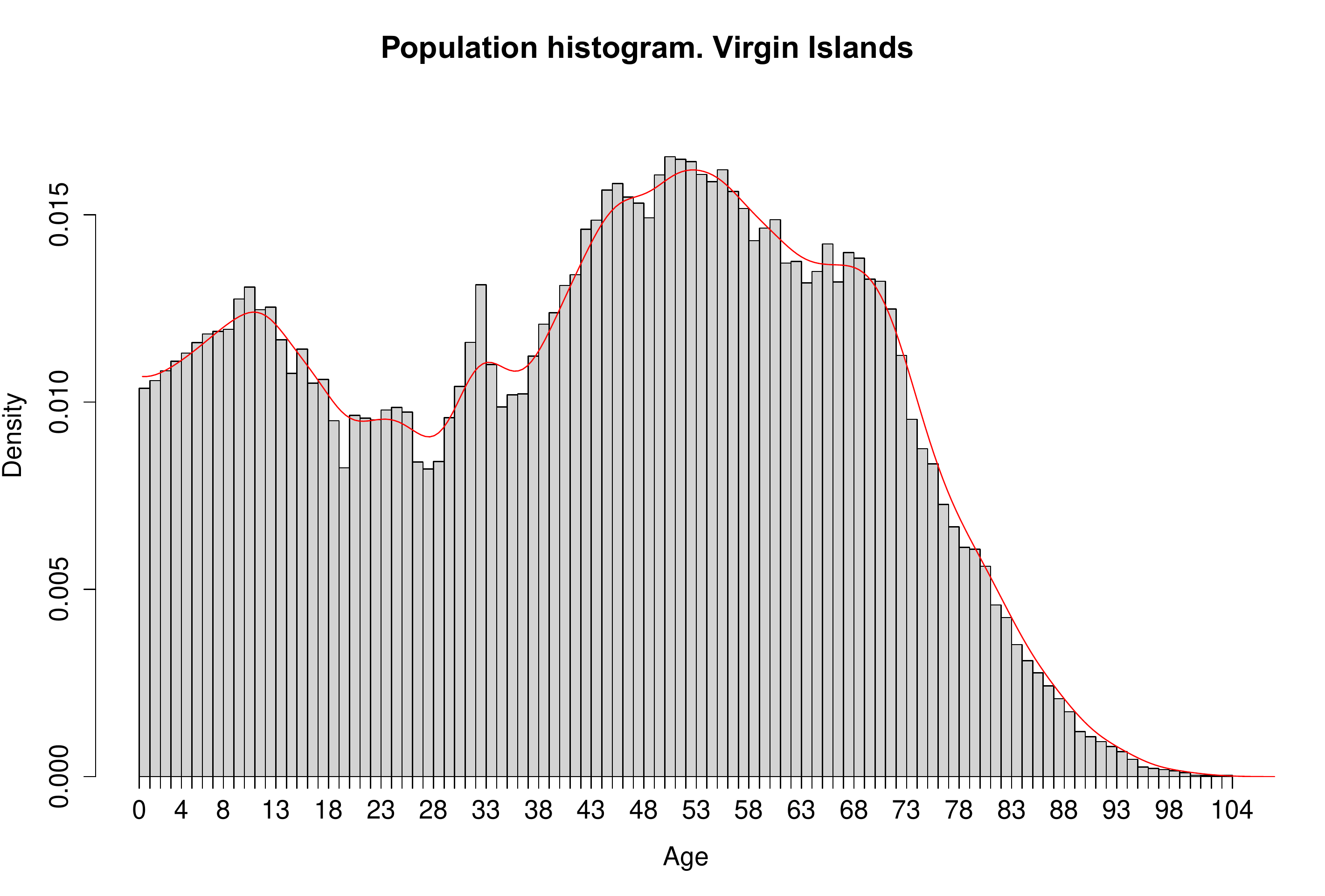}
\vspace{-1cm}
\caption{Histograms of some populations. }
\label{Figure1}
\end{figure}
\end{center}

In this section we illustrate the application of trimmed $k$-barycenters in Wasserstein space to
clustering age distributions of countries in the Americas. Our dataset has been downloaded (13/08/2015) 
from {\tt  http://www.census.gov}, which provides population estimates with real-time updates. This
database has been used in \cite{Delicado}  for illustration of functional PCA of densities and in 
\cite{Bigot2} to show different PCA techniques, including geodesic PCA in Wasserstein space. We have 
excluded very small countries from our analysis and considered the 36 countries with population
at least 100000. After some preprocessing (see \cite{TR} for details) for each country in the study our dataset 
consists of the total population numbers by age, with age ranging from 0 to 104. 
From this we compute an approximation to the sample quantile function for each country, which is all the information required by
our methods. In Figure \ref{Figure1} we include three (pre-processed) population histograms. Although our computations involve
only quantile functions we have added a kernel density estimate and this is what we show in subsequent 
figures in this section.

Our goal is to obtain a simple, but comprehensive enough description of the types of age distributions of
the countries in the dataset, a task that we carry out by looking for a (small) number of age distributions
that play the role of `main profiles' of `representative types'. In the case of multivariate data 
this is often done with the aid of clustering techniques (see e.g. Flury \cite{Flury} for an example).
Here we will use trimmed $k$-barycenters for clustering this dataset of distributions. 
Since we are just interested in the different age-distribution profiles we will 
assume equal weights for all countries. In our analysis we will deal simultaneously
with the choice of the number of centers, $k$, needed for a succesful summary of the profiles
and of the number of outliers. Intuitively, if one distribution represents a true sector of the 
population, increasing the level of trimming should not lead to sharp changes of the center of this sector.
In contrast, if a cluster were artificial it would consist of points belonging to different true clusters
and the trimming process would result in a greater displacement of the barycenter.
Also, if a point $x$ is an outlier, it should be separated 
from the true clusters and, once the points in the dataset which are more outlying than $x$ have been trimmed, 
$x$ himself should be trimmed for every reasonable choice $k$. Thus, if $x$ is really an outlier, 
there should exist a trimming threshold $\alpha_0$ such that $x$ should be trimmed in the $\alpha$-trimmed $k$-barycenter
problem for $\alpha>\alpha_0$, independently of $k$, and if $k$-barycenters are to give a good description the dataset 
then the $k$ barycenters should not change much with small changes of the trimming level.

We have used the algorithm introduced in Section \ref{computation} with $\alpha\in\{\frac 1 {36},\ldots, 
\frac 6 {36}\}$ and $k=2,\ldots,6$. 
Table \ref{Tabla.Trimmings} reports the total number of times that countries were trimmed (if any) 
in the process. Note that for each $\alpha$ there are 5 choices of $k$, hence, 
the maximum number times that a country can be trimmed is 30.

\begin{table}[htb]
\begin{center}
\begin{tabular}{lr|lr}
Country & \# of times  & Country & \# of times 
\\
\hline
Virgin Islands  & 28 & Canada & 22
\\
Argentina & 7 &  Nicaragua      &     7
\\
Trinidad and Tobago & 7 & Uruguay & 6
\\
Jamaica & 5 & Guatemala & 4
\\
Belize & 3 &  Haiti       &    3
\\
Honduras    &    3 &       St Vincent \&   Grenadines    &      3 
\\
 Puerto Rico & 2 & Grenade & 2
\\
 Bolivia & 1 & Chile & 1  
\\
Saint Lucia & 1 &
\\
\hline
\end{tabular}
\end{center}
\vspace{-.5cm}
\caption{Number of removals from analysis, by country}
\label{Tabla.Trimmings}
\end{table}

For $\alpha=1/36$, the trimmed countries were Virgin Islands (V.I.), four times, and Argentina (once). 
For $\alpha=2/36$ only  Canada and V.I. were trimmed (five times each). With $\alpha=3/36$, 
apart of Canada and V.I. (5 times each), Argentina (twice), Guatemala, Nicaragua and Trinidad 
and  Tobago (T.T.)  were also trimmed. At level $\alpha=4/36$   Canada and V.I. were trimmed  
(four times each), Argentina and Nicaragua were trimmed twice and there were eight 
countries trimmed once. For $\alpha=5/36$, V.I. was trimmed five times;  Canada four;  Jamaica,
Nicaragua, T.T. and Uruguay twice and there were 8 countries trimmed only once.
Finally, for $\alpha=6/36$,   V.I. was trimmed five times and  Canada four; T.T. 
and Uruguay were trimmed three times and Jamaica   and Nicaragua were trimmed twice. 11 countries were trimmed once.
From this table we see that the age distributions of V.I. and Canada
deviate from the distributions of all the other countries and should be trimmed. 
The next candidates for trimming are Argentina, Nicaragua  and T.T. However,  for 
$\alpha \in \{3/36,\ldots,6/36\}$ these countries are trimmed only  one third of times and 
we, therefore, decided to trim only two countries, that is, choose $\alpha =2/36$.

Turning to the choice of $k$,   we would accept that there exist at least $k$ groups in our dataset if the variation of the $k$ barycenters, 
when $\alpha$ changes is small.

\begin{center}
\begin{figure}[htb]
\includegraphics[width=8cm]{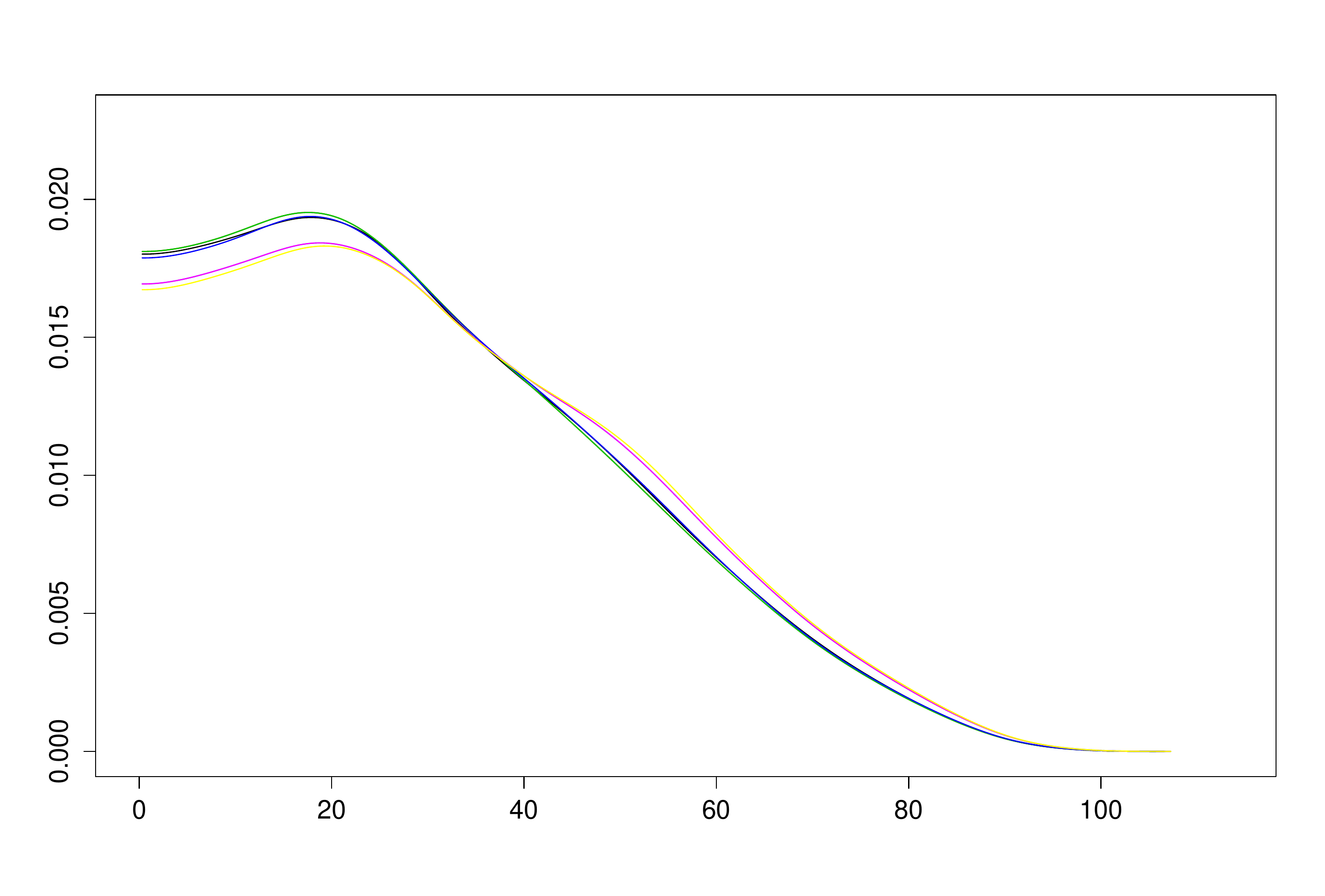}
\includegraphics[width=8cm]{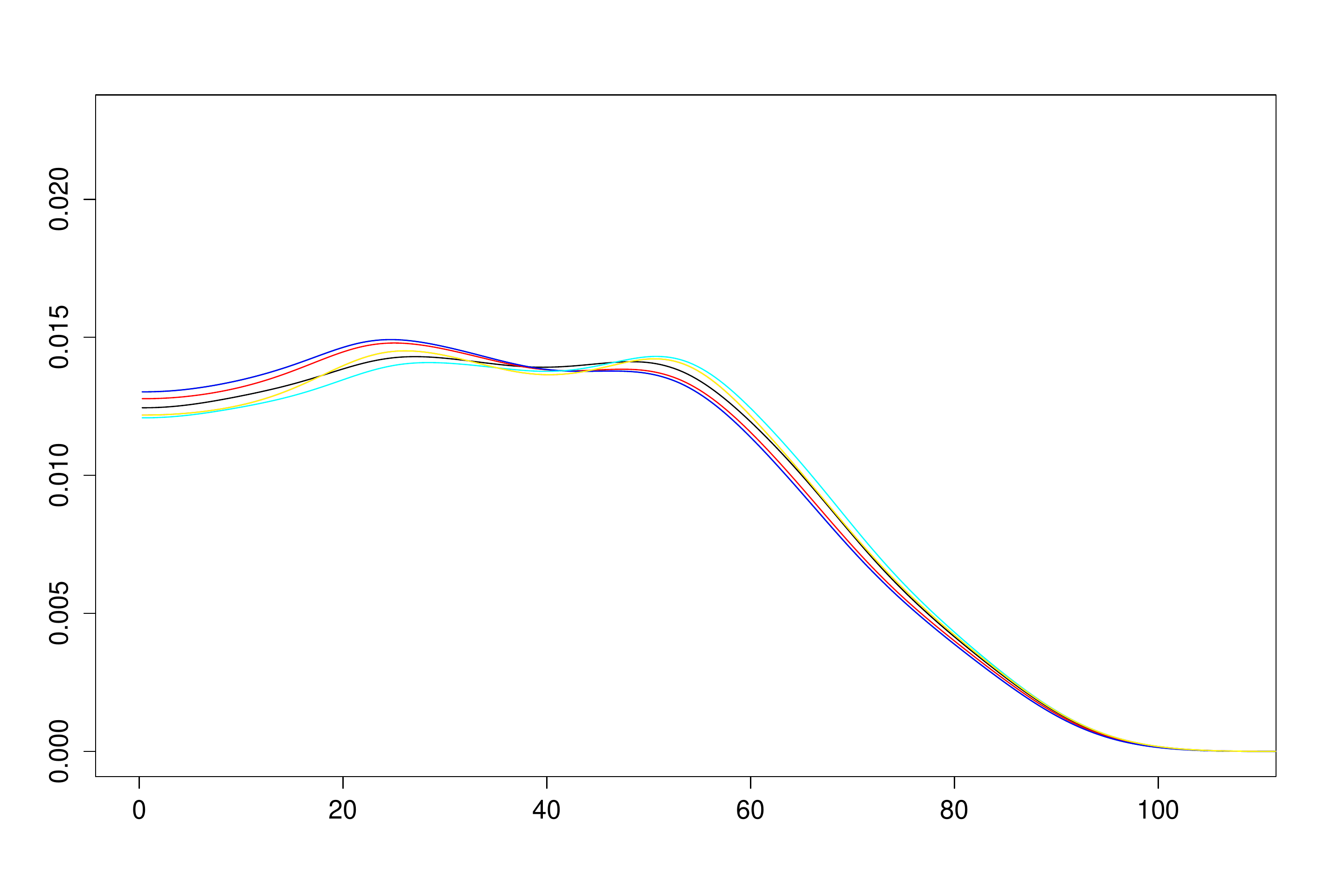}
\vspace{-1cm}
\caption{$2$-barycenters when $\alpha= 0,1/36,\ldots, 6/36$. Colors refer to different values of $\alpha$}
\label{Figure2}
\end{figure}
\end{center}

\begin{center}
\begin{figure}[htb]
\includegraphics[width=8cm]{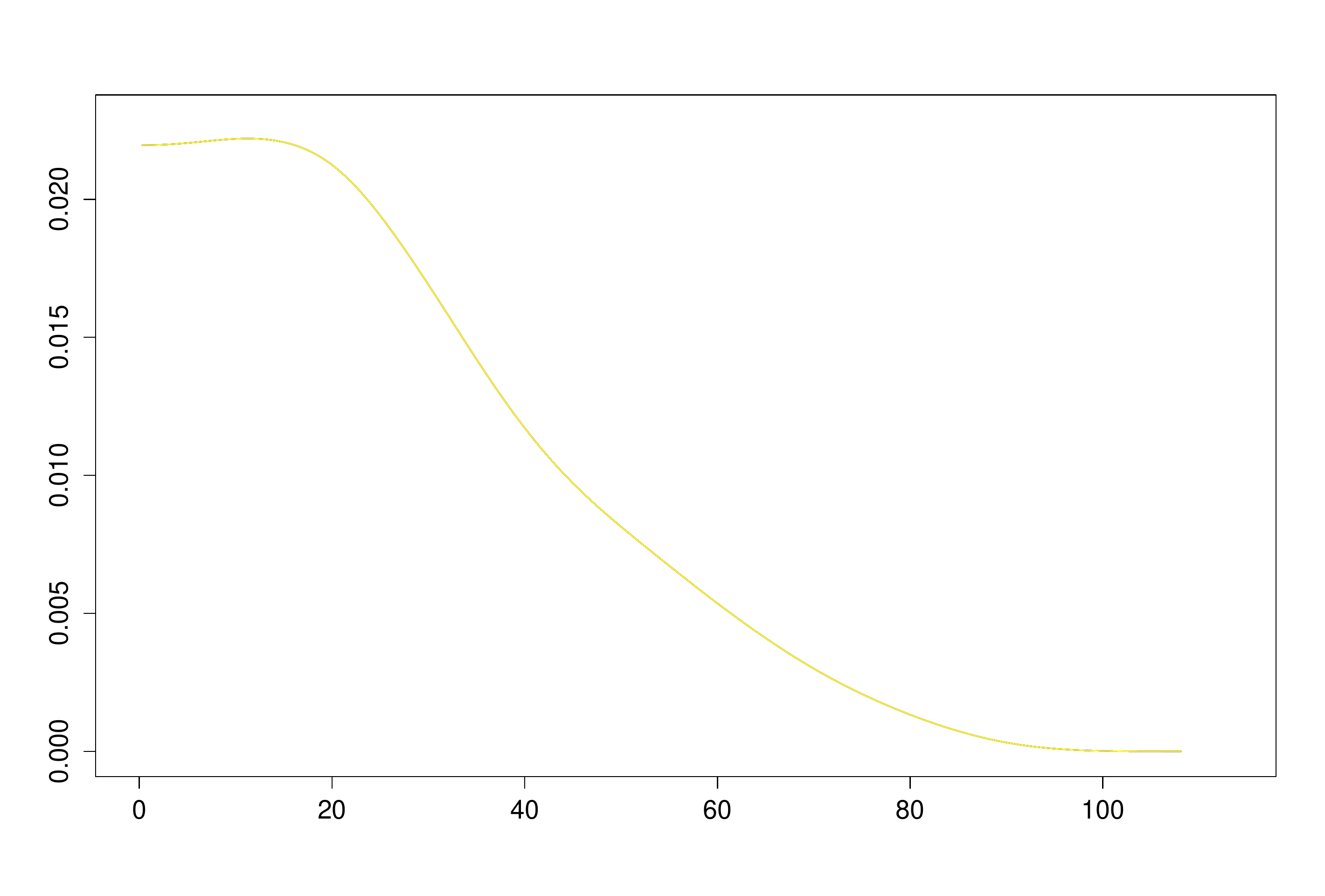}
\includegraphics[width=8cm]{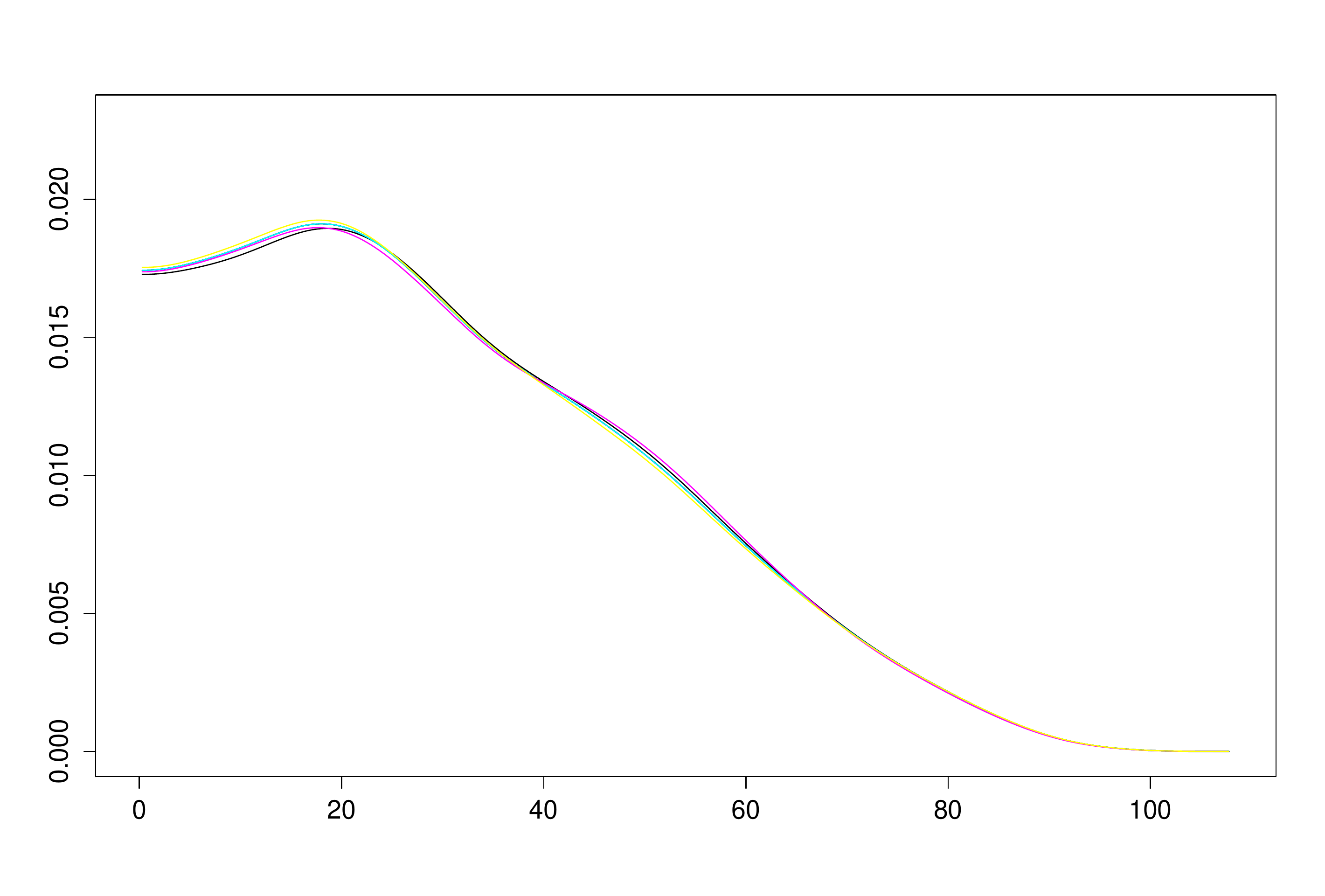}
\includegraphics[width=8cm]{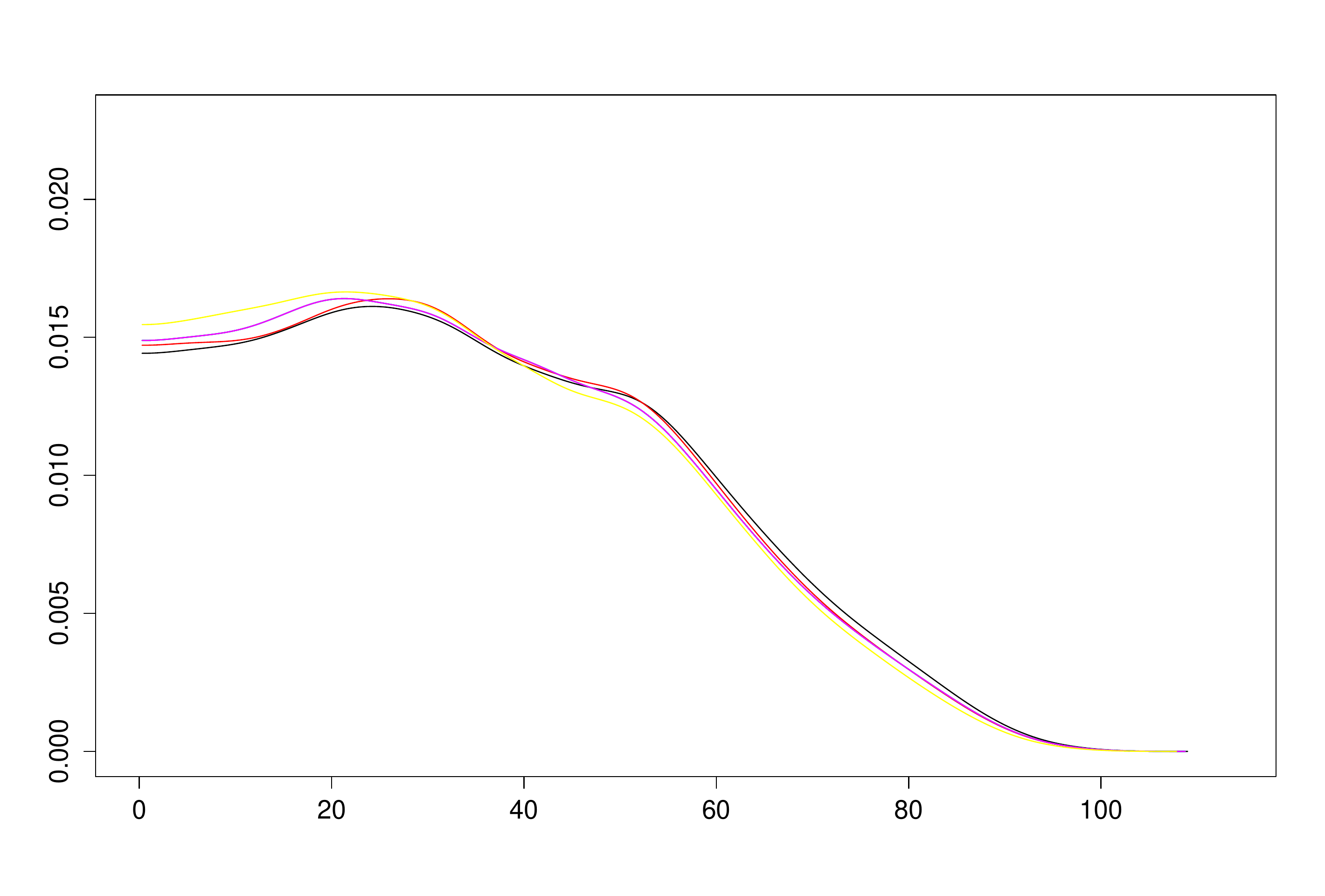}
\includegraphics[width=8cm]{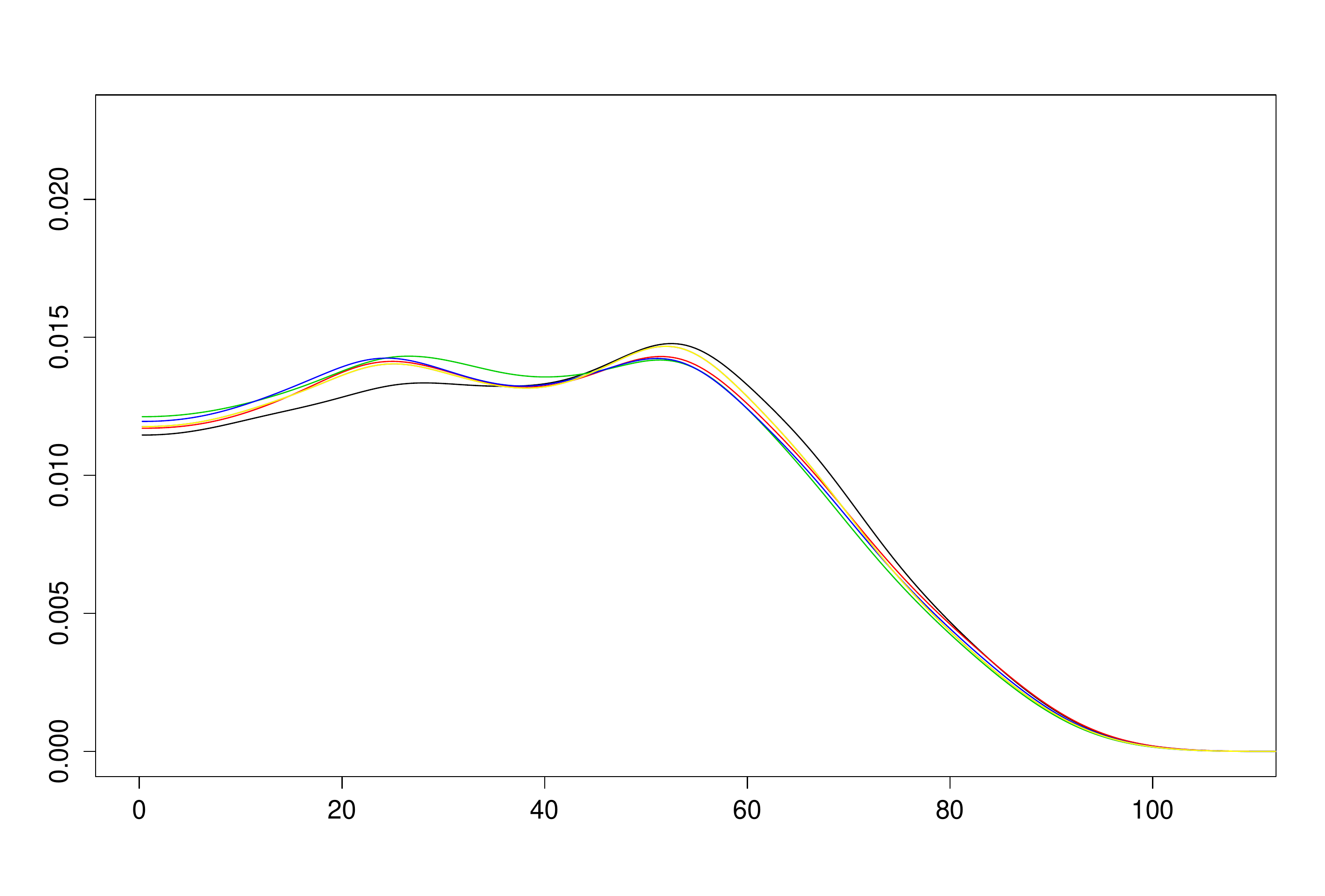}
\vspace{-1cm}
\caption{$4$-barycenters when $\alpha= 0,1/36,\ldots, 6/36$. Colors refer to different values of $\alpha$}
\label{Figure3}
\end{figure}
\end{center}

\begin{center}
\begin{figure}[htb] \label{Fig.Histo1}
\includegraphics[width=8cm]{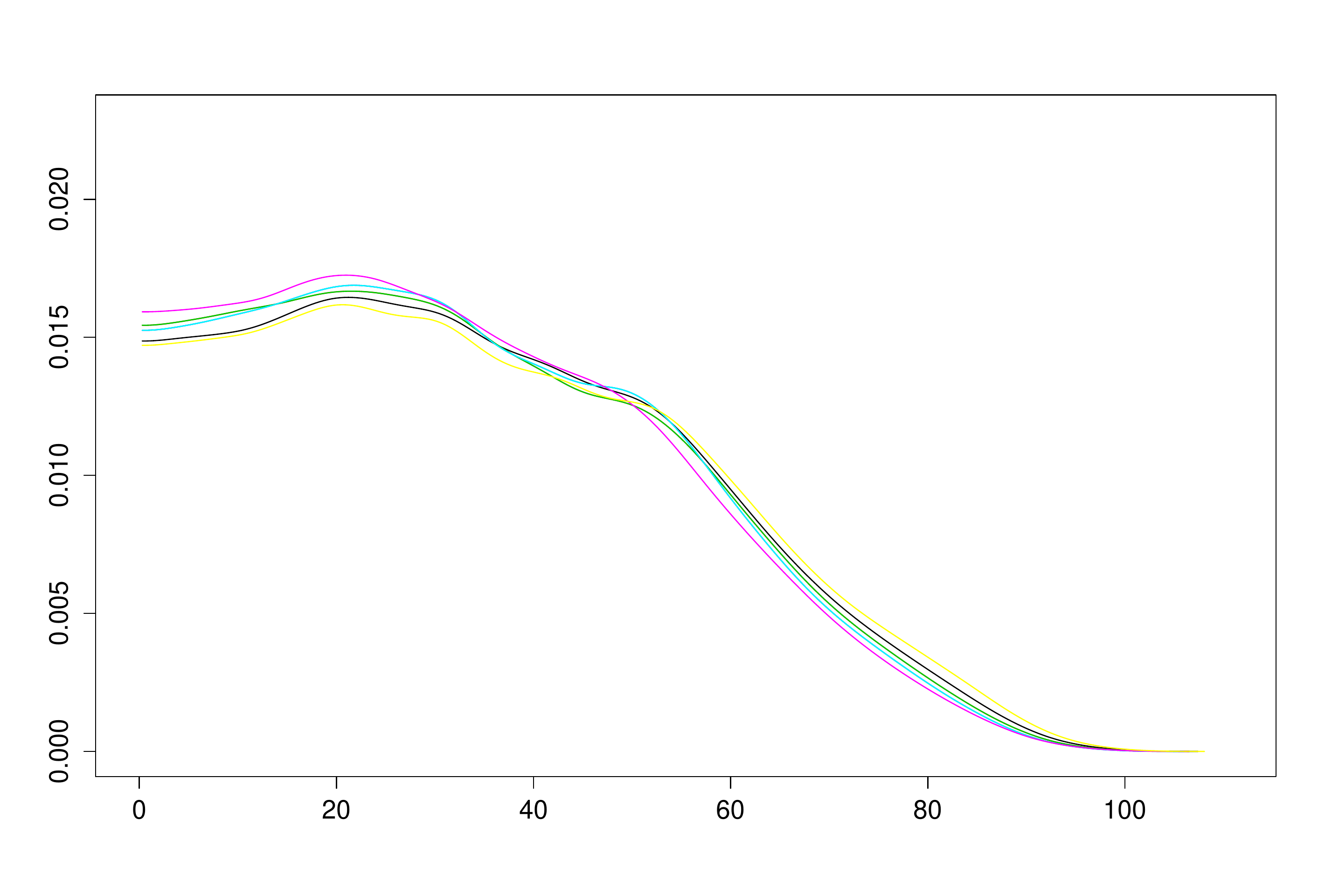}
\includegraphics[width=8cm]{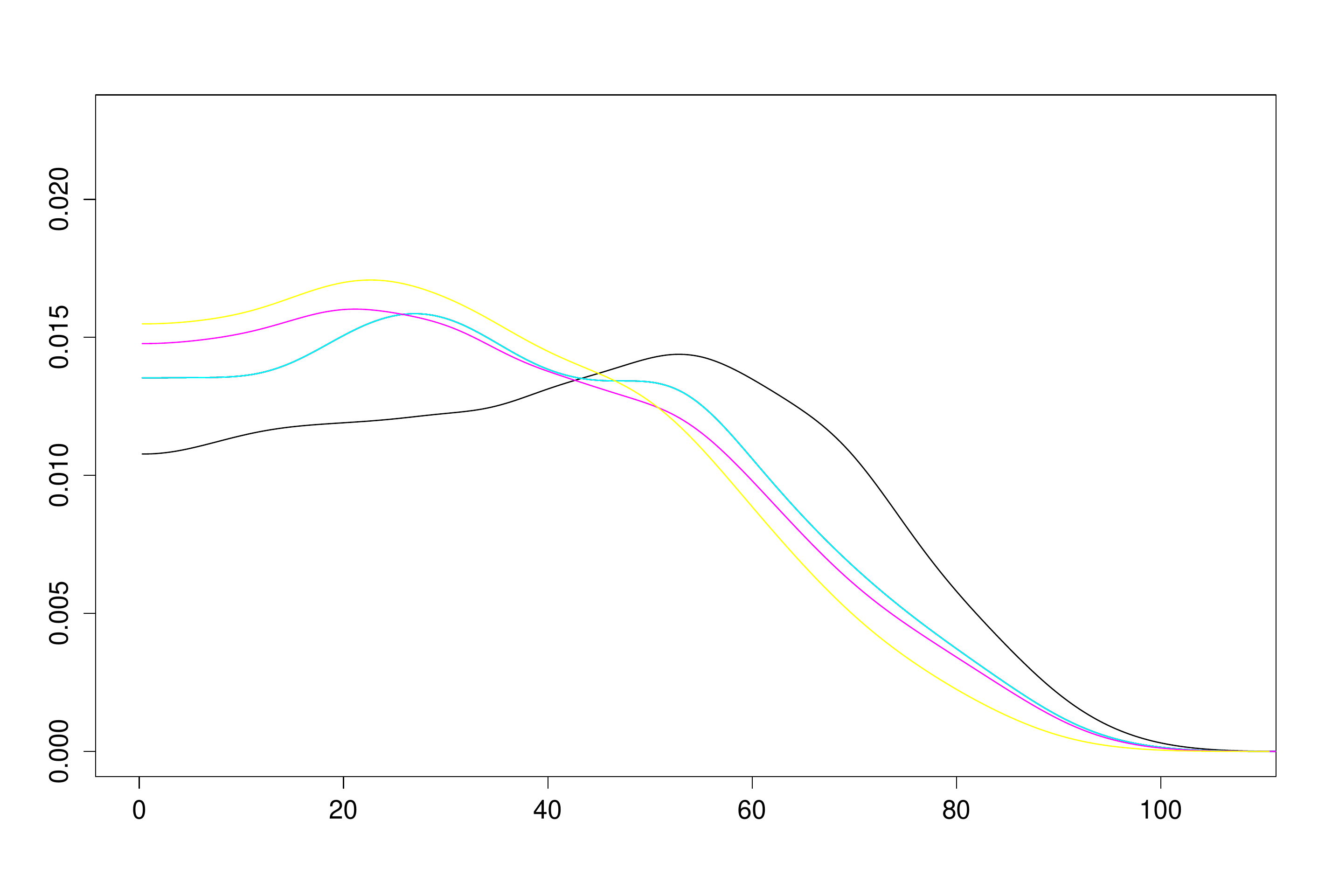}
\includegraphics[width=8cm]{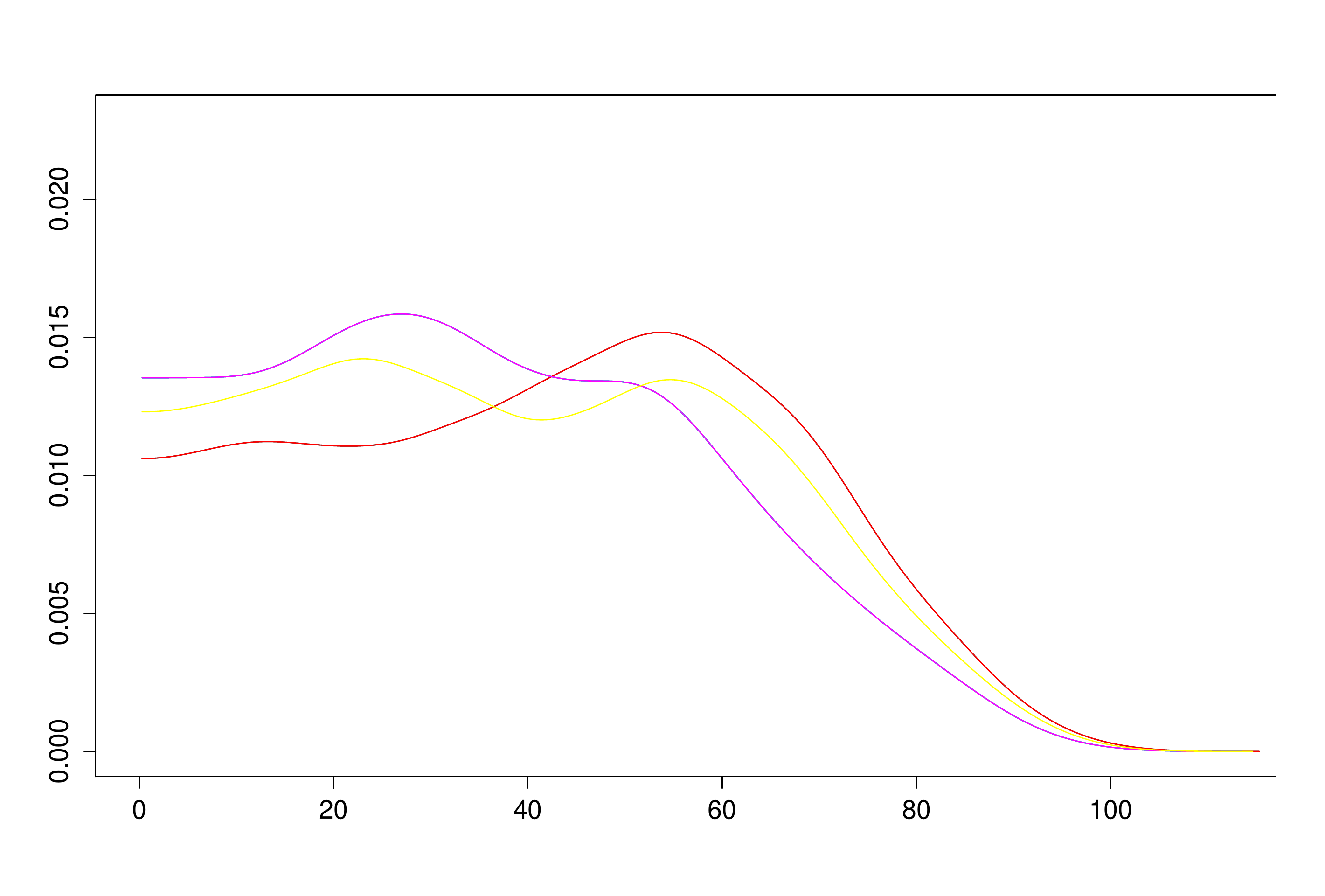}
\includegraphics[width=8cm]{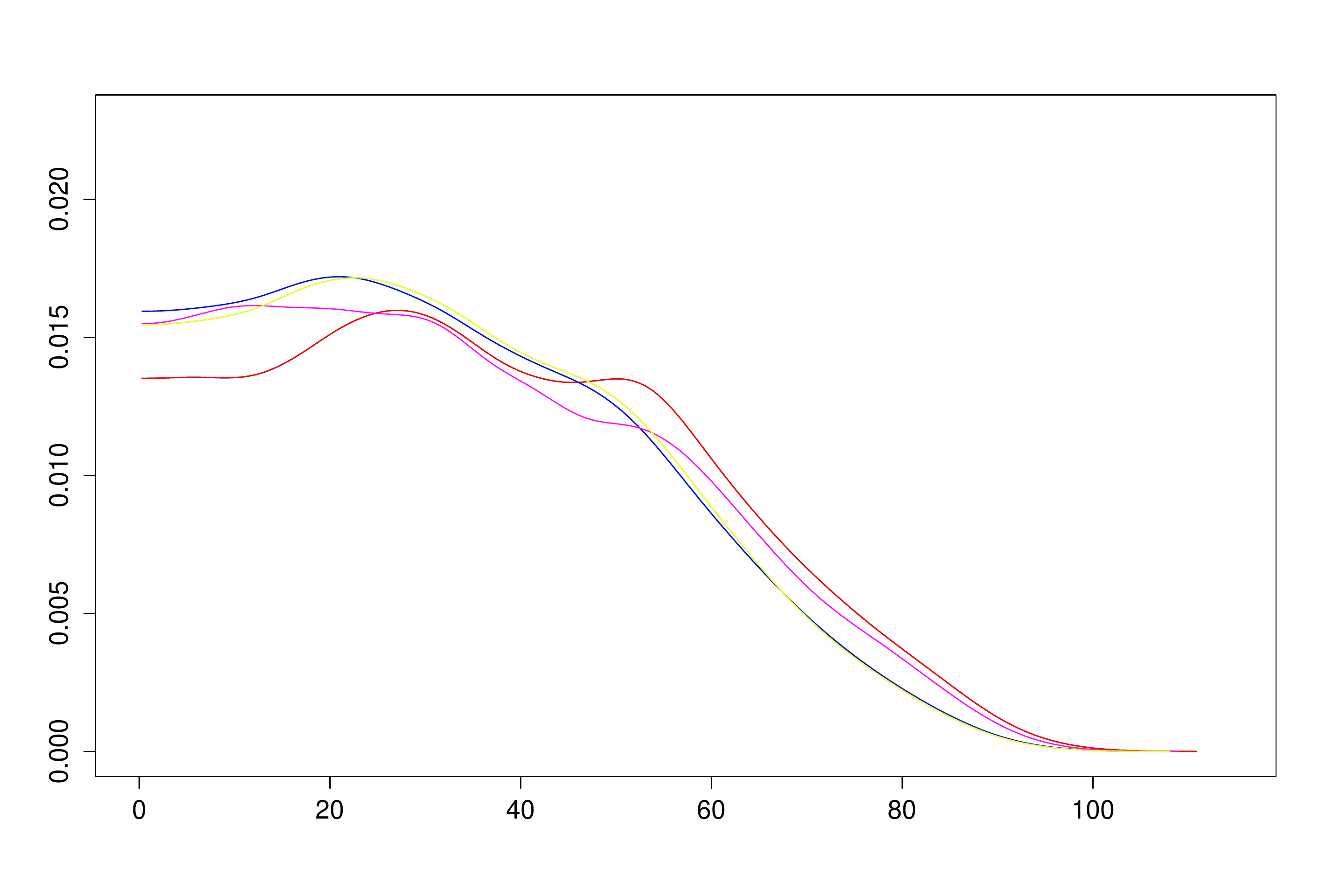}
\vspace{-1cm}
\caption{Two $5$-barycenters (resp. $6$ barycenters), first  (resp. second) 
row, when $\alpha= 0,1/36,\ldots, 6/36$. Colors refer to different values of $\alpha$}
\label{Figure4}
\end{figure}
\end{center}

For $k=2$, letting $\alpha = 0,1/36,\ldots, 6/36$, we obtain seven 
pairs of 2-barycenters. We have grouped them by similarity. This is shown in
Figure \ref{Figure2}. Each graph includes seven barycenters, corresponding to the
different trimming levels and stability becomes apparent.
 We conclude that $k\geq 2$. 
The cases $k=3$ and $k=4$ produce similar ouputs. We only comment the case $k=4$. 
Figure \ref{Figure3} contains the 4-barycenters obtained for  $\alpha = 0,1/36,\ldots, 6/36$, 
with younger countries represented in the upper left corner, moving to older countries as we move 
left to right and from the first to the second row.
We notice the very small variation among profiles in each group (in particular, there is no 
variation at all in the young countries group) and we conclude that $k\geq 4$.
The picture changes if we take $k=5$ or $k=6$. In the first case, four of the five groups remain 
stable (an example appears in the upper left hand side graph in Figure \ref{Figure4}), but 
we observe a new group  (see the upper right graph in Figure \ref{Figure4}) which is not 
stable at all. Similarly, with $k=6$ we have two unstable groups (bottom row  in Figure \ref{Figure4}). 
This suggests that 
$k=4$ is a good choice for this dataset. 

\begin{center}
\begin{figure}[htb]
\includegraphics[width=8cm]{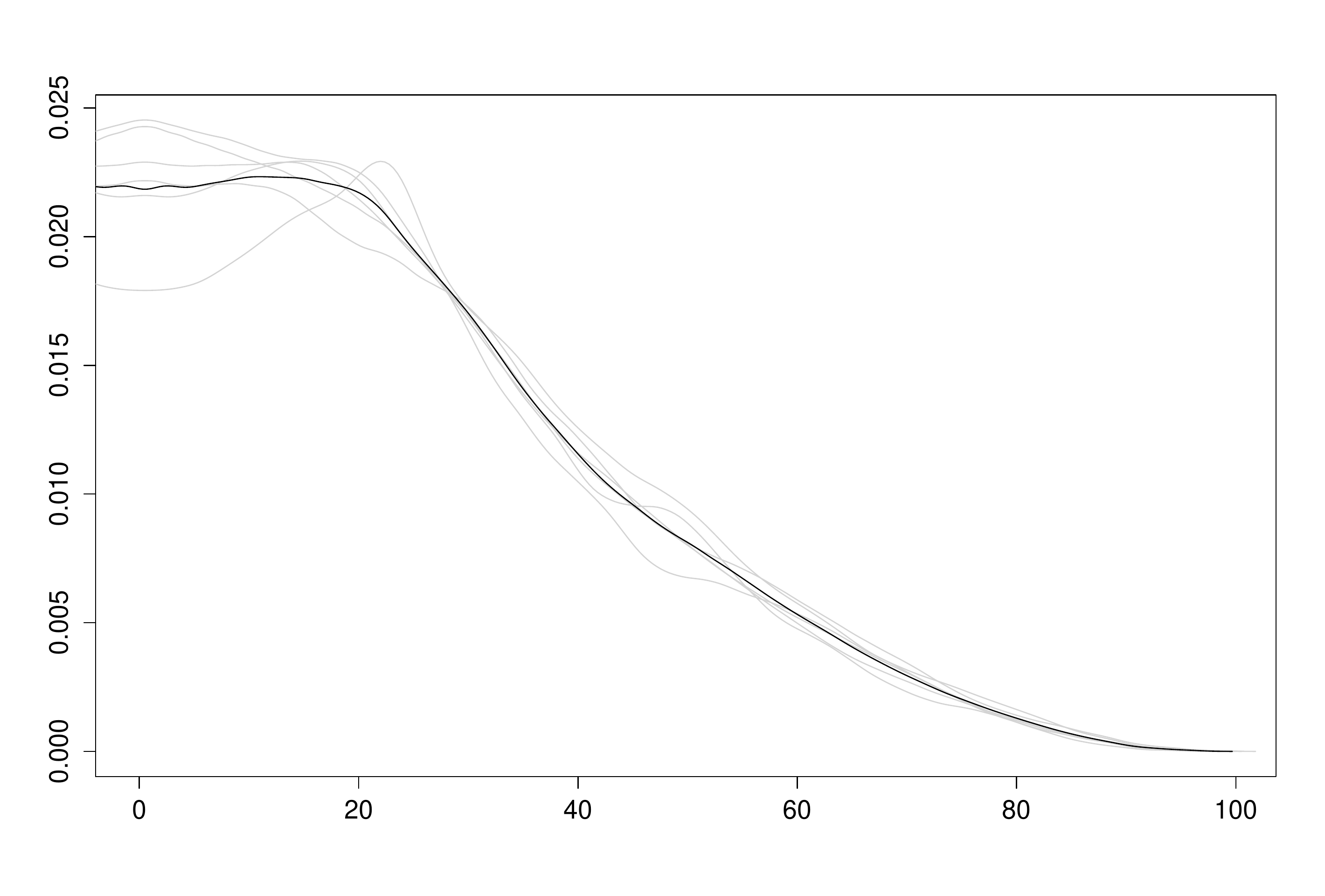}
\includegraphics[width=8cm]{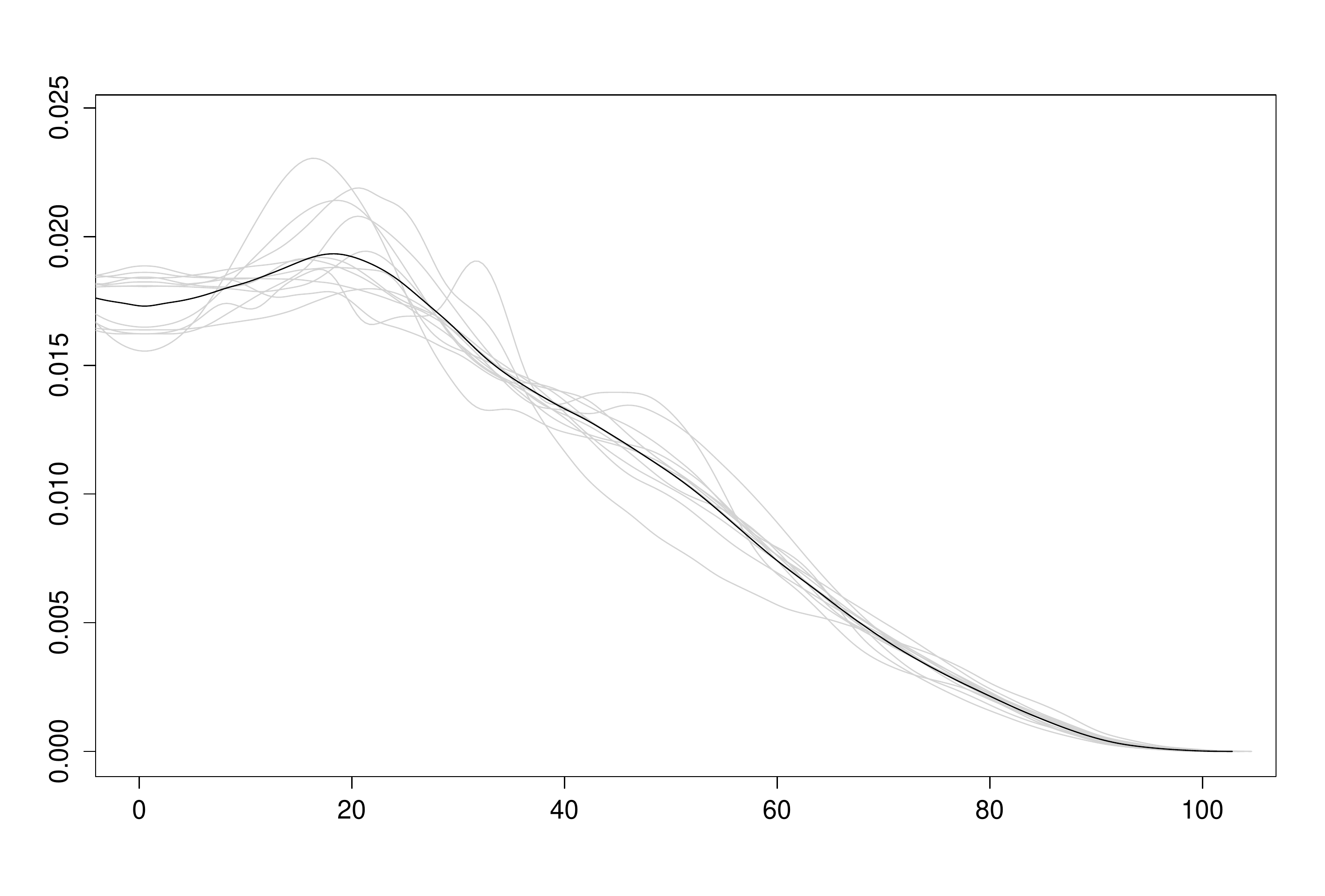}

\includegraphics[width=8cm]{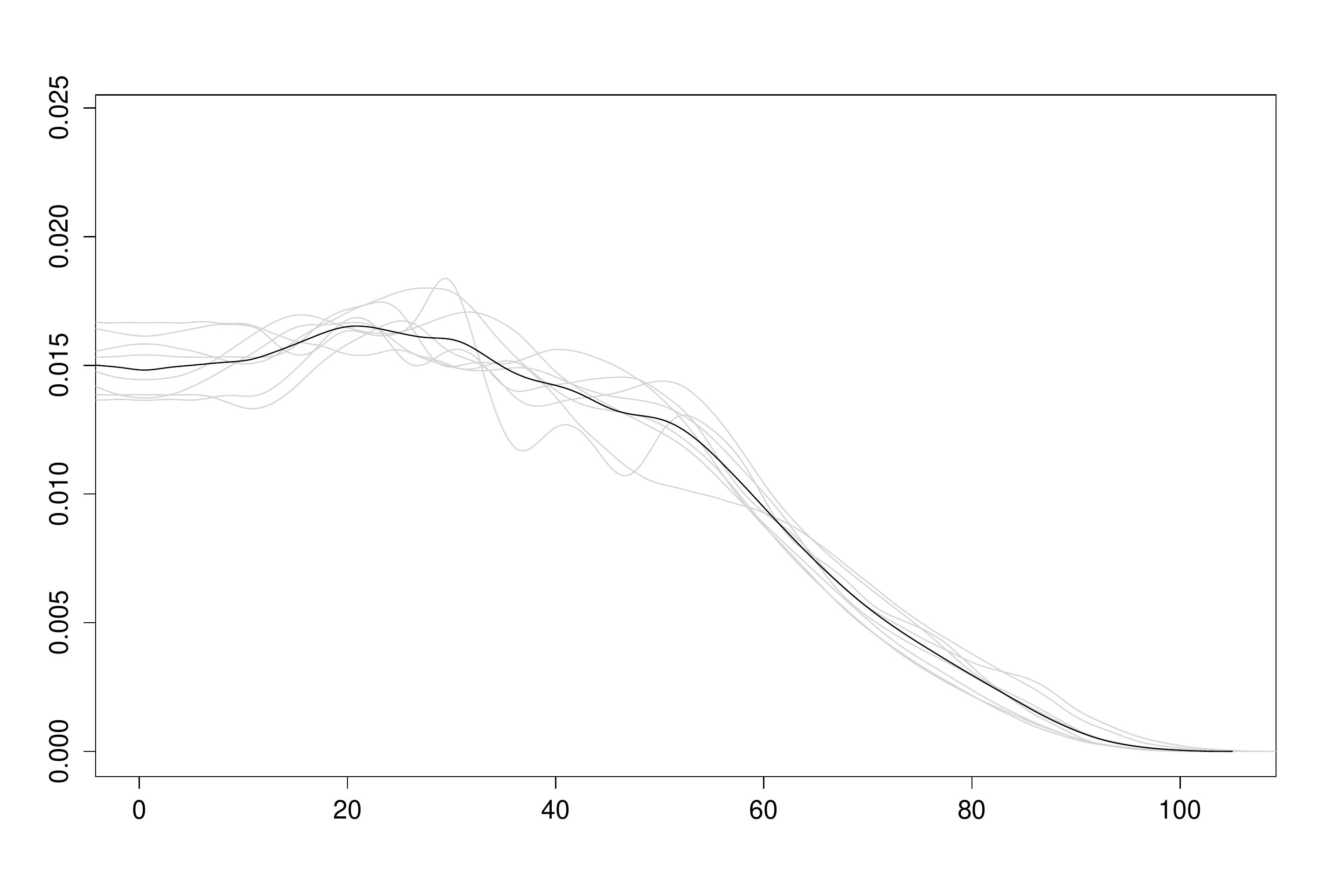}
\includegraphics[width=8cm]{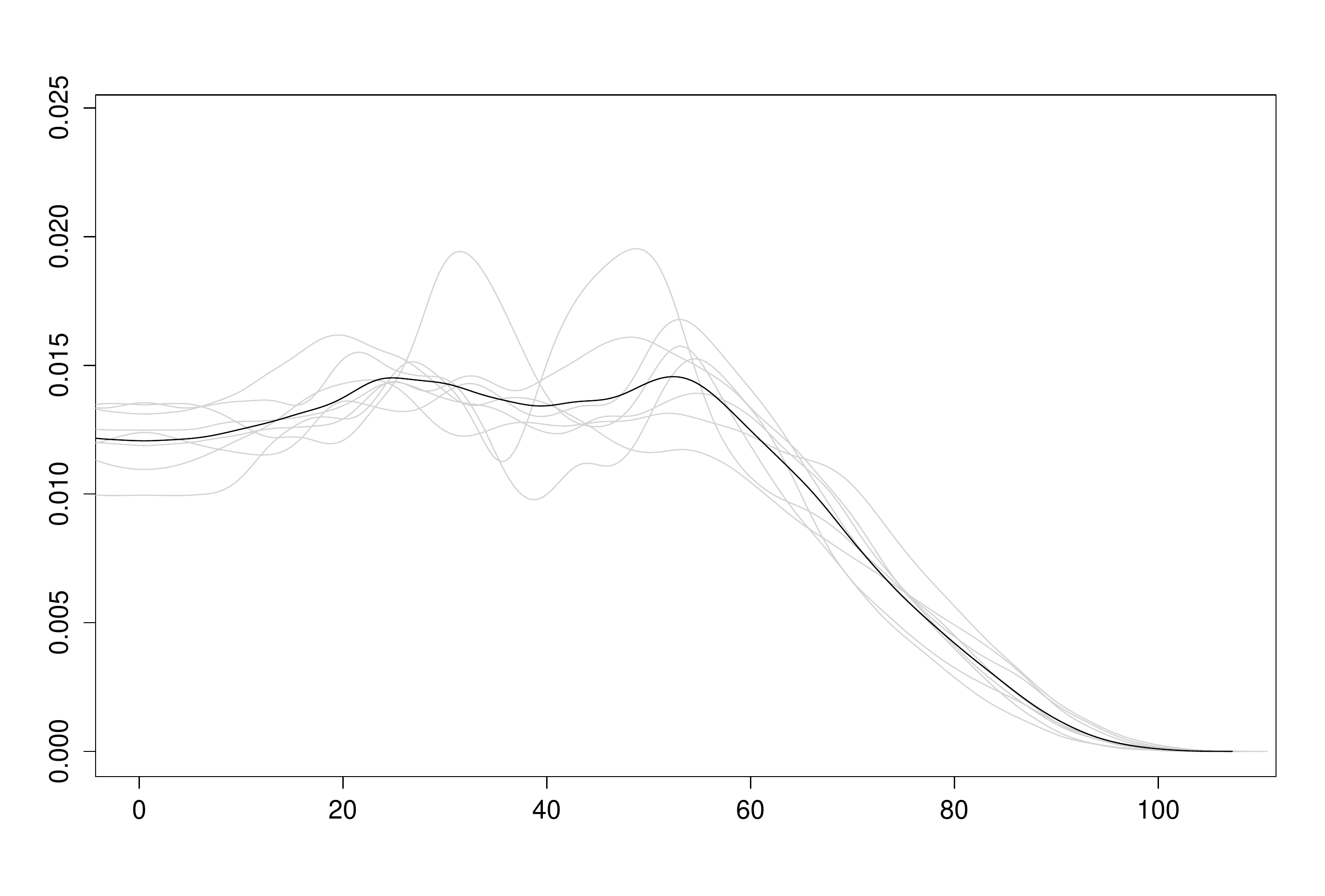}
\vspace{-1cm}
\caption{The $4$-barycenters (black curves) and the countries associated to each of them (grey curves)}
\label{Figure5}
\end{figure}
\end{center}

As a final summary, for the choice $\alpha= 2/36$ and $k=4$  
the composition of the clusters  is shown in Table \ref{Tabla.Grupos}, 
while  Figure \ref{Figure5} shows the density functions of the four 
barycenters, and those of the countries 
included in each group (groups I and II in top row, III and IV in bottom).
We recall that  trimmed countries  are shown in
 Figure \ref{Figure1}.

\vspace{-2cm}
\begin{table}[htb]
\begin{center}
\begin{tabular}{l|l}
\hline
{\bf Group I} &Belize, Bolivia, Guatemala, Haiti,         Honduras,  Nicaragua  
\\
[1mm]
\hline
{\bf Group II} &  Colombia,          Dominican Republic, Ecuador,           El Salvador,        Guyana,            Jamaica,                               
 \\
[1mm]
 & Mexico,  Panama, Paraguay,         Peru,              Suriname,         Venezuela
\\
[1mm]
\hline
{\bf Group II} & Argentina,                   Bahamas,                  Brazil ,                     
Chile,                      Costa Rica,                   Grenada,                     
St Lucia,                   
\\
[1mm]
& St Vincent and Grenadines
\\
[1mm]
\hline
{\bf Group IV} & Aruba,            Barbados,          Cuba,              Curacao,           Puerto Rico,        Trinidad and Tobago,
\\
[1mm]
&United States,      Uruguay 
\\
[1mm]
\hline
{\bf Trimmed} & Canada, Virgin Islands
\\
[1mm]
\hline
\end{tabular}
\caption{Barycenters and members of the groups. $k=4$ and $\alpha = 2/36$}
\label{Tabla.Grupos}
\end{center}
\end{table}

\vspace{-2cm}\section{Clustering aggregation in location-scatter}\label{loc-scatsection}
We present now some applications of trimmed $k$-barycenters in Wasserstein space in the aggregation of  model
based clustering. It is widely known that the success of a model-based clustering procedure 
depends strongly on the number, $k$, of `clusters' (underlying distributions)
and the degree of separation among them, the dimension of the space, $d$, 
and the sample size. Our proposal can be used with any model-based clustering
statistical engine. We emphasize that our goal here is not to provide arguments 
for or against any particular statistical engine, but, rather, to present applications 
showing the positive effects that the trimmed $k$-barycenter approach can provide.

Our choice for the statistical engine in this section is TCLUST (see
\cite{TCLUST} and \cite{Fritz} for details). It is a consistent procedure that applied to a 
sample of $d$- dimensional data yields  clusters obtained from
estimates of centers and shapes of the clusters based on ellipsoidal
regions. It involves the maximization of a pseudo-likelihood function
through a natural generalization of the spurious-outlier model introduced in
\cite{r11}. It is based on $k$ Gaussian distributions, allowing different
scatter matrices, and assumes the presence of some underlying positive
weights  associated to the distributions generating the set of `regular'
Gaussian observations. Additionally, to avoid degeneration of solutions, TCLUST includes
contraints on the eigenvalues of the covariance matrices to control the
relative shapes of the clusters. For a given trimming level $\alpha$, the method discards 
the proportion $\alpha$ of data consisting of points with a worse fit to the model
and reports the $k$ normal distributions 
(determined by their mean vectors, covariance matrices and associated weights)
that best summarize the remaining data. We recall that the aggregation of the involved weights will be carried through (\ref{pesos}).

{We will measure the deviation between outputs of the TCLUST algorithm in terms of 
\begin{equation}\label{alterHaus}
D^2(\{P_1,\dots,P_k\}, \{Q_1,\dots,Q_k\})=  \min_\sigma \Big \{\frac 1 k \sum_{j=1}^k 
\mathcal{W}_2^2(P_j,Q_{\sigma(j)})\Big\},  
\end{equation}
where $\sigma$ ranges in the set of permutations of $\{1,\dots,k\}$. We note
that when $A_n$, $A\subset \Pd$ are subsets with $k$ elements $d_H(A_n,A)\to 0$ if and only if $D(A_n,A)\to 0$,
but $D$ seems more appropriate from a computational point of view.

\subsection{Parallelization}

In this setup we compare the performance of TCLUST
applied to a large sample versus that
obtained when we apply trimmed $k$-barycenters to the set of $m$ solutions
obtained by TCLUST from $m$ subsamples produced through a partition of the sample.

Figure \ref{fig:1000000} shows the output of TCLUST with $k=5$ and trimming level $\gamma=0.05$
on a simulated sample of size $n=10^6$. This sample has been obtained from 5 normal distributions $N_i,
i=1,\dots,5$ with respective proportions of the data  15\%, 15\%, 15\%, 20\%
and 33\% and parameters:
\begin{eqnarray}\label{modelXX}\nonumber
 &&m_1=(0,0), \Sigma_1=(4,2,2,4),  m_2=(-3, 4), \Sigma_2=(2,-1,-1,4),
m_3=(6,6),\\ &&\Sigma_3=(2,0,0,3),m_4=(5,0), \Sigma_4=(2,0,0,2), m_5=(1,5),
\Sigma_5=(2,-1,-1,1).
\end{eqnarray}
The remaining 2\%, to be considered merely as noise, has been obtained from
a normal distribution with parameters $m_0=(2,2.5)$, $\Sigma_0=(4,0,0,4),$
playing a troubling `bridge-effect' among clusters. Additionally,
to explore the effect of higher dimension, we added independent
observations of a standard normal distribution to fill $8$ additional
variables, completing a data set of dimension $10$. We notice that the
configuration of this problem presents difficulties for TCLUST, because the
underlying normal distributions present a big overlap, being better suited
for a mixture model. A succesful clustering analysis requires very large sample sizes.
This will allow us to stress on the usefulness of parallelized computation.

We have also applied TCLUST 100 times, with the same parameters, on the 100
subsamples  of equal sizes, $10^4$, obtained through a partition of the
large sample, thus we are just considering a parallelization of the above
problem based on 100 units. To get a solution based on the ones given by the
100 units, we resort to the 0.1-trimmed $5$-barycenters. The procedure has
been applied to the meta-sample of the $500=5\times 100$ provided Gaussian
distributions, giving the consensus solution. The labels which relate those
5 distributions reported by the same unit have been used only to design the
(non-random) initialization steps of the algorithm. This is a natural and much more
efficient alternative to random initializations because we are guaranteeing
that in most cases we are choosing one distribution of every detected
cluster (at least by  one of the units).

In Figure \ref{fig:10000} we show (in gray) the ellipses corresponding
to the solutions, as well as those based on the  consensus solution (solid dark)
and on the large sample solution (dotted dark). We note
the instability of the solutions given by TCLUST for the two clusters on the
left, that for a very large sample has less effect, but is clear for
several solutions provided for the $10^4$ sized problems. We must stress the
role of trimming in the $k$-barycenter step, being able to discard the most
outlying solutions. In this sense, we would suggest to analyze the
sensibility of the final solutions against changes in the trimming level, 
with stability as evidence of a succesful clustering.

As  Figure \ref{fig:10000}  shows, there is complete agreement between the solution
obtained from the full data set and that obtained from the parallelized
version (with squared distance, $D^2=0.00175$). It is worth to mention that
the computation time in the parallelized case is considerably
shorter than that for the complete sample (33m vs 2h 8m in a MacBook
Pro with a 2,5 Ghz processor Intel Core i7, using the library ``parallel" in R with 8 cores). Regarding the reported weights and the aggregated obtained through (\ref{pesos}),
the greatest difference was 0.0035, corresponding to the SE cluster.

\begin{figure}
\centering
\includegraphics[width=14cm]{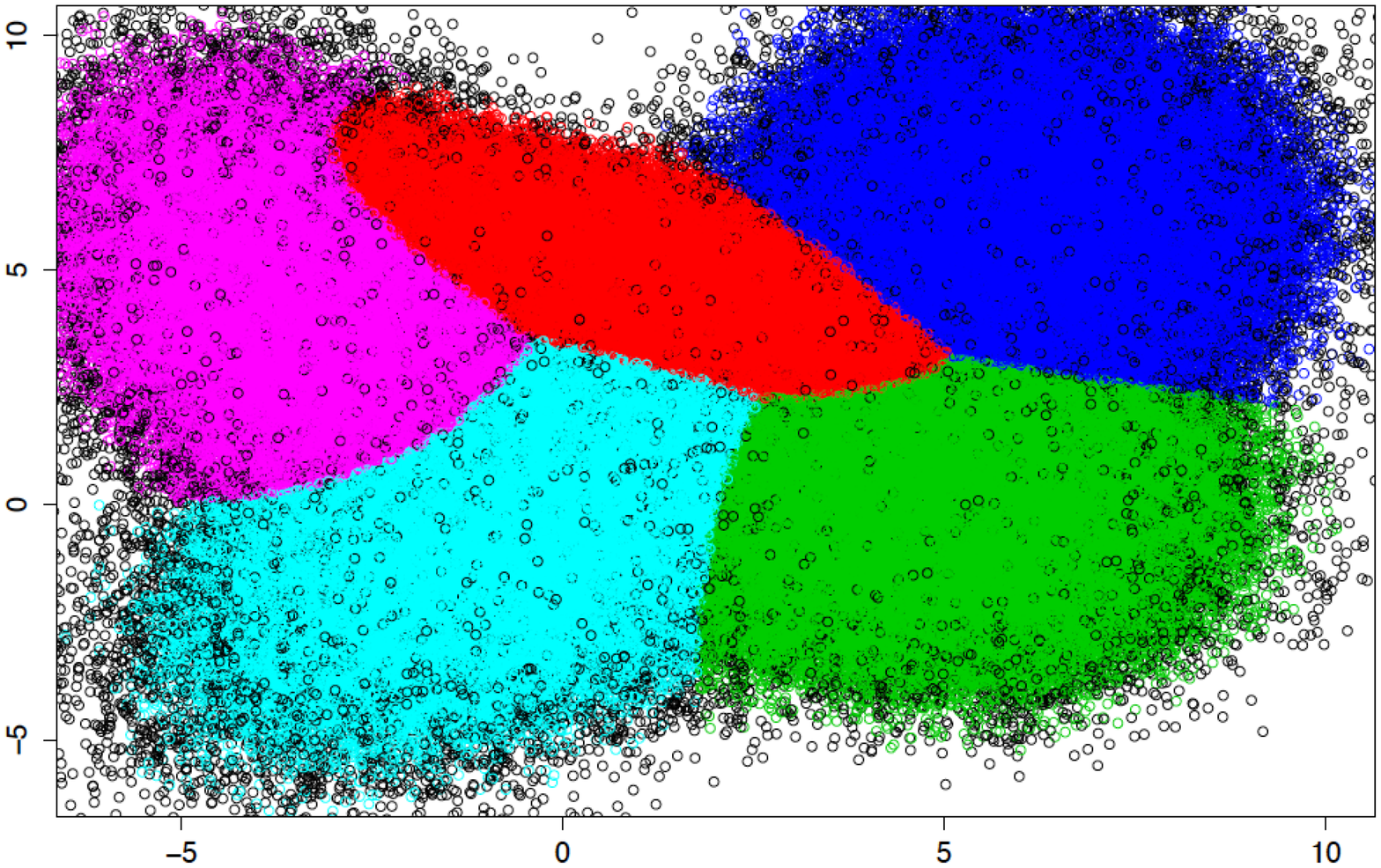}
\vspace{-1.5cm}
\caption{First and second coordinates of the TCLUST classification
($k=5$, $\alpha=0.05$, nstart = 150, restr.fact = 50, iter.max = 100, equal.weights = F) 
applied to simulated sample of size $10^6$ from model (\ref{modelXX}) in dimension 10. Black circles
represent trimmed points.}
\label{fig:1000000}
\end{figure}

\begin{figure}
\centering
\includegraphics[width=12cm]{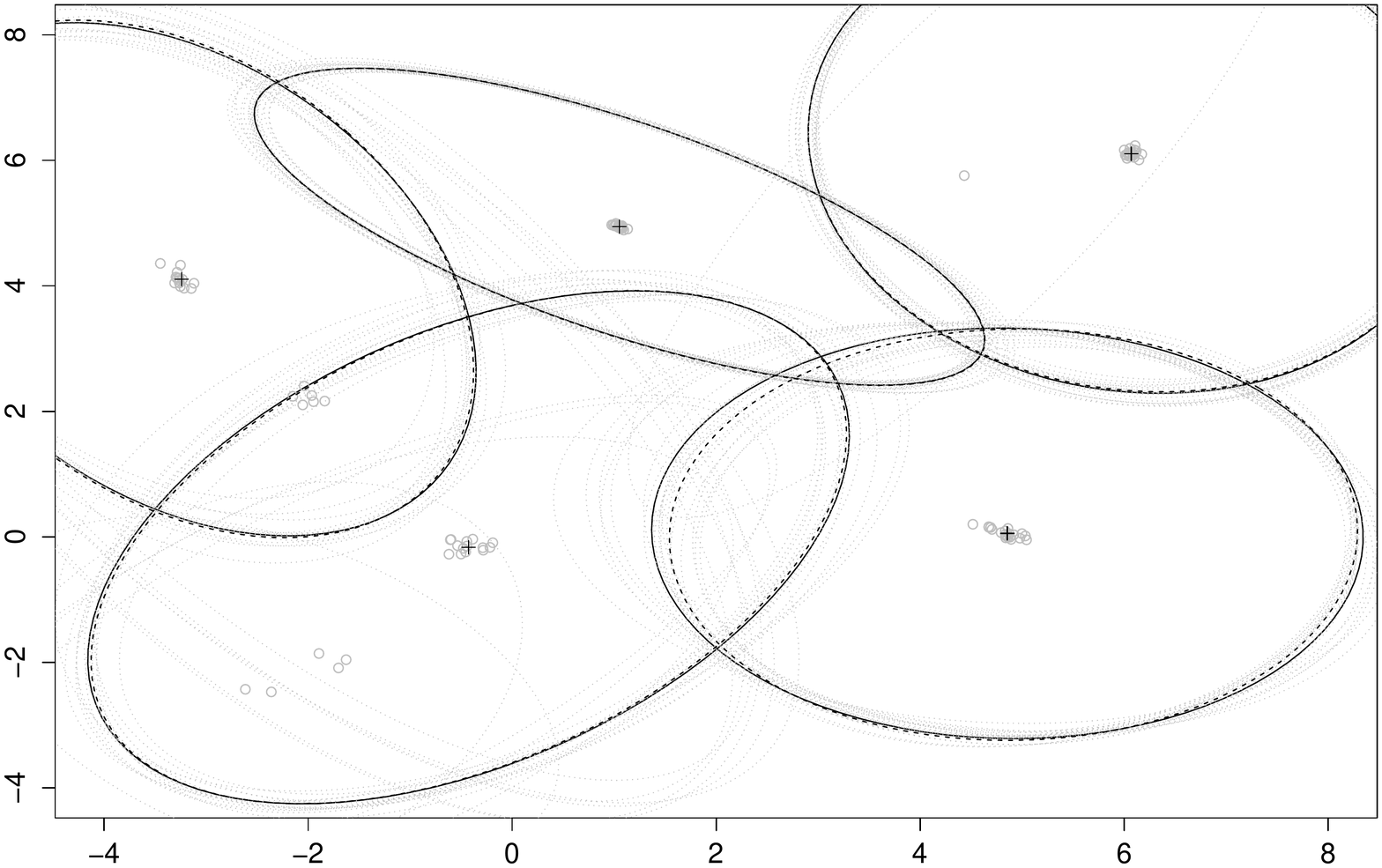}
\vspace{-1cm}
\caption{Gray: first two coordinates of 95\% level ellipses (with
their centers) of normal distributions reported by TCLUST on 100 subsamples of
size $10^4$ (a random partition of the full sample in Figure
\ref{fig:1000000}); solid black:
ellipses for 0.1-trimmed $5$-barycenters of the $5\times 100$ normal distributions
reported by TCLUST; dotted dark: TCLUST output for the full sample}
\label{fig:10000}
\end{figure}

\subsection{Resampling aggregation}\label{subragging}

The use of resampling methods to improve accuracy of statistical
prediction begins with the seminal works by Breiman (see e.g.
\cite{Breiman}).  In the clustering setting,  `bagged clustering' was
initiated in Leisch \cite{Leisch}, using an aggregation of the bootstrapped
solutions based on a combination of partitioning and hierarchical methods.
Later, Dudoit and Fridlyand \cite {Dudoit} introduced other variants,
resorting to plurality voting or to modifications of the dissimilarity
matrix to get the final cluster solution. Subagging is a term considered in
B\"uhlmann \cite{Buhlmann} as a `sobriquet for subsample aggregating where
subsampling is used instead of the bootstrap for the aggregation', while
`subragging' is a robust version which chooses the median instead of the
mean in the aggregation step. Since our aggregation procedure is based on a
robustified version of the $k$-barycenter, in the spirit, the following
examples could be considered also as a bragging and subragging approaches.

We have applied these principles to a simulated dataset generated from 
 model (\ref{modelXX}), but now we consider 18 additional independent N(0,1) 
variables to get a data set in dimension 20.  
We consider a sample of size 15000 and 100 samples of size 8000 obtained by
resampling in each of two scenarios: with  and without replacement. 
Figure \ref{resampling} is similar to the display in
Figure \ref{fig:10000}. In each case we applied TCLUST ($k=5$, $\alpha=0.05$, nstart = 150,
restr.fact = 10, iter.max = 100, equal.weights = F) to get the solution 
based on the full sample (dotted lines) and on 100 resamples (gray lines)
of size 8000 in each case with (left) or without (right) replacement. The aggregated solutions 
(solid lines) are given by the trimmed $k$-barycenter procedure ($\alpha=0.1$, $k=5$) 
applied to the $5\times 100$ estimations.

We see in Figure \ref{resampling} that subsampling looks more stable than bootstrap, 
but the aggregated solutions are similar by the effect of trimming. We note that the 
maximum difference between global and aggregated (through (\ref{pesos})) weights was 0.006 (resp. 0.002) for resampling with (resp.
without) replacement, while the squared distances between the $k$-sets
obtained between the solution obtained for the complete sample and the
obtained by aggregation were respectively 0.023 and 0.012. By considering
bootstrap samples (i.e. resampling with replacement and resample size equal
to the original sample size), $D^2$ changed from 0.023 to 0.010.

\begin{center}
\begin{figure}[htb]
\includegraphics[width=8cm]{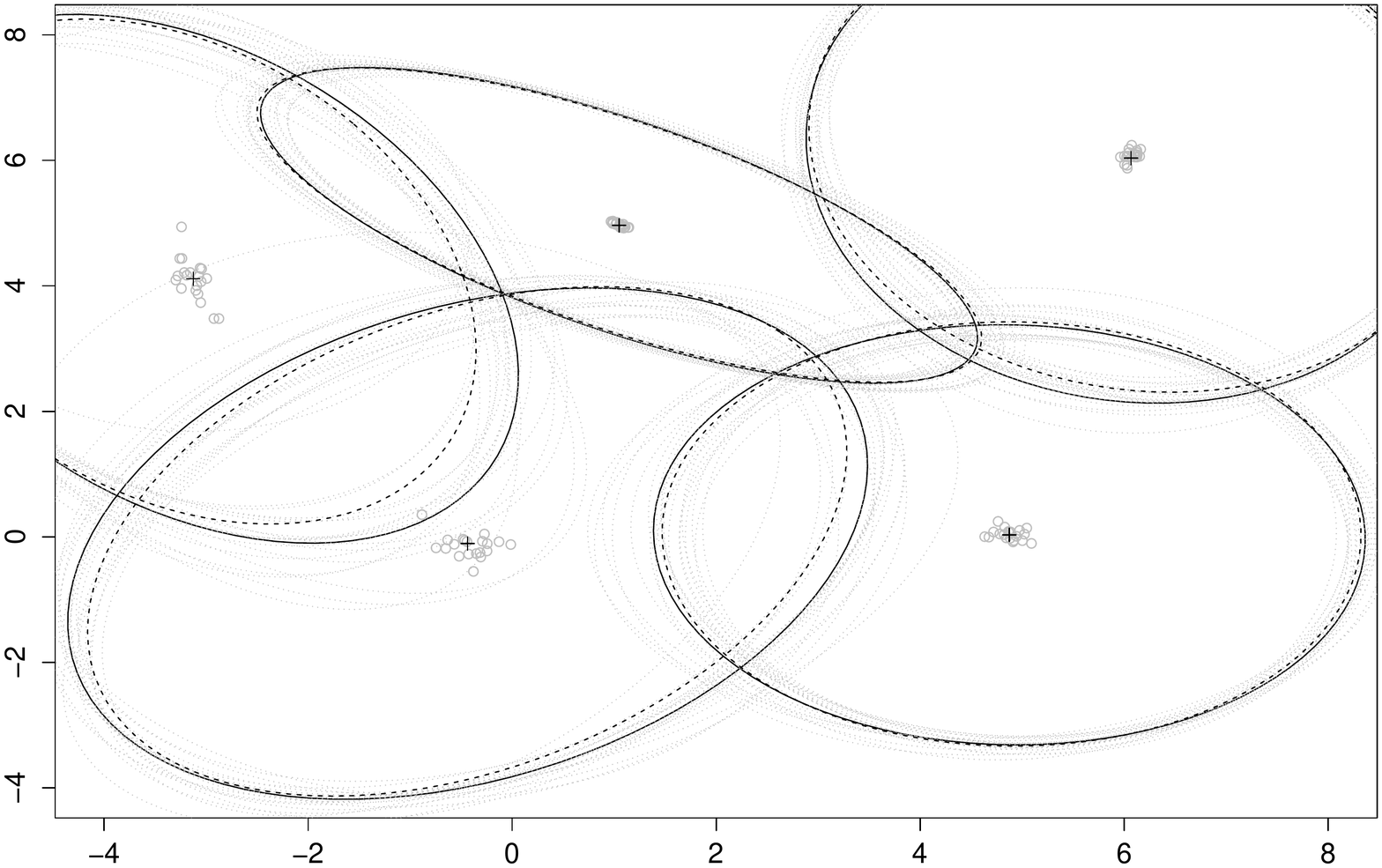}
\includegraphics[width=8cm]{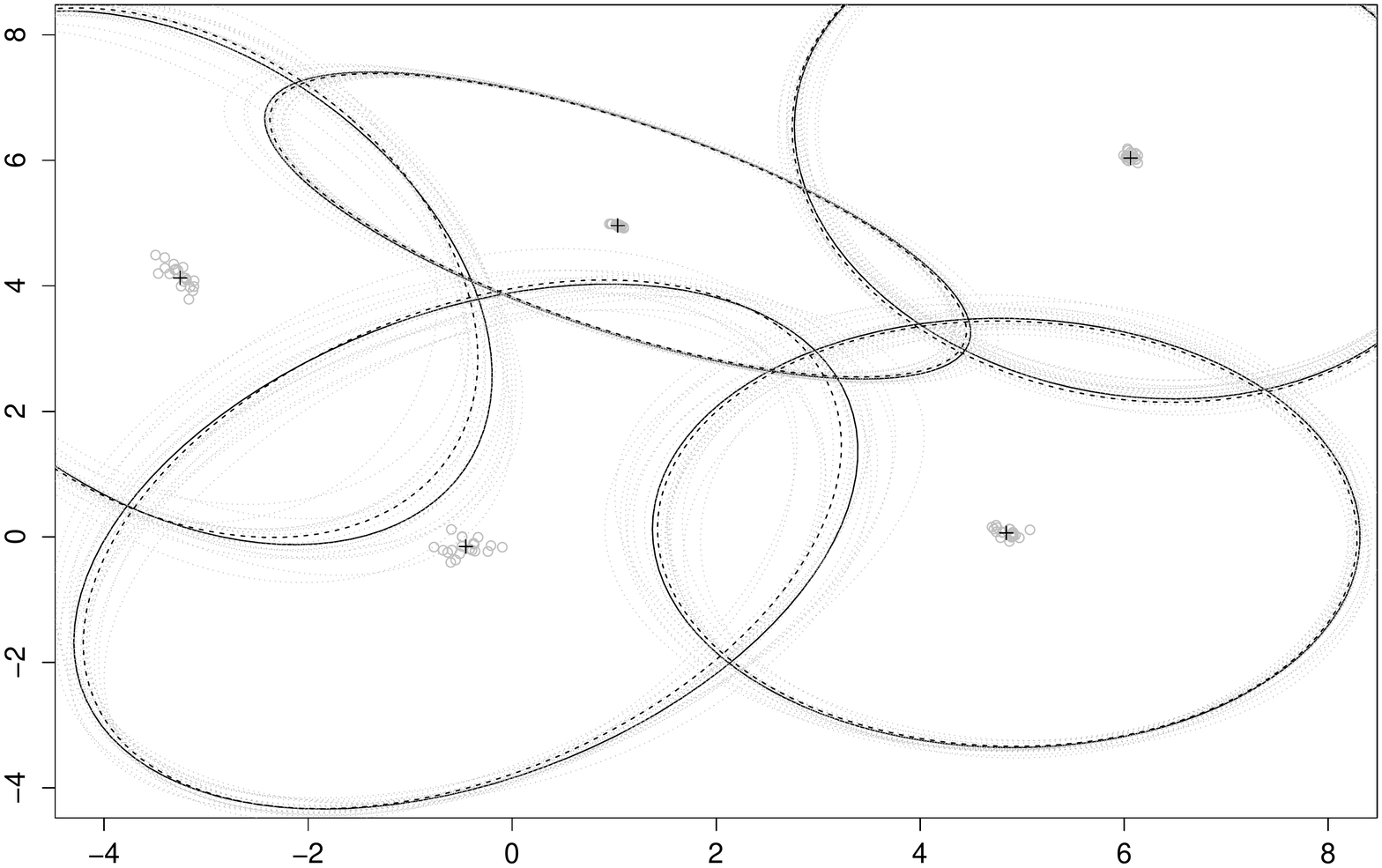}
\vspace{-1cm}
\caption{Dotted: ellipses provided by TCLUST ($k=5$, $\alpha=0.05$, nstart = 150,
restr.fact = 10, iter.max = 100, equal.weights = F) for the full dataset 
in subsection \ref{subragging}; Gray: 100 estimations
provided by TCLUST (with the same parameters) for resamples of sizes 8000
Solid: bragging (left) and
subragging (right) solutions given by the
$\alpha$-trimmed $k$-barycenters procedure (for $k=5$ and $\alpha=0.1$)}
\label{resampling}
\end{figure}
\end{center}

\vspace{-1cm}\subsection{Improving clustering initializations}\label{initializations}

Performance of model based clustering procedures depends on the initializations. Since most 
of these  clustering methods involve random initializations, 
the number of initial random starts should be dramatically increased to 
get guarantee of a successful identification of
the components. As we will see, our robust aggregation proposal can be
also helpful in this task by combining even unfortunate solutions. We will
present their effects in a simulated example which is not based on an 
isolated data set, but in a batch of them, which share components with common patterns.
We will show how trimmed $k$-barycenters could provide smart
initial solutions to clustering procedures, allowing to increase the number
of well-identified populations.

We consider 100 data sets of same sizes generated in the same way. Any data set contains 100 observations from 9 fixed normal distributions on $\Rea^{25}$. They also include a 5\% of contamination data. Only the two first coordinates of the involved distributions present differences. The distribution for each observation in the other 23 variables is
$N(0,I_{23})$. In Figure \ref{Agus1} we include the representation of a data set, where the different colors show the distributional origin of the observations. The central component is $N(0,2.25\times I_2)$, the other main components  are centered at the vertices of
a regular octagon centered at the origin with
radius  7. Their covariance matrices are characterized by their eigenvalues,
$(\sqrt3, \sqrt2)$, and their eigenvectors, one corresponding  to the
direction joining the center and the origin and the other to its
orthogonal direction. The  contaminated observations correspond to
data obtained from a mixture of 8 normal distributions with equal weights and covariance matrices $0.2\times I_2$,
whose centers are uniform perturbations of the coordinates of the vertices of a regular  centered octagon. For this  octagon  the radius is the maximum distance observed
from the regular observations to the center.

The dependence of TCLUST solutions 
with respect to the input parameters (trimming level and restriction factor) 
has been studied in \cite{Garcia2011} and \cite{Garcia2015} providing evidence
on stability as well as practical tools to assist in the choice of parameters. 
The choice of trimming level is a compromise between robustness and efficiency. 
It should be big enough for eliminating the effect of outliers, but not so big as to
eliminate an entire cluster. For small  sample sizes and  high dimensions,
low levels of restriction factor (controlling the relative shapes between clusters) are required for
avoiding spurious solutions.  These considerations lead to our choices of
0.05 for the trimming level and 4 for the restriction factor for applying TCLUST ($k=9$, $\alpha=0.05$, nstart = 50, restr.fact = 4,
iter.max = 20, equal.weights = F) to each data set in this batch. After we
computed the trimmed $k$-barycenters (for $k=9$) corresponding to the $9\times
100$ TCLUST estimations, for different trimming levels ranging
values from 0.50 to 0.05.  We find
also stability with respect to the trimming level 
for the $k$-barycenter estimation, but, as this trimming level approaches 
50\% the $k$-barycenter is no longer able to find the central population 
and with low trimming levels the inaccuracies of TCLUST in identifying 
the true components are inherited by trimmed $k$-barycenter. 
The best performance of $k$-barycenters corresponded to trimming levels 
around 20\%. A summary of the results is shown in Figure \ref{Agus1}.
Later we use this solution as the initial solution for TCLUST.

\begin{center}
\begin{figure}[htb]
\includegraphics[width=16cm]{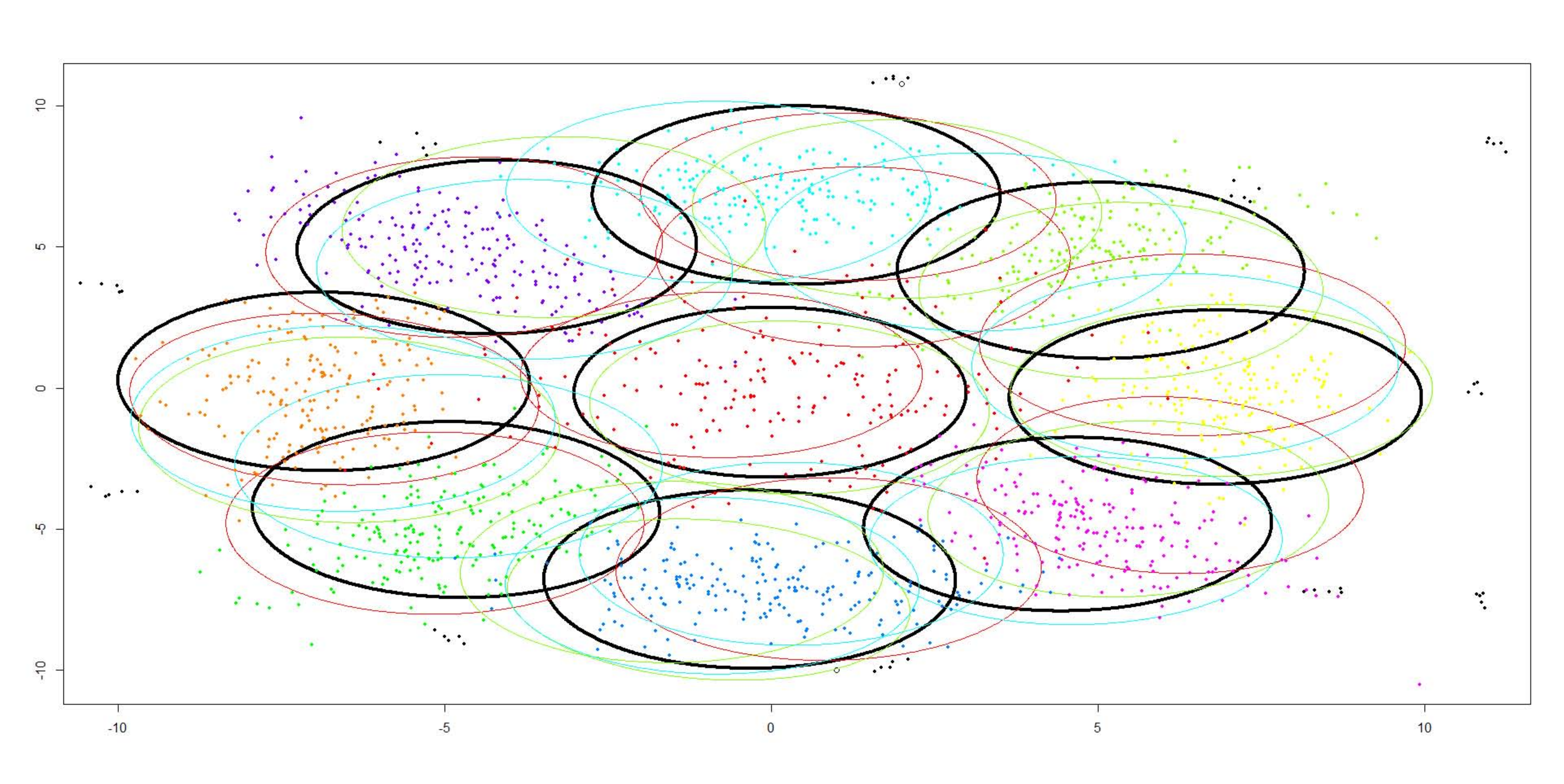}
\vspace{-1cm}
\caption{Black lines: ellipsoids of 20\% trimmed $9$-barycenter of the  
$9\times 100$  TCLUST estimations from the 100 samples described in subsection
\ref{initializations};  other colors: ellipsoids of TCLUST solutions
for first 5 datasets in batch (ellipses with same color come from the
TCLUST solution for the same dataset)}
\label{Agus1}
\end{figure}
\end{center}

\vspace{-1cm}Now we will compare the behavior of standard TCLUST estimations based on random initial starts, with those based on the start given by the aggregation just obtained through the  trimmed $k$-barycenter (kB TCLUST). For a better comparison, we also include (Oracle TCLUST) the behavior when the initial start is based just on the true parameters, a kind of oracle
information not available when applying clustering in real applications.  For the evaluation of  these  TCLUST proposals, we  reutilize the batch of 100 data sets and three criteria. The
first one corresponds to the overall percentage of rightly classified
observations (overall \%). For the second we consider the percentage of
datasets in which there is a bijection between estimated clusters and real
components, in the sense than more than 50\% of the observations from each
true component are rightly identified by the associated cluster
(bijection$>50\%$). The third one is related with the overall percentage of
true components, whose associated clusters identify more than the 75\% of
the true observations from the component (\% components$>75\%$).
Table \ref{tabla} gives the values obtained by the mentioned proposals when
applied to the batch of 100 datasets. In relation with the performance of
the mentioned proposals, kB TCLUST reached a similar performance than the
reference given by oracle TCLUST corresponding to right identification of
the 9 components and better performance than standard TCLUST that failed in
two of the 9 main populations.

\begin{table}[htp]
\begin{center}
\begin{tabular}{|c|c|c|c|}
\hline
\multicolumn{1}{|c}{}&\multicolumn{1}{|c|}{Overall}&{Bijection  $>
50\%$}&{\% (Components $>75\%$)}\\
\hline
{kB TCLUST}&{    86.34\%}&{98\%}&{88.33\%}\\
\hline
{Standard TCLUST}&{63.10\%}&{1\%}&{41\%}\\
\hline
{Oracle TCLUST}&{94.70\%}&{100\%}&{99.22\%}\\
\hline
\end{tabular}
\end{center}
\vspace{-.5cm}
\caption{Performance comparison of TCLUST for different initial starts}
\label{tabla}
\end{table}

\subsection{Application to cytometric analysis}

The  automated analysis of flow cytometry data is an active field in several research areas. While traditional methods of analysis   rely on subjective manual gating, in the last years, different groups have developed computational methods for identifying cell populations in multidimensional  data. However, the task is by no means easy, because to the expected variability between individuals we must add noises from  diverse sources. In fact, ``the lack of statistical and bioinformatics tools to parallel recent high-throughput technological advancements has hindered this technology from reaching its full potential" (sic, see \cite{Lo}).  

Our goal here is to provide evidence of the usefulness of statistical tools, arising from the  trimmed $k$-barycenters approach,  in connection with this topic. 
We  use the T cell phosphorylation dataset
analyzed in Pyne et al. \cite{Pyne2}, \cite{Pyne1} and available in the
Genepattern website ({\small http://www.broadinstitute.org/cancer/software/genepattern/FLAME-view-publish
ed-data}). This data set contains cytometric samples of 30 subjects in CD4,
CD45RA, SLP76 and ZAP70 before (B - 13 samples) and 5 min after stimulation
(A - 30 samples). There are differences in these
variables  for B and A samples as shown in \cite{Pyne1}, \cite{Pyne2} and references therein. It
is worth noting the recent availability of multilevel
methodologies designed for simultaneous modeling and registration of cytometric
data as the JCM proposal considered in \cite{Pyne1}.

 We applied trimmed $k$-baricenters (for k=5) to TCLUST estimation  corresponding to each
available sample in A and B sets. The three plots in Figure
\ref{cytometries} represent observations in two samples contained in A
(left and center panels) and one sample in B (right panel).
It
is possible appreciate in them the high variability between samples, even
when they belong to the same set, as it is the case for the
two samples on the left and the center panels. The plots show blue
ellipsoids for representing the TCLUST estimation and black ellipsoids
corresponding to the 30\% trimmed $k$-barycenter based on the 30 cytometries of group A. Since the data are
4-dimensional, notice that in all the graphics in this subsection, we use the plots based on the 2 first canonical
components corresponding to this trimmed $k$-barycenter and the black ellipsoids are kept  as a reference. 

Since we have two batches of 30 (A) and 13 (B) samples, we expect that an
application of trimmed $k$-barycenter with a high level of trimming (50\%),
allows to identify the most common pattern in the A set, which represents
nearly 70\% of the samples.  Trimming was able to identify most of the A
samples when we applied 50\% trimmed $k$-barycenters to the  full set of estimations provided from the samples. The procedure was able to eliminate 89.2\% of
samples from B, while 67.3\% of samples from A survived as non-trimmed.
In Figure \ref{finalcytom}, the left plot shows the trimmed k-barycenters of
the B samples obtained, for $k=5$ at different levels of trimming. We should
stress on the stability for most of the components estimations when using
different trimming levels and the changes in the configuration of these
components between  A and B. The plot on the right shows the ellipsoids 
corresponding to 50\% trimmed $k$-barycenters (gray) summarizing samples in 
B. The plot also includes the ellipsoids corresponding to
the TCLUST solutions (non-trimmed in blue and trimmed in red).

\begin{center}
\begin{figure}[htb]
\includegraphics[width=5cm]{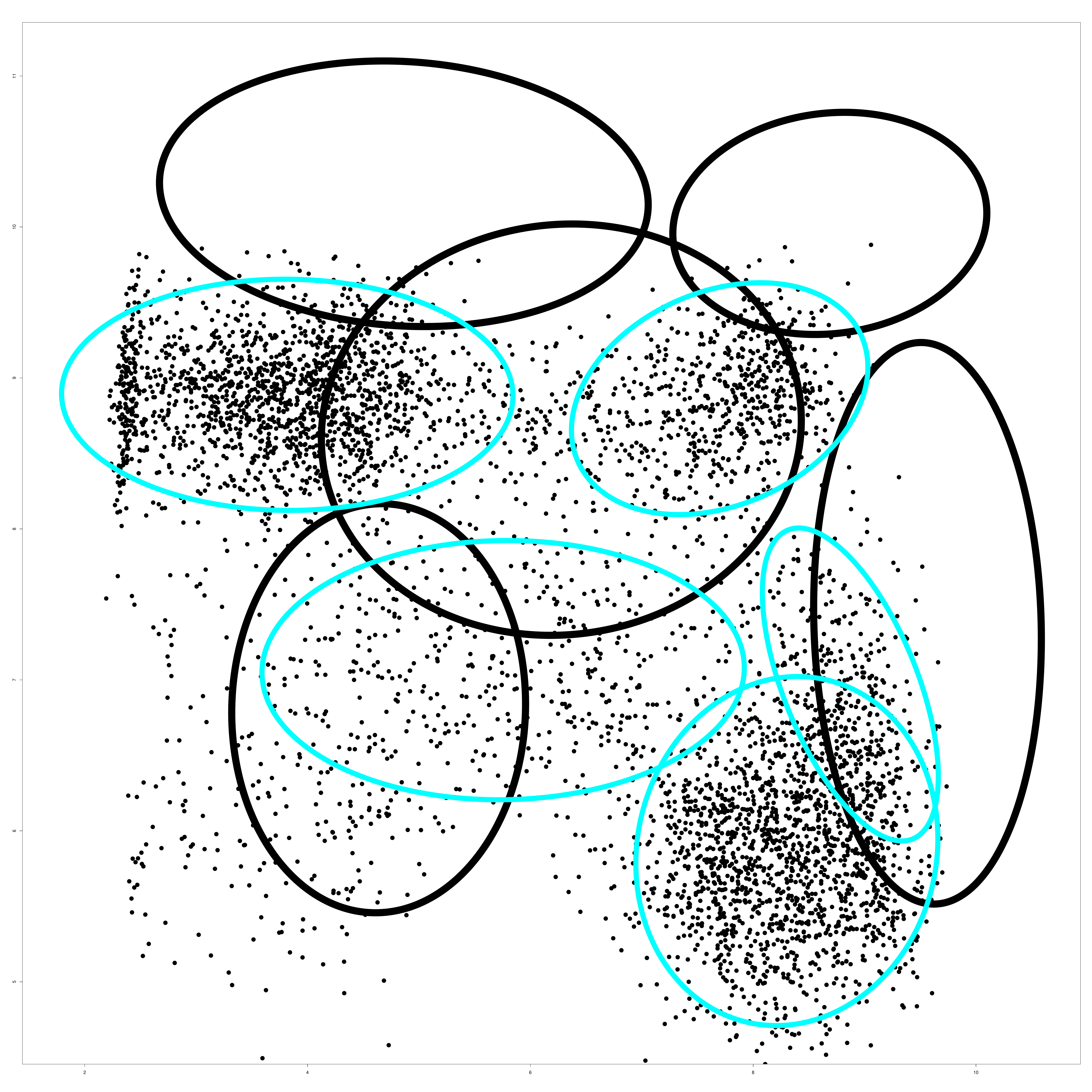}
\includegraphics[width=5cm]{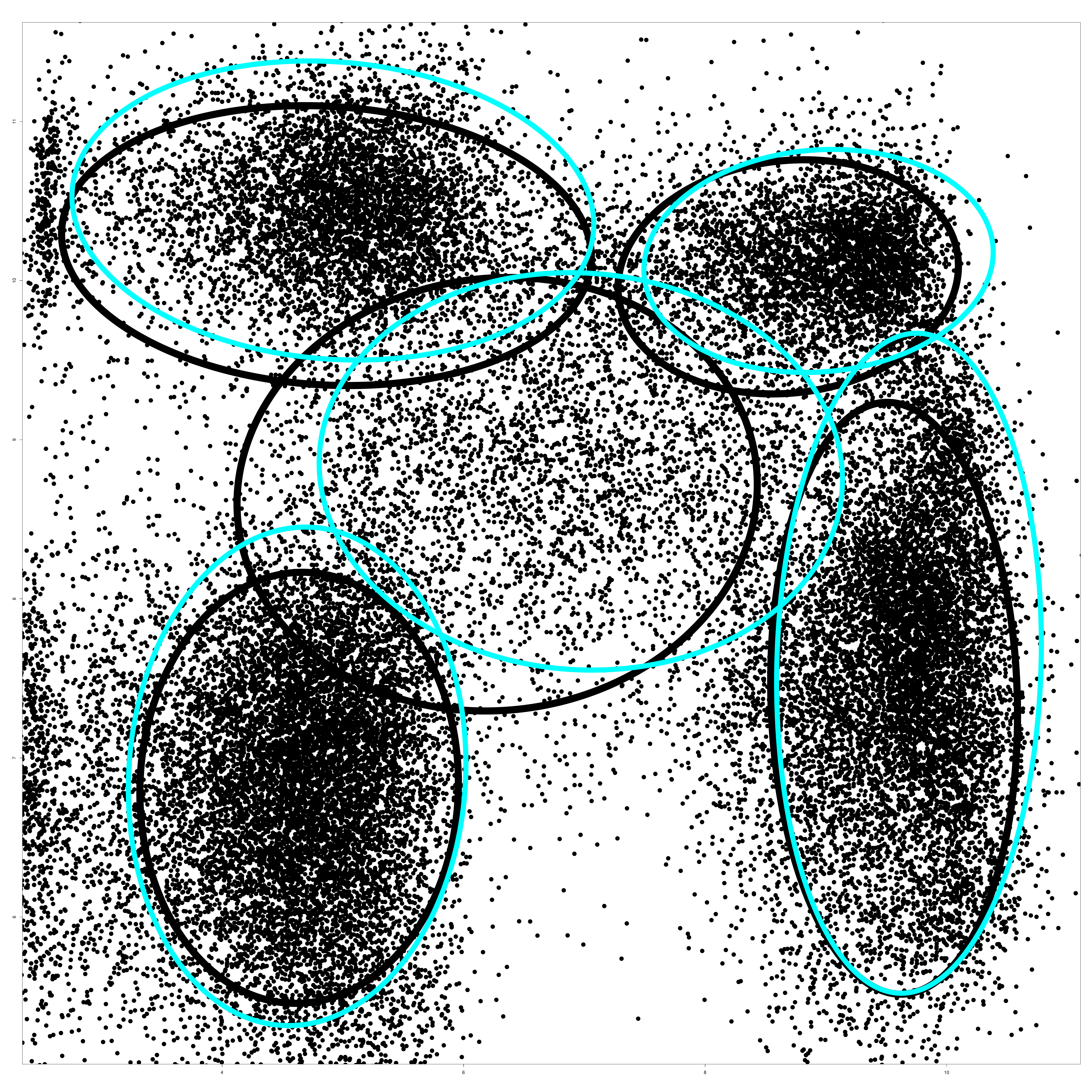}
\includegraphics[width=5cm]{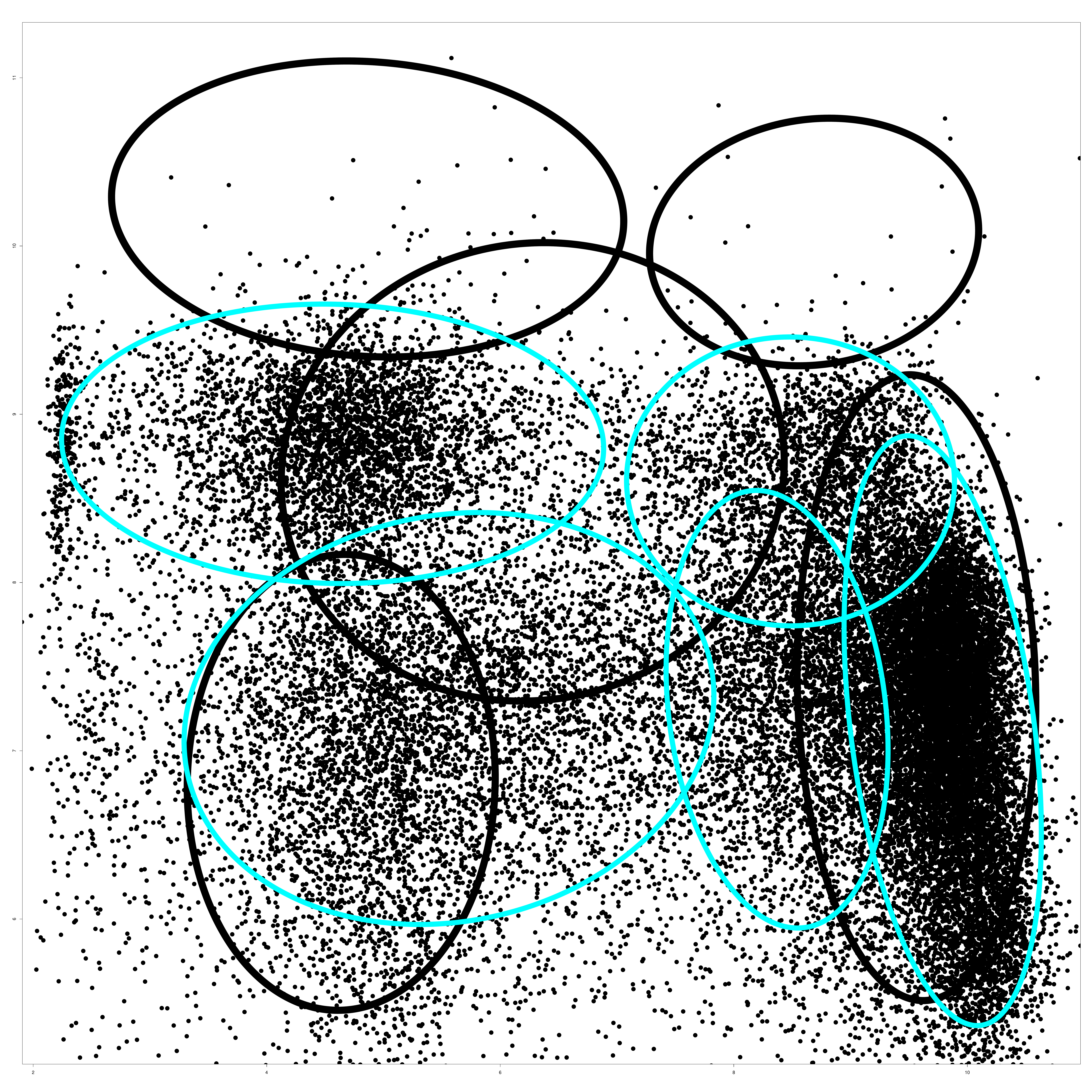}
\caption{Left and middle: cytometries of
two individuals after stimulation (group A); right:
cytometry of an individual in group B. Blue ellipsoids: TCLUST (k=5, $\alpha=0.15$, nstart=400, iter.max=50,
restr.fact =25); black ellipsoids: trimmed
$k$-barycenter ($k=5$, $\alpha=0.3$) based on the $30 \times 5$ estimations
obtained through  TCLUST from the 30 cytometries of group A.}
\label{cytometries}
\end{figure}
\end{center}

\begin{center}
\begin{figure}[htb]
\includegraphics[width=8cm]{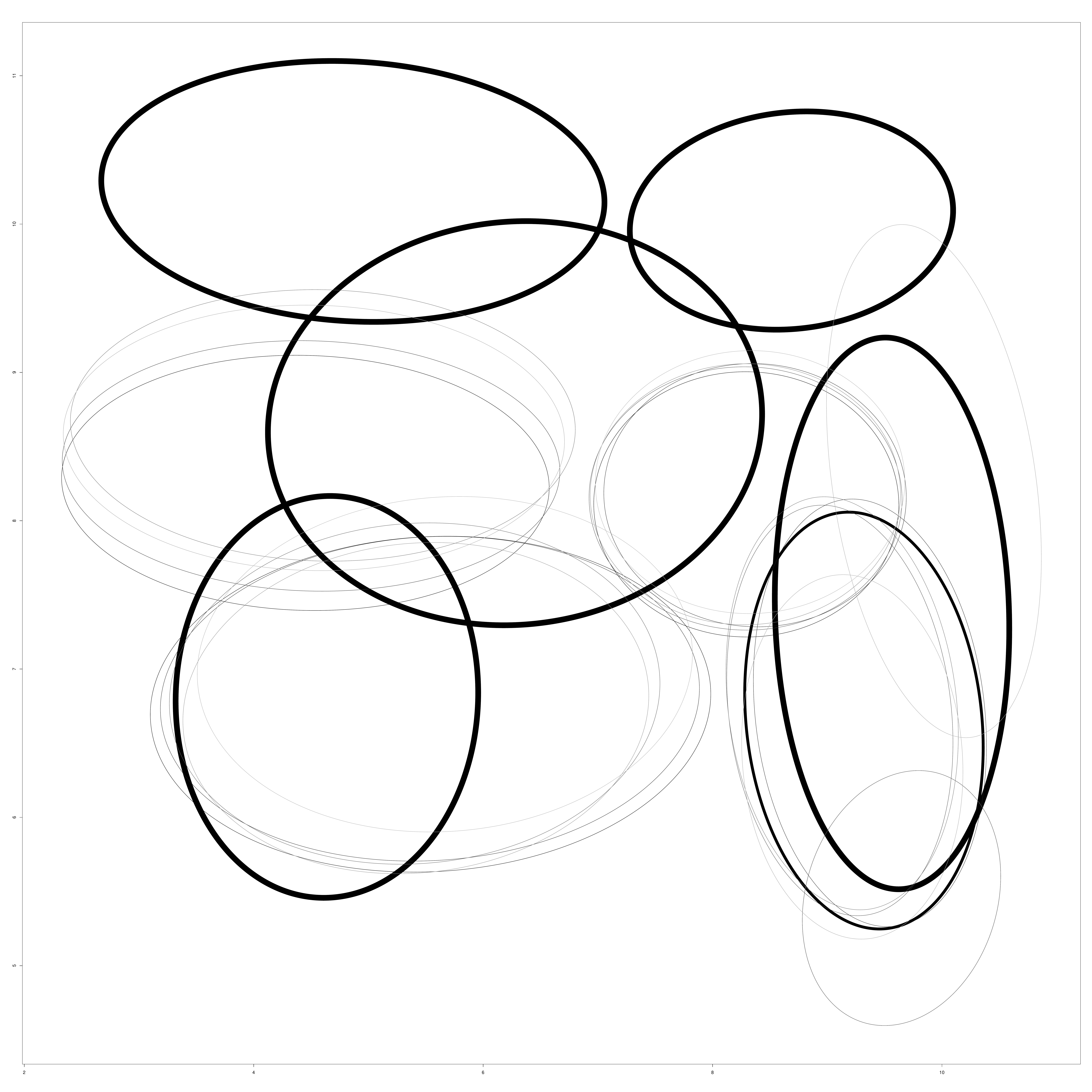}
\includegraphics[width=8cm]{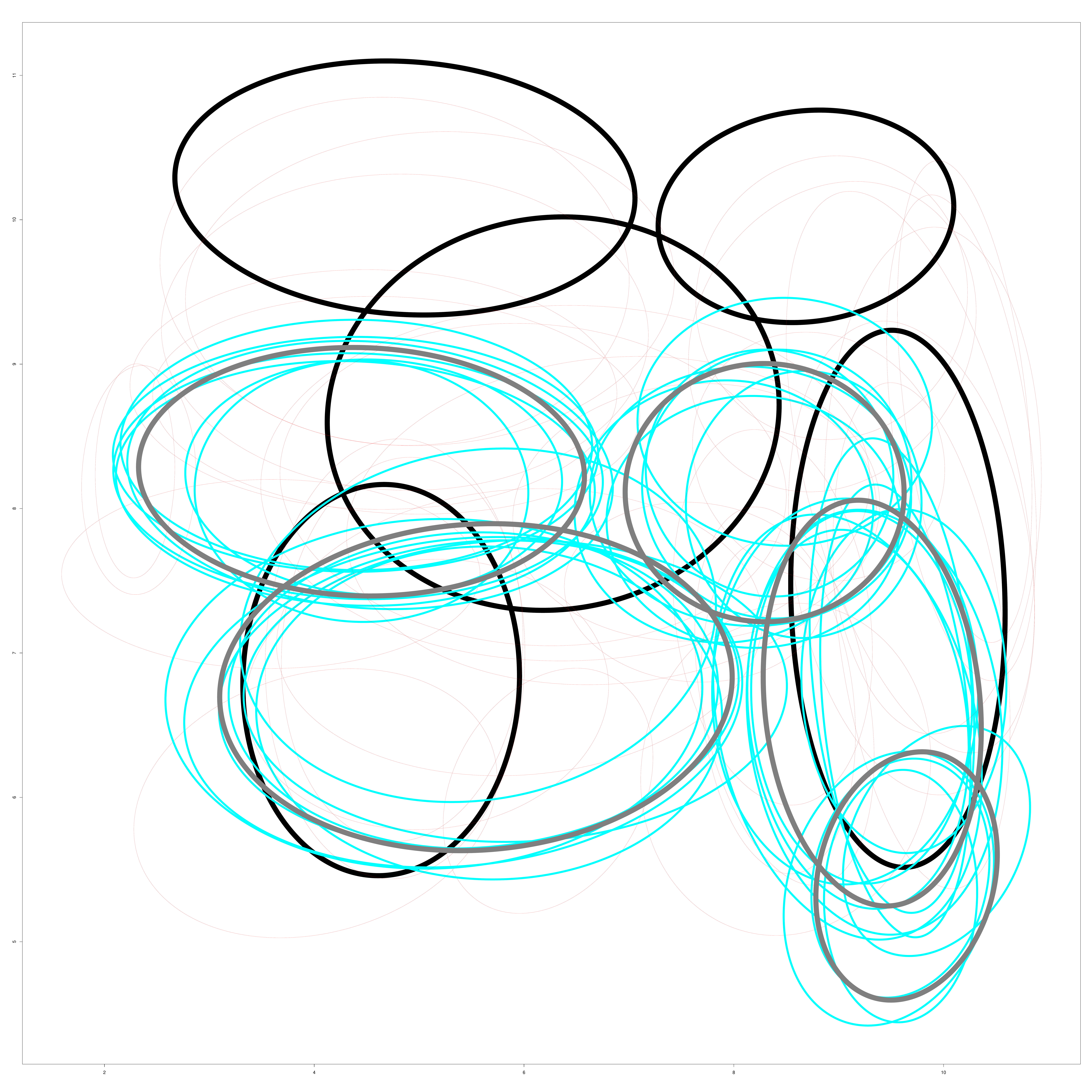}
\caption{In the canonical coordinates of the
30\% trimmed $5$-barycenters corresponding to the A samples. Left: trimmed $5$-barycenters of B samples
for different trimming levels (50\%, 40\%, 30\%, 20\%, 10\% and
5\%; gray level ranging from black to white). Right: 
trimmed $5$-barycenters for both sets of estimations and trimming level
$50\%$, and different estimations provided by TCLUST, in red the trimmed ones.}
\label{finalcytom}
\end{figure}
\end{center}

\vspace{-2cm}\section{Appendix}\label{AppWass}

Here we present the proofs of the main results in the paper. Most of the proofs make use of some 
well-known features of transportation cost metrics. For the sake of readability we include also some 
relevant facts concerning the $L_2$-Wasserstein distance defined in (\ref{int1}) in connection with the 
present work. We refer to \cite{Villani2} for a comprehensive approach.

We note first that the infimum in (\ref{int1}) is attained, that is, there exists a pair $(X,Y)$, 
defined on some probability space, with ${\cal L}(X)=P$ and ${\cal L}(Y)=Q$ such that $\Exp\|X-Y\|^2=\mathcal{W}_2^2(P,Q)$.
Such a pair $(X,Y)$ is called a  optimal transportation plan (o.t.p.) or optimal coupling 
for $(P,Q)$.
If the probability $P$ has a density, the o.t.p. $(X,Y)$ for $(P,Q)$
can be represented as $(X,T(X))$ for some suitable 
map $T$. This {\it optimal transport map}, minimizing the transportation cost  for $(P,Q)$  
is the $P$-a.s. unique cyclically monotone map that transports $P$ to $Q$.
Thus, optimality is a feature of the map itself and cyclically monotone maps are always 
optimal maps from $P$ to the image measure. As an example, 
an  affine map, $T:\Read \to \Read$, written in matrix notation as $T(x)=Ax+b$, is an optimal map if 
and only if $A$ is a (symmetric) positive semidefinite matrix. Optimality of maps 
is not generally preserved by composition. However, some kind of operations 
like positive linear combinations and point-wise limits of optimal maps keep optimality. 

For probabilities on the real line, if  $F_P^{-1}$ and $F_Q^{-1}$ are the quantile functions 
associated to $P$ and $Q$, they are an \Wd-o.t.p., that is,
\begin{equation}\label{casoreal}
\Wd(P,Q)=\left(\int_0^1\left(F_P^{-1}(t)-F_Q^{-1}(t)\right)^2dt\right)^{1/2}.
\end{equation}
In higher dimension there is no equivalent simple expression. However, if $m_P,m_Q$ are the means of $P$ and 
$Q$, and $P^*,Q^*$ are the corresponding centered in the mean probabilities, then we can focus on the case of centered probabilities because
\begin{equation}\label{location}
\Wdd(P,Q)=\|m_P-m_Q\|^2+\Wdd(P^*,Q^*).
\end{equation}

Recall that $\Pd$ is the set of probabilities on $\mathbb{R}^d$ with finite
second moment. Convergence in the $\Wd$  metric on $\Pd$, $\Wd(P_n,P) \conv 0$, is characterized by
\begin{equation}\label{convergences}
  P_n \convw P \ \ \mbox{ and } \ \int_{\Read}\|x\|^2P_n(dx) \conv \int_{\Read}\|x\|^2P(dx).
\end{equation}
Finally, we mention that if $(P_n)_n, (Q_n)_n$ are sequences in $\Pd$, such that
$P_n \convw P$ and $Q_n \convw Q$, then $\Wd(P,Q)\leq \liminf\Wd(P_n,Q_n)$.

A main tool for proving convergence results in the space  $W_2(\Pd)$ (recall the definition (\ref{W2space}) in Section 2)
is given by the next couple of results, extending (\ref{convergences}) and the subsequent comment 
for $\Wd$ (see e.g. Theorem 6.9 and Remark 6.12 in \cite{Villani2}) to the Wasserstein distance, $\mathcal{W}_{\mathcal{P}_2}$, defined on $W_2(\Pd)$ . 
\begin{Theo}\label{convW2}
Assume $(\mu_n)_n, \mu \in W_2(\Pd)$, and consider the probability concentrated at zero, $\delta_{\{0\}}$ 
(that can be substituted by any other fixed probability in $\Pd$). Convergence $\mathcal{W}_{\mathcal{P}_2}(\mu_n ,\mu)\conv 0$ 
holds if and only if 
\begin{equation}\label{condW2}
\mu_n \convw \mu \ \ \mbox{ and } \lim_{R\conv \infty} \limsup_{n\to \infty} \int_{\mathcal{W}_2(\delta_{\{0\}},P)>R}\Wdd(\delta_{\{0\}},P)\mu_n(dP)=0.
\end{equation}

\end{Theo}

\begin{Prop}\label{propiedades} (Lower semicontinuity).
If the sequences $(\mu_n)_n, (\nu_n)_n$ in $W_2(\Pd)$ are such that $\mu_n \convw \mu$ and $\nu_n 
\convw \nu$, then $\mathcal{W}_{\mathcal{P}_2}(\mu,\nu)\leq \liminf\mathcal{W}_{\mathcal{P}_2}(\mu_n,\nu_n)$. 
\end{Prop}

Now we prove the main results concerning existence and consistency of $k$-barycenters.

\begin{Theo}\label{existkbar} (Existence of $k-$barycenters)
Let $\mu \in W_2(\Pd)$ and, for $k\geq 1$, define \begin{equation}\label{k-var}
V_k(\mu):= \inf\left\{\int \min_{i\in\{1,\dots,k\}}\Wdd(P,Q_i)\mu(dP) , \{Q_1,\dots,Q_k\} \subset \Pd \right\}.
\end{equation}
Then $V_k(\mu)\leq V_{k-1}(\mu)\leq \dots \leq V_1(\mu)<\infty$ and all the inequalities are strict 
unless $\mu$ is supported on less than $k$ elements of $\Pd$. Moreover, there exists a 
$k$-barycenter of $\mu$, say ${\bf \bar M}=\{\bar M_1,\ldots,\bar M_k\}\subset \Pd,$ that satisfies
\begin{equation}\label{k-bar}
\int \min_{i\in\{1,\dots,k\}}\Wdd(P,\bar M_i)\mu(dP) =V_k(\mu).
\end{equation}
\end{Theo}
\medskip
\noindent
\textbf{Proof: } From the assumption $\mu \in W_2(\Pd)$  we see that 
$$0\leq V_k(\mu)\leq V_{k-1}(\mu)\leq \ldots \leq V_1(\mu)\leq \int \Wdd(P,\delta_{\{0\}})\mu(dP)<\infty.$$
Let $\{{\bf H}_n\}_n=(\{Q_1^n,\ldots,Q_k^n\})_n$ be a minimizing sequence of $k$-sets, namely, such that 
$$\int \Wdd(P,{\bf H}_n)\mu(dP) \conv V_k(\mu).$$
We first prove that the sequence $\{{\bf H}_n\}_n$ is bounded. After a rearrangement, if necessary,
we can assume that $\mathcal{W}_2(Q_1^n,\delta_0)\leq \cdots \leq \mathcal{W}_2(Q_k^n,\delta_0)$. Now, since
$\mathcal{W}_2^2({\bf H}_n,\delta_0)\leq 2\mathcal{W}_2^2({\bf H}_n,P) +2\mathcal{W}_2^2 \Wd(P,\delta_{\{0\}})$
we see, integrating, that $\mathcal{W}_2^2({\bf H}_n,\delta_0)$ is a bounded sequence, that is,
$\mathcal{W}_2^2(Q_1^n,\delta_0)$ is bounded, and, consequently, $\{Q_1^n\}_n$ is a tight sequence. Taking subsequences
we can assume that for some $l\in \{2,\ldots,k\}$, $Q_i^n\to_w Q_i$, $i=1,\ldots,l$, while $\mathcal{W}_2^2(Q_i^n,\delta_0)\to\infty$
for $i\geq l+1$. By lower semicontinuity and Fatou's Lemma we find that
\begin{eqnarray*}
&&V_l(\mu)\leq \int \min_{i=1,\dots,l}\Wdd(P,Q_i)\mu(dP)\leq \int \liminf_{n\to \infty}\min_{i=1,\dots,k}\Wdd(P,Q_i^n)\mu(dP)\leq \\ && \liminf_{n\to \infty} \int \min_{i=1,\dots,k}\Wdd(P,Q_i^n)\mu(dP)=\liminf_{n\to \infty} \int \Wdd(P,{\bf H}_n)\mu(dP)=V_k(\mu)\leq V_l(\mu).
\end{eqnarray*}
Therefore $\{Q_1,\ldots,Q_l\}$ is a $l$-barycenter of $\mu$ and, for any choice of $\{Q_{l+1},\ldots,Q_{k}\}$,
$\{Q_1,\ldots,Q_k\}$ is a $k$-barycenter of $\mu$. If $\mu$ is not supported on an $l$-set then given an 
$l$-barycenter of $\mu$ there exists $Q_{l+1}$ and $r>0$ such that $\mathcal{W}_2(Q_i,Q_{l+1})>2r$, $i=1,\ldots,l$
and $\mu(B_{\mathcal{W}_2}(Q_{l+1},r))>0$. But then, on the ball $B_{\mathcal{W}_2}(Q_{l+1},r)$, we have
$\min_{i=1,\dots,l+1}\Wdd(P,Q_i)<\min_{i=1,\dots,l}\Wdd(P,Q_i)$ and, as a consequence $V_{l+1}(\mu)<V_{l}(\mu)$.
\FIN

We note at this point that for $\mu \in W_2(\Pd)$ and $Q\in \Pd$, 
$$\mathcal{W}_{\mathcal{P}_2}^2(\mu,\delta_Q)=\int \Wdd(P,Q)\mu(dP).$$
In particular, for $P,Q \in \Pd$, $\mathcal{W}_{\mathcal{P}_2}(\delta_P,\delta_Q)=\Wd(P,Q)$. 
We note also that the set of probabilities with support on the $k$-set $\bar{\bf Q}=\{Q_1,\dots,Q_k\}$, 
that we will denote by $\Prob(\bar{\bf Q})$, is a closed convex set in $W_2(\Pd)$ and 
\begin{equation}\label{H0}
\mathcal{W}^2_{\mathcal{P}_2}(\mu, \Prob(\bar{\bf Q})):=\inf\{\mathcal{W}^2_{\mathcal{P}_2}(\mu,\nu), 
\nu \in \Prob(\bar{\bf Q})\} =\int \min_{i=1,\dots,k}\Wdd(P,Q_i)\mu(dP).
\end{equation}
In particular, when ${\bf \bar M}$ is a $k$-barycenter of $\mu$, this and (\ref{k-bar}) yield the characterization
\begin{equation}\label{clconv}
V_k(\mu)=\mathcal{W}^2_{\mathcal{P}_2}(\mu, \Prob(\bar{\bf M})).
\end{equation}

\begin{Theo}(Consistency of $k$-barycenters)\label{consistencykbar}
Let $(\mu_n)_n, \mu $ be  probabilities { on $W_2(\Pd)$} such that  $\mathcal{W}_{\mathcal{P}_2}(\mu_n, \mu)\to 0$.  
Then  the  $k$-variations of $\mu_n$  converge, $V_{k}(\mu_n)\conv V_{k}(\mu)$. 
If  $\mu$ is not supported on a $(k-1)$-set of $W_2(\Pd)$ and ${\bf \bar M}_n$ is any 
$k$-barycenter of $\mu_n$,   the sequence  $({\bf \bar M}_n)_n$ is sequentially compact  and 
any limit is a  $k$-barycenter of $\mu$. { If}  $\mu$ has a unique  $k$-barycenter, ${\bf \bar M}$, 
then  ${\bf \bar M}_n$ converges to ${\bf \bar M}$ in Hausdorff distance.
\end{Theo}
\medskip
\noindent
\textbf{Proof:} If ${\bf \bar M_0}$ is a $k$-barycenter of $\mu$, the convergence 
$\mathcal{W}_{\mathcal{P}_2}(\mu_n, \mu)\to 0$ implies the convergence of the 
distances to the closed set $\Prob(\bar{\bf M}_0)$, $\mathcal{W}^2_{\mathcal{P}_2}(\mu_n, 
\Prob(\bar{\bf M}_0)) \conv \mathcal{W}^2_{\mathcal{P}_2}(\mu, \Prob(\bar{\bf M}_0))$. Hence, recall (\ref{clconv})
\begin{equation}\label{H2}
\int \min_{i=1,\dots,k}\Wdd(P,M_i^0)\mu_n(dP)\conv \int \min_{i=1,\dots,k}\Wdd(P,M_i^0)\mu(dP)=V_k(\mu).
\end{equation}
It the degenerate case of $\mu$ supported on a $(k-1)$-set, we know that $V_k(\mu)=V_{k-1}(\mu)=0$ 
and the argument leading to (\ref{H2})  would give $V_k(\mu_n)\leq V_{k-1}(\mu_n)\to 0$. Therefore, 
let us  assume that $V_{k-1}(\mu)>0$ and  denote by $M_i^n$, $i=1,\dots,k$ the probabilities 
in ${\bf \bar M}_n$. Arguing as in the proof of Theorem \ref{existkbar}, the assumption $\mathcal{W}_{\mathcal{P}_2}(\mu_n, \mu)\to 0$ 
easily leads to guarantee that the sequences $\{M_i^n\}_n$ are tight. 
Therefore, any subsequence has a subsequence (for which we keep the same notation) with weakly convergent components 
$M_i^n \convw M_i$ for $i=1,\dots,k$. We write ${\bf \bar M}=\{M_1,\ldots,M_k\}$.

On the other hand, since $\mu_n \convw \mu$, we can apply Skorohod's Representation Theorem 
(see, e.g., Theorem 11.7.2 in Dudley \cite{Dudley}), and assume that there are $W_2(\Pd)$-valued random elements $Z_n,Z$, 
defined on some probability space $(\Omega, \sigma, \varUpsilon)$ with laws 
$\mathcal L(Z_n)=\mu_n, \mathcal L(Z)=\mu,$ such that $Z_n \conv Z \ \varUpsilon-$a.s. 
By lower semi-continuity (Proposition \ref{propiedades}), this leads to 
\begin{equation}\label{H6}
\Wd(Z(\omega),M_i) \leq \liminf_{n\to \infty}\Wd(Z_n(\omega), M_i^n), \ i=1,\dots,k \mbox{ for } \varUpsilon-\mbox{a. e. } \omega,
\end{equation}
thus
\begin{equation}\label{H3}
\min_{i=1,\dots,k}\Wd(Z, M_i)\leq \liminf_{n\to \infty} \min_{i=1,\dots,k}\Wd(Z_n, M_i^n) \ \varUpsilon-\mbox{ a.s..}
\end{equation} 
From this, Fatou's theorem, (\ref{H0}) and (\ref{H2}) we get
\begin{eqnarray}\nonumber
\mathcal{W}^2_{\mathcal{P}_2}(\mu, \Prob(\bar{\bf M}))=\int \min_{i\in\{1,\dots,k\}}\Wdd(P,M_i)\mu(dP)\leq 
\liminf_{n\to \infty}\int \min_{i=1,\dots,k}\Wd(P, M_i^n)\mu_n(dP) \\ 
\label{H5} \leq \limsup_{n\to \infty}\int \min_{i=1,\dots,k}\Wd(P, M_i^n)\mu_n(dP)\leq  
\lim_{n\to \infty}\int \min_{i=1,\dots,k}\Wd(P, M_i^0)\mu_n(dP)=V_k(\mu),
\end{eqnarray}
hence $V_k(\mu_n) \to V_k(\mu)$ and any weak limit ${\bf \bar M}$ of a weakly 
convergent subsequence of ${\bf \bar M}_n$ is a $k$-barycenter of $\mu$.

It only remains to show that, in fact, these weakly convergent subsequences are convergent (through
subsequences) in the $\Wd$-sense. 
For this, observe that inequalities in (\ref{H5}) are, in fact, equalities. But then (\ref{H3}) must be also an equality.
From this, taking into account that the support of $\mu$ is not degenerated in a $(k-1)-$set and (\ref{H6}),  
we conclude that the sets $\Omega_i:=\left\{\omega: \Wd(Z(\omega),M_i) =\liminf_{n\to \infty} 
\Wd(Z_n(\omega),M_i^n)\right\}$ have positive probability, $\varUpsilon(\Omega_i)>0, i=1,\dots,k.$
Choose any subsequence and take $\omega \in \Omega_i$ satisfying also $Z_n(\omega)\to Z(\omega)$. 
Then there exists a new subsequence for which we additionally have $\Wd(Z_n(\omega),M_i^n)$ $
\to \Wd(Z(\omega),M_i)$. But then, by Lemma 14 in \cite{Le Gouic}, $\mathcal{W}_2 (M_i^n,M_i)\to 0$. This shows that
from every subsequence we can extract a further subsequence such that $d_H({\bf \bar M}_n, {\bf \bar M})\to 0$. All the other
claims follow from this fact.
\FIN

When $\mu_n$ are the sample distributions obtained from $n$ realizations, $P_1,\dots,P_n$, of the random probability measure $\mu \in W_2(\Pd),$  
Varadarajan's Theorem guarantees that $\mu_n \convw \mu$ almost surely. Taking the probability degenerated at zero, 
$\delta_{\{0\}},$  the classical Strong Law of Large Numbers applied to the  i.i.d. random variables $\Wdd(P_i, \delta_{\{0\}})$ states 
$$
\int_{\Pd}\Wdd(P, \delta_{\{0\}})\mu_n(dP)=\frac 1 n \sum_{i=1}^n\Wdd(P_i, \delta_{\{0\}})\convs \int_{\Pd}\Wdd(P, \delta_{\{0\}})\mu(dP),
$$
hence the characterization in Theorem \ref{convW2} of convergence in the $\mathcal{W}_{\mathcal{P}_2}$ sense, and 
Theorem \ref{consistencykbar}, prove the  Strong Law of Large Numbers for $k$-barycenters. 

\begin{Theo}\label{consistency}
Assume that $\mu \in W_2(\Pd)$.  If $\mu_n$ is the sample probability giving mass $1/n$ to the 
probabilities $P_1,\ldots,P_n$ obtained as independent realizations of $\mu$, then $V_k(\mu_n) 
\conv V_k(\mu)$ a.s.. If  the $k$-barycenter of $\mu$, ${\bf \bar M}$, is unique and ${\bf \bar M}_n$ is a 
sample $k$-barycenter, then the $k$-barycenters are consistent, i.e. ${\bf \bar M}_n\convs {\bf \bar M}$ in Hausdorff
distance.
\end{Theo}

Once the existence and consistency for $k$-barycenters have been proved, the adaptation to cover the trimmed versions relies on the same arguments as those employed in Section 5.5 of \cite{preprint2}. Therefore we omit the proofs for Proposition \ref{caract} and Theorems \ref{consistencytypeLoubes} and \ref{coro}. In particular it must be stressed that, in the trimmed setting, the integrability condition on the $\mu_n$ and $\mu$ probability measures is unnecessary. In contrast we  give the following proof because it involves a different way of looking at clustering of clusters.
\vspace{4mm}

\noindent
\textbf{Proof of Theorem \ref{consist2}.}
We write $r=r(\eta)$ and note that $H>2$. Since 
$$\bigcup_{i=1}^k\left(\mathbb F(  {\hat P}_j)\cap B(N_i,r)=\emptyset\right)=
\Big(\max_{1\leq i\leq k}\min_{1\leq l\leq k} \mathcal{W}_2(\hat{N}_l^j,N_i)>r\Big)\subset \Big(d_H(\mathbb{F}(\hat{P}_j),\mathbb{F}({P}))>r \Big),$$
we see that, 
\begin{equation}\label{cover}
\mbox{Pr}\left[\bigcup_{i=1}^k\left(\mathbb F( {\hat P}_j)\cap B(N_i,r)=\emptyset\right)\right]\leq \mbox{Pr}\left[\mathcal{W}_{2}( {\hat P}_j ,  
P)>\eta\right]<\frac{\alpha}{2}, \ j=1,\dots.
\end{equation} 
Let us set now $X_{i}^j=I\{\sharp\big(\mathbb F( \mathbb {\hat P}_j)\cap B(N_i,r)\big)=1\}$. Observe that, for fixed
$i$, $X_1^1,\ldots,X_i^{m}$ are independent Bernoulli random variables. Call $p_i^j=\mbox{Pr}(X_i^j=1)$. The balls $B(N_i,r)$ are
pairwise disjoint (recall that $H>2$) and therefore, $\left[\bigcup_{i=1}^k\left(\mathbb F( {\hat P}_j)\cap B(N_i,r)=\emptyset\right)\right]^C
\subset (X_i^j=1)$, which implies that $p_{i}^j> 1-\frac{\alpha}{2}$. But then, if $B_k$ denotes a binomial r.v. with parameters
$m$ and $1-\frac \alpha{2},$ we see from Hoeffding's inequality that $\mbox{Pr}\big(\sum_{j=1}^m X_i^j<m(1-\alpha ) \big)\leq 
\mbox{Pr}\big(B_k<m(1-\alpha) \big)\leq e^{-\frac{\alpha^2 }{2}m}$. As a consequence,
\begin{eqnarray}\label{enoughalt}
\lefteqn{\mbox{Pr}\bigg\{\bigcup_{i=1}^k\bigg[\#\bigg(\bigcup_{j=1}^m \mathbb F(  {\hat P}_j)\cap B(N_i,r)\bigg)<m(1-\alpha)\bigg]  
\bigg\}}\hspace*{6cm}\\
\nonumber
&\leq & \mbox{Pr}\big(\sum_{j=1}^m X_i^j<m(1-{\textstyle \alpha}) \big)\leq  ke^{-\frac{\alpha^2 }{2}m}.
\end{eqnarray}

From (\ref{enoughalt}) we see that, with probability at least $1-ke^{-\frac{\alpha^2 }{2}m}$, each ball $B(N_i,r)$ contains,
at least, $m(1-\alpha)$ sample points $\hat{N}_i^j$ and, in particular, the number of points outside $\cup_{i=1}^k
B(N_i,r)$ is less than $\alpha k m$. Hence, the optimal $\alpha$-trimmed $k$-variation is upper bounded by $r^2$ 
(take a trimming of the empirical measure concentrated on $\cup_{i=1}^k
B(N_i,r)$ and use the $N_i$'s as centers). To ease notation let us write $\mathbf{M}=\mathbf{M}_{n_1,\ldots,n_m; m}$
and $\hat{r}_\alpha$ for the associated optimal trimming radius as in Proposition \ref{caract}. Since, by assumption, $\alpha k m$
is an integer, we can assume that the optimal trimming function, $\tau_\alpha$ takes values in $\{0,1\}$. We write $\tilde{B}(\mathbf{M},\hat{r}_\alpha)
=\{x\in\bar{B}(\mathbf{M},\hat{r}_\alpha): \, \tau_\alpha (x)=1 \}$ and similarly for
$\tilde{B}(N,\hat{r}_\alpha)$ for $N\in \mathbf{M}$. Now, the number of points outside $\tilde{B}(\mathbf{M},\hat{r}_\alpha)$ is $\alpha k m$.
Hence, each ball $B(N_i,r)$ contains, at least, $m(1-\frac\delta k -\alpha k)=m(1-(k+1)\alpha)$ points in $\tilde{B}(\mathbf{M},\hat{r}_\alpha)$.
Let us focus on the ball $B(N_1,r)$. There exists $N\in\mathbf{M}$ such that $\tilde{B}(N,\hat{r}_\alpha)$ contains at least a fraction
$\frac 1 k$ of the points in $B(N_1,r)\cap  \tilde{B}(\mathbf{M},\hat{r}_\alpha)$. But then $C:=B(N_1,r)\cap \tilde{B}(N,\hat{r}_\alpha) $ contains at least 
$\frac m k (1-(k+1)\alpha)$ points and, from equation (\ref{descomp}), denoting $s=\mathcal{W}_2(N,B(N_1,r))$ we conclude
$$r^2\geq 
\frac 1 {km(1-\alpha)}\sum_{i,j: \hat N_{i}^j \in C} {\cal W}_{2}^2 (\hat N_i^j,N) \geq s^2 \frac{(1-\alpha (k+1))m}{k^2m(1-\alpha)}$$
and, therefore,
$$\mathcal{W}_2(N,B(N_1,r)) \leq rk\left(\frac{1-\alpha}{1-\alpha (k+1)} \right)^{1/2}.$$
This, in turn, implies that $\mathcal{W}_2(N,N_1)\leq r \frac H 2 $ (recall that $H=2 \big(1+k\big(\frac{1-\alpha}{1-\alpha (k+1)} \big)^{1/2}\big)$).
Observe that the choice of $r$ guarantees that the balls $\bar{B}(N_i, r\frac H 2)$ are disjoint. 
Since the choice of $N_1$ was arbitrary, we conclude that for every $i=1,\ldots,k$ there exists $\tilde{N}_i\in \mathbb{M}$ such that
$W_2(N_i,\tilde{N}_i)\leq r\frac H 2$. The fact that the balls $\bar{B}(N_i, r\frac H 2)$ are disjoint ensures that 
$\{\tilde{N}_1,\ldots,\tilde{N}_k\}$ is just a relabeling of $\mathbf{M}$ and, as a consequence that, with probability at least $1-k e^{-\frac {\alpha^2} 2 m}$,
$$d_H(\mathbf{M}, \{N_1,\ldots,N_k\})\leq \frac {rH}2.$$
This completes the proof.
 \FIN
 
 Next we prove that the algorithm  introduced in Section \ref{computation} converges when applied to a finite set of absolutely continuous or to discrete (with finite support) probabilities. This will be the consequence of the following propositions.

\begin{Prop} \label{Prop.cont}
Let $P_i, \ i = 0,\ldots,r$ be probability measures in ${\cal P}_2(\Rea^d)$ with associated weights $w_i, \ i = 0,\ldots,r$, with $\sum_{i=0}^rw_i=1.$ Let  $\overline P$ be the barycenter of  $P_i, \ i = 1,\ldots,r$ with weights $w_i^*=w_i(\sum_{i=1}^rw_i)^{-1}, \ i = 1,\ldots,r$ and $\overline P_0$  be the barycenter of  $P_i, \ i = 0,\ldots,r$. If we assume that $w_0>0 $, that $P_0\neq \overline P$, and that $\overline P$ is absolutely continuous, then $\overline P_0 \neq \overline P$.
\end{Prop}
\medskip
\noindent
{\bf Proof:}
Since $\overline P$ is absolutely continuous,   there exist maps $T_i: \Rea^d \to \Rea^d, i=0,\ldots, k$ pushing forward $\overline P$ to $P_i$, such that
$
{\cal W}^2(\overline P,P_i)= \int_{\mathbb{R}^d} \| x - T_i(x)\|^2 d \overline P .
$
Moreover, by Proposition 3.3  in \cite{preprint}, it must be 
$$
 {\sum_{i=1}^r w_i^* T_i(x)}=x, \ \overline P-\mbox{a.s.}
$$
 On the other hand, 
$$
{\sum_{i=0}^r w_i T_i(x)}=
\sum_{i=1}^r w_i  x
+
{w_0}T_0(x) =: T^*(x), 
\ \overline P-\mbox{a.s.}
$$
 
 Since $T_0$ is not the identity $\overline P$-a.s., and $w_0>0$, it is clear that $\overline P\{T_0(x) \neq x\} >0$, and consequently (see the  proof of Proposition 3.3 in \cite{preprint}) we get:
 \begin{eqnarray*}
 \sum_{i=0}^r  w_i{\cal W}^2(\overline P,P_i) 
 &=&
 \sum_{i=0}^r  w_i \int \|x -T_i(x) \|^2 d\overline P
 >
 \sum_{i=0}^r  w_i \int \| T^* (x) -T_i(x) \|^2 d\overline P
 \\
 &\geq&
 \sum_{i=0}^r  w_i  {\cal W}^2(P_{T^*(x)},P_i)
 \geq
 \sum_{i=0}^r  w_i  {\cal W}^2(\overline P_0,P_i). \hspace{4.5cm}
\bullet
 \end{eqnarray*}

Recall that the condition $\overline P$ is absolutely continuous is verified if we know that $P_i, \ i = 1,\ldots,r$ are absolutely continuous. Concerning  the discrete case, let us retain the notation of the previous proposition and let $\{a^i_1\ldots, a^i_{h_i}\}$ be the support of $P_i, i=0,1,\dots,r$. As stated in  Lemma \ref{Lem.discreto} below,    $\overline P$ also has a finite support, say $\{b_1, \ldots, b_h\}$.  In this case, the optimal coupling between $\overline P$ and $P_i, \ i=0,\ldots,r$, is determined by a family of positive  numbers $p^i_{s,t} , \ s=1,\ldots, h; \ t=1,\ldots, h_i$ which satisfy
\[
 \sum_{i=1}^r p^i_{s,t} = \overline P[b_s]
 \mbox{ and }
 \sum_{s=1}^h p^i_{s,t} = P_i[a_t^i] .
\]

Slight modifications of the proof of  Theorem 1 in  \cite{Anderes} lead to the  following  lemma.

\begin{Lemm}Ê\label{Lem.discreto}
With the notation above, we have that, 
\begin{enumerate}
\item
The support of $\overline P$ is finite.

\item
For every $s=1,\ldots,h$, and $i=1,\ldots, r$, it holds: $\# \{s: p^i_{s,t} >0 \}=1$. Moreover, if 
 $k_i$ denotes the only index such that $p^i_{s,k_i} >0$, then
\[
b_s = \sum_i w_i^* a^i_{k_i}, \ \mbox{ and } P_i[a^i_{k_i}] \geq \overline P[b_s].
\]
\item
There exist maps $T_i: \Rea^d \to \Rea^d$ such that, if the distribution of $X$  is $\overline P$, then the distribution of $T_i(X)$ is $P_i$ and $
{\cal W}^2(\overline P,P_i)= \int_{\mathbb{R}^d} \| x - T_i(x)\|^2 d \overline P .
$
\end{enumerate}

\end{Lemm}

Item 3 in Lemma \ref{Lem.discreto} allows to repeat the proof of Proposition \ref{Prop.cont} to obtain:

\begin{Prop} \label{Prop.discret}
Assume the same notation as in the previous proposition, and let  $P_i, \ i = 0,\ldots,r$ be  probability measures with finite support. If  $w_0>0 $ and $P_0\neq \overline P$, then $ \overline P_0 \neq \overline P$.
\end{Prop}

\begin{Prop}\label{convalg}
Let $k\in \Nat$ fixed, and consider a family of probabilities $P_1, \ldots, P_r$ with weights $w_i\geq 0$ and $\sum_i w_i=1$. The proposed algorithm to compute trimmed $k$-barycenters with weights converges to a local minimum of the objective function given in (13) either if the involved probabilities are absolutely continuous or if all their supports are finite.
\end{Prop}
\medskip
\noindent
{\bf Proof:} 
Notice that   Step 4 in the algorithm provides a partition of the set $\{1,\ldots, r\}$ in the finite family of  subsets
\begin{eqnarray*}
T^n & = & \{ i: w^n_i=0 \}
\\
R_j^n & = & \{ i: w^n_i=1 \mbox{ and }  g_i^n = j\}, \ j=1,\ldots,k
\\
Q_j^n & = & \{ i: 0 < w^n_i < 1 \mbox{ and }  g_i^n = j\},\ j=1,\ldots,k.
\end{eqnarray*}
Thus, $T^n$ contains the indices of the probabilities which are completely trimmed and $R_j^n $ those of the probabilities which are not trimmed and are associated to  group $j$. After this, at most one index  $i_q$ remains. In such a case, it belongs to one of the sets $Q_j^n$, while the remaining sets are empty. Moreover, once the two sets $\cup_j R_j^n$ and $T^n$ have been fixed, the value of $w_{i_q}^n$  is also fixed, and, then, only the index $g_{i_q}^n$ can vary. Thus, Step 4 only has a finite number of possibilities. Therefore, if we show that each time we run steps 2 and 3 the value of the objective function strictly decreases, the result will be proved because this implies that we cannot visit twice the same solution.
But this is trivial because, if the probabilities are absolutely continuous, then Proposition \ref{Prop.cont} implies that if the stopping condition is not fulfilled, then at least a barycenter $P_j^{n+1}$ will vary and we will have a reduction of the objective function. The same happens using Proposition \ref{Prop.discret} if the supports of the probabilities are finite. 
\FIN


\vspace{-1cm}\begin{thebibliography}{99}
\small

\bibitem{Agueh10}
\textsc{Agueh, M.} and  \textsc{Carlier, G.} (2011).  Barycenters in the Wasserstein space. {\it SIAM
J. Math. Anal.}, 43 (2), 904--924.

\bibitem{Anderes}
\textsc{Anderes, E., Borgwardt, S., and Miller, J.} (2016). Discrete Wasserstein Barycenters: Optimal Transport for Discrete Data. \textit{Mathematical Methods of Operations Research}, 84, 389--409. 




\bibitem{preprint}
\textsc{\'Alvarez-Esteban, P. C., del Barrio, E., Cuesta-Albertos, J. A.,} and \textsc{Matr\'an, C.} (2016). 
A fixed-point approach to barycenters in Wasserstein space. {\it Jour. Math. Anal. and Appl.}  441(2), 744--762



\bibitem{preprint2}
\textsc{\'Alvarez-Esteban, P. C., del Barrio, E., Cuesta-Albertos, J. A.,} and \textsc{Matr\'an, C.} (2016). 
A Wide  Consensus approach to aggregation in the Wasserstein Space. Preprint.

\bibitem{Tasio}
\textsc{del Barrio, E., Lescornel, H.,} and \textsc{Loubes, J.M.} (2015). A statistical analysis of a 
deformation model with Wasserstein barycenters: estimation procedure and goodness of fit test. 
Preprint. http://arxiv.org/abs/1508.06465

\bibitem{TR}
\textsc{del Barrio, E., Cuesta-Albertos, J. A.,} and \textsc{Matr\'an, C.} (2016). Profiles of Pyramid Ages in American countries: A trimmed $k$-barycenters approach. Technical Report.

\bibitem{Benamou}
\textsc{Benamou, J. D., Carlier, G., Cuturi, M., Nenna, L.,} and \textsc{Peyre, G.} (2015). Iterative Bregman 
projections for regularized transportation problems. {\it SIAM J. Sci. Comput.}, 37(2), 1111--1138.


\bibitem{Bigot} \textsc{Bigot, J.} and \textsc{Klein, T.} (2015). Consistent estimation of a population barycenter 
in the Wasserstein space. ArXiv e-prints, arXiv:1212.2562v5, March 2015.

\bibitem{Bigot2}
\textsc{Bigot, J., Gouet, R., Klein, T., } and \textsc{ L\'opez, A. }(2013). Geodesic PCA in the 
Wasserstein space by Convex PCA. To appear in {\it Ann. Inst. Henri Poincar\'e, Probab. Statist.}

\bibitem{Bois15}
\textsc{Boissard, E., Le Gouic, T. }and \textsc{Loubes, J-M.} (2015).  Distribution's template 
estimate with Wasserstein metrics. {\it Bernoulli}, 21(2), 740--759. 

\bibitem{Breiman}
\textsc{Breiman, L.} (1996) Bagging predictors. {\it Machine Learning}, 24,
123--140.

\bibitem{Buhlmann}
\textsc{B\"uhlmann, P.} (2012). Bagging, Boosting and Ensemble Methods. In
{\it Handbook of Computational Statistics: Concepts and Methods.} Eds.
Gentle, E.J.,  H\"ardle, K.W. and Mori, Y., pp. 985--1022. Springer. Berlin



\bibitem{Carlier}
\textsc{Carlier, G., Oberman, A.} and \textsc{Oudet, E.} (2015). Numerical methods for 
matching for teams and Wasserstein barycenters, {\it ESAIM Math. Model. Numer.Anal.}, 49(6), 1621--1642.

\bibitem{Carlier2}
\textsc{Carlier, G., Chernozhukov, V., and Galichon, A.} (2015). Vector Quantile Regression: 
An Optimal Transport Approach. Ann. Statist., to appear.

\bibitem{Depth}
\textsc{Chernozhukov, V., Galichon, A., Hallin, M.,} and \textsc{Henry, M.} (2014). 
Monge-Kantorovich Depth, Quantiles, Ranks, and Signs. Ann. Statist., to appear.

\bibitem{Cuesta97}
\textsc{Cuesta-Albertos, J. A.}, \textsc{Gordaliza, A.,} and \textsc{Matr\'an, C.} (1997). 
Trimmed k-means: An attempt to robustify quantizers. {\it Ann. Statist.}, 25(2), 553--576.

\bibitem{Cuesta-Fraiman}
\textsc{Cuesta-Albertos, J. A.,} and \textsc{Fraiman, R.} (2007). Impartial trimmed k-means 
for functional data. {\it Computational Statistics and Data Analysis}, 51(10), 4864--4877. 

\bibitem{Cues88}
\textsc{Cuesta-Albertos, J. A.} and \textsc{Matr\'an, C.} (1988)
The Strong Law of Large Numbers for $k$-means and best possible nets of Banach valued random variables.
{\it Probab. Theo. Related Fields} 78, 523--534 




\bibitem{Cuturi}
\textsc{Cuturi, M.} and \textsc{Doucet, A.} (2014). Fast computation of Wasserstein barycenters, in {\it Proceedings of the 31st International Conference on Machine Learning, Beijing, China, 2014}. JMLR: W\&CP vol 32.

\bibitem{Delicado}
\textsc{Delicado, P.} (2011). Dimensionality reduction when data are density functions. {\it Computational Statistics and Data Analysis}, 55(1), 401--420. 


\bibitem{DobricYukich95} \textsc{Dobric, V.} and \textsc{Yukich, J.E.} (1995). 
Asymptotics for transportation cost in high dimensions. \textit{J. Theor. Probab.}, \textbf{8} 97--118.

\bibitem{Dudley}
\textsc{Dudley, R. M.} (1989). {\it Real Analysis and Probability}. Wadsworth \& Brooks.


\bibitem{Dudoit}
\textsc{Dudoit, S.}, and \textsc{Fridlyand, J.} (2003). Bagging to improve
the accuracy of a clustering procedure. {\it Bioinformatics}, 19(9),
1090--1099. 


\bibitem{Flury}
\textsc{Flury, B. }(1993) 
Estimation of principal points.
{\it Appl. Statist.}, 42(1), 139--151


\bibitem{Fritz}
\textsc{Fritz, H., Garc\'{\i}a-Escudero, L. A.,} and \textsc{Mayo-Iscar, A.} (2012). tclust: An R Package for a Trimming Approach to Cluster Analysis. \textit{Journal of Statistical Software}, 47(12). 




\bibitem{r11} \textsc{Gallegos, M.T. and Ritter, G.} (2005), A robust method for cluster analysis,
\textit{Ann. Statist.}, {33}, 347-380.

\bibitem{splines}
\textsc{Garc\'{\i}a-Escudero, L. A., }and \textsc{Gordaliza, A.} (2005). A Proposal for Robust Curve Clustering. 
{\it Journal of Classification,} 22(2), 185--201.

\bibitem{tools}
\textsc{Garc\'{\i}a-Escudero, L. A., Gordaliza, A., and Matr\'an, C.} (2003). Trimming Tools in Exploratory Data Analysis. {\it Journal of Computational and Graphical Statistics}, 12(2), 434--449. 

\bibitem{TCLUST}
\textsc{Garc\'{\i}a-Escudero, L. A., Gordaliza, A., Matr\'an, C.,} and \textsc{Mayo-Iscar, A.} (2008). 
A general trimming approach to robust cluster analysis. {\it Ann. Statist.}, 36(3), 1324--1345.

\bibitem{Garcia2011}
\textsc{Garc\'{\i}a-Escudero, L. A., Gordaliza, A., Matr\'an, C.,} and
\textsc{Mayo-Iscar, A.} (2011). Exploring the number of groups in robust
model-based clustering. {\it Statistics and Computing},  21, 585--599

\bibitem{Garcia2015}
\textsc{Garc\'{\i}a-Escudero, L. A., Gordaliza, A., Matr\'an, C.,} and
\textsc{Mayo-Iscar, A.} (2015). Avoiding Spurious Local Maximizers in
Mixture Modeling. {\it Statistics and Computing} 25,  619--633











\bibitem{Gasser}
\textsc{Kneip, A.}, and \textsc{Gasser,} T. (1992). 
Statistical Tools to analyze Data Representing a Sample of Curves. {\it Ann. Statistics}, 20(3), 1266--1305.

\bibitem{Hennig}
\textsc{Hennig, C; Meila, M.; Murtagh, F.; and Rocci, R.} Eds. (2016). {\it Handbook of Cluster Analysis.}  Chapman and Hall/CRC



\bibitem{Le Gouic}
\textsc{Le Gouic, T.} and \textsc{Loubes, J.M.} (2016). Existence and consistency of Wasserstein barycenters. To appear in  {\it Probab. Theo. Related Fields } 



\bibitem{Leisch}
\textsc{Leisch, F.}(1999) Bagged clustering. Technical report. 
http://www.ci.tuwien.ac.at/ ?leisch/papers/ fl-techrep.html

\bibitem{Lember}
\textsc{Lember, J.} (2003). On minimizing sequences for k-centres. {\it Journal of Approximation Theory, }120(1), 20--35. 

\bibitem{Lo}
\textsc{Lo, K., Brinkman, R. R., and Gottardo, R.} (2008). Automated gating of flow cytometry data via robust model-based clustering. {\it Cytometry. Part A : The Journal of the International Society for Analytical Cytology}, 73(4), 321Ð32. doi:10.1002/cyto.a.20531

\bibitem{Luschgy}
\textsc{Luschgy, H.,} and \textsc{Pag\`es, G.} (2002). Functional quantization of Gaussian processes. Journal of Functional Analysis, 196, 486--531.

\bibitem{Parna2}
\textsc{P\"arna, K.} (1986). Strong consistency of k-means clustering criterion, {\it Acta Comm. Univ. Tartuensis 733} 86--96.



\bibitem{Parna}
\textsc{P\"arna, K.} (1990). On the existence and weak convergence of k-centres in Banach spaces, {\it Acta Comm. Univ. Tartuensis  893}, 17--28.


\bibitem{Pyne2}
\textsc{Pyne, S., Hu, X., Wang, K.,} et al. (2009). Automated
high-dimensional flow cytometric data analysis. {\it Proceedings of the
National Academy of Sciences of the United States of America}, 106(21),
8519--8524. 


\bibitem{Pyne1}
\textsc{Pyne, S., Lee, S. X., Wang, K., Irish, J., Tamayo, P., Nazaire, M.
D., ... \& Nolan, G. P.} (2014). Joint modeling and registration of cell
populations in cohorts of high-dimensional flow cytometric data. {\it PloS
one}, 9(7), e100334.


\bibitem{Sverdrup}
\textsc{Sverdrup-Thygeson, H.}, 1981. Strong law of large numbers for measures of central tendency and dispersion of random variables in compact metric spaces. {\it Ann. Statist.} 9 (1), 141--145.



\bibitem{Villani2} \textsc{Villani, C.} (2008). {\it Optimal Transport: Old and New}, Vol. 338. Springer Science \&
Business Media.

\end{thebibliography}
 \end{document}